\documentclass[twocolumn]{aastex63}

\newcommand{\todo}{\ifmmode \text{\color{red}\Huge{\(\bullet\)}} \else {\color{red}{\Huge$\bullet$}}\fi}
\newcommand{\tido}{\ifmmode {{\color{red}\bullet}} \else {\color{red}$\bullet$}\fi}

\newcommand{\E        }[1]{\ifmmode 10^{#1} \else $10^{#1}$\fi}
\newcommand{\tE        }[1]{\ifmmode \times10^{#1} \else $\times10^{#1}$\fi}
\newcommand{\til}{\ifmmode \sim \else $\sim$\fi}

\newcommand{\et}{et al.\ }

\newcommand{\pc}	{\ifmmode {\rm pc} \else pc\fi}
\newcommand{\kpc}	{\ifmmode {\rm kpc} \else kpc\fi}
\newcommand{\ld}	{\ifmmode {\rm l.d.} \else l.d.\fi}
\newcommand{\kms}	{\ifmmode {\rm km\,s}^{-1} \else km\,s$^{-1}$\fi}
\newcommand{\cc}	{\ifmmode {\rm cm}^{-3}    \else cm$^{-3}$\fi}
\newcommand{\cmii}	{\ifmmode {\rm cm}^{-2}    \else cm$^{-2}$\fi}
\newcommand{\ergs}	{\ifmmode {\rm erg\,s}^{-1} \else erg s$^{-1}$\fi}
\newcommand{\ergcms}	{\ifmmode {\rm erg\,cm}^{-2}\,{\rm s}^{-1} \else erg\,cm$^{-2}$\,s$^{-1}$\fi}
\newcommand{\ergcmsA}	{\ifmmode {\rm erg\,cm}^{-2}\,{\rm s}^{-1}\,{\rm\AA}^{-1}
\else erg\,cm$^{-2}$\,s$^{-1}$\,\AA$^{-1}$\fi}
\newcommand{  \ergcmsHz  }{\ifmmode{\rm erg\,cm}^{-2}\,{\rm s}^{-1}\,{\rm Hz}^{-1}
                       \else ergs\,cm$^{-2}$\,s$^{-1}$\,Hz$^{-1}$\fi}
\newcommand{\kev}	{\ifmmode {\rm keV} \else keV\fi}

\newcommand{\mic}	{\ifmmode {\rm \mu m} \else $\mu$m\fi}
\newcommand{\vFWHM}	{\ifmmode v_{\mbox{\tiny FWHM}} \else $v_{\mbox{\tiny FWHM}}$\fi}
\newcommand{\vBLR}	{\ifmmode v_{\mbox{\tiny BLR}} \else $v_{\mbox{\tiny BLR}}$\fi}
\newcommand{\sigBLR}	{\ifmmode \sigma_{\mbox{\tiny BLR}} \else $\sigma_{\mbox{\tiny BLR}}$\fi}
\newcommand{\vNLR}	{\ifmmode v_{\mbox{\tiny NLR}} \else $v_{\mbox{\tiny NLR}}$\fi}
\newcommand{\tauBLR}	{\ifmmode \tau_{\mbox{\tiny BLR}} \else $\tau_{\mbox{\tiny BLR}}$\fi}

\newcommand{\Hubble}	{\ifmmode {\rm km\,s}^{-1}\,{\rm Mpc}^{-1} \else km\,s$^{-1}$\,Mpc$^{-1}$\fi}
\newcommand{\NDunit}	{\ifmmode {\rm Mpc}^{-3} \else Mpc$^{-3}$\fi}
\newcommand{\LFunit}	{\ifmmode {\rm Mpc}^{-3}\,{\rm mag}^{-1} \else Mpc$^{-3}$\,mag$^{-1}$\fi}
\newcommand{\MFunit}	{\ifmmode {\rm Mpc}^{-3}\,{\rm dex}^{-1} \else Mpc$^{-3}$\,dex$^{-1}$\fi}

\newcommand{\Msun}{\ifmmode M_{\odot} \else $M_{\odot}$\fi}
\newcommand{\Lsun}{\ifmmode L_{\odot} \else $L_{\odot}$\fi}
\newcommand{\Zsun}{\ifmmode Z_{\odot} \else $Z_{\odot}$\fi}
\newcommand{\mpyr}{\ifmmode \Msun\,{\rm yr}^{-1} \else $\Msun\,{\rm yr}^{-1}$\fi}

\newcommand{\Msol}{\Msun}

\newcommand{\qnote}{\ifmmode q_{0} \else $q_{0}$\fi}
\newcommand{\Hnote}{\ifmmode H_{0} \else $H_{0}$\fi}
\newcommand{\hnote}{\ifmmode h_{0} \else $h_{0}$\fi}
\newcommand{\anote}{\ifmmode a_{0} \else $a_{0}$\fi}
\newcommand{\tnote}{\ifmmode t_{0} \else $t_{0}$\fi}

\def\gsim{\;\rlap{\lower 2.5pt \hbox{$\sim$}}\raise 1.5pt\hbox{$>$}\;}
\def\lsim{\;\rlap{\lower 2.5pt \hbox{$\sim$}}\raise 1.5pt\hbox{$<$}\;}

\newcommand{  \Halpha   }{\ifmmode {\rm H}\alpha \else H$\alpha$\fi}

\newcommand{  \ha       }{\Halpha}
\newcommand{  \Hbeta    }{\ifmmode {\rm H}\beta \else H$\beta$\fi}

\newcommand{  \hb       }{\Hbeta}
\newcommand{  \Hgamma   }{\ifmmode {\rm H}\gamma \else H$\gamma$\fi}
\newcommand{  \Hdelta   }{\ifmmode {\rm H}\delta \else H$\delta$\fi}
\newcommand{  \Lya      }{\ifmmode {\rm Ly}\alpha \else Ly$\alpha$\fi}
\newcommand{  \Lyb      }{\ifmmode {\rm Ly}\beta \else Ly$\beta$\fi}
\newcommand{  \Pa       }{\ifmmode {\rm P}\alpha \else P$\alpha$\fi}
\newcommand{  \Pb       }{\ifmmode {\rm P}\beta \else P$\beta$\fi}
\newcommand{  \Bra      }{\ifmmode {\rm Br}\alpha \else Br$\alpha$\fi}
\newcommand{  \Brg      }{\ifmmode {\rm Br}\gamma \else Br$\gamma$\fi}
\newcommand{  \hii      }{\ifmmode {\rm H}\,\textsc{ii} \else H\,\textsc{ii}\fi}
\newcommand{  \hei      }{\ifmmode {\rm He}\,\textsc{i} \else He\,\textsc{i}\fi}
\newcommand{  \heii     }{\ifmmode {\rm He}\,\textsc{ii} \else He\,\textsc{ii}\fi}
\newcommand{  \HeIIuv   }{\ifmmode {\rm He}\,\textsc{ii}\,\lambda1640 \else He\,\textsc{ii}\,$\lambda1640$\fi}
\newcommand{  \HeIIop   }{\ifmmode {\rm He}\,\textsc{ii}\,\lambda4686 \else He\,\textsc{ii}\,$\lambda4686$\fi}
\newcommand{  \CII	}{\ifmmode \left[{\rm C}\,\textsc{ii}\right]\,\lambda157.74\,\mu{\rm m} \else [C\,{\sc ii}]\ $\lambda157.74\,\mu{\rm m}$\fi}
\newcommand{  \cii	}{\ifmmode \left[{\rm C}\,\textsc{ii}\right] \else [C\,{\sc ii}]\fi}

\newcommand{  \ciii     }{\ifmmode {\rm C}\,\textsc{iii}\right] \else C\,\textsc{iii}]\fi}
\newcommand{  \CIII     }{\ifmmode {\rm C}\,\textsc{iii}\right]\,\lambda1909 \else C\,\textsc{iii}]\,$\lambda1909$\fi}
\newcommand{  \civ      }{\ifmmode {\rm C}\,\textsc{iv}  \else C\,\textsc{iv}\fi}
\newcommand{  \CIV      }{\ifmmode {\rm C}\,\textsc{iv}\,\lambda1549 \else C\,\textsc{iv}\,$\lambda1549$\fi}
\newcommand{  \NIIopt   }{\ifmmode \left[{\rm N}\,\textsc{ii}\right]\,\lambda6584 \else [N\,\textsc{ii}]\,$\lambda6584$\fi}
\newcommand{  \nii      }{\ifmmode \left[{\rm N}\,\textsc{ii}\right]  \else [N\,\textsc{ii}]\fi}
\newcommand{  \niii     }{\ifmmode {\rm N}\,\textsc{iii} \else N\,\textsc{iii}\fi}
\newcommand{  \NIII     }{\ifmmode {\rm N}\,\textsc{iii}\,\lambda4640 \else N\,\textsc{iii}\,$\lambda4640$\fi}
\newcommand{  \niv      }{\ifmmode {\rm N}\,\textsc{iv}  \else N\,\textsc{iv}\fi}
\newcommand{  \NIVuv    }{\ifmmode {\rm N}\,\textsc{iv}\,\lambda1486 \else N\,\textsc{iv}\,$\lambda1486$\fi}
\newcommand{  \nv       }{\ifmmode {\rm N}\,\textsc{v}   \else N\,\textsc{v}\fi}
\newcommand{\oi}{\ifmmode \left[{\rm O}\,\textsc{i}\right] \else [O\,{\sc i}]\fi}
\newcommand{\OI}{\ifmmode \left[{\rm O}\,\textsc{i}\right]\,\lambda6300 \else [O\,{\sc i}]$\,\lambda6300$\fi}
\newcommand{\oii}{\ifmmode \left[{\rm O}\,\textsc{ii}\right] \else [O\,{\sc ii}]\fi}
\newcommand{\OII}{\ifmmode \left[{\rm O}\,\textsc{ii}\right]\,\lambda3727 \else [O\,{\sc ii}]\,$\lambda3727$\fi}
\newcommand{\oiii}{\ifmmode \left[{\rm O}\,\textsc{iii}\right] \else [O\,{\sc iii}]\fi}
\newcommand{\OIII}{\ifmmode \left[{\rm O}\,\textsc{iii}\right]\,\lambda5007 \else [O\,{\sc iii}]\,$\lambda5007$\fi}
\newcommand{  \OIIIbf   }{\ifmmode {\rm O}\,\textsc{iii}\,\lambda3133 \else O\,\textsc{iii}\,$\lambda3133$\fi}
\newcommand{  \OIIIuv   }{\ifmmode {\rm O}\,\textsc{iii}\,\lambda1663 \else O\,\textsc{iii}\,$\lambda1663$\fi}
\newcommand{  \oiv      }{\ifmmode {\rm O}\,\textsc{iv}  \else O\,\textsc{iv}\fi}
\newcommand{  \OIVuv    }{\ifmmode {\rm O}\,\textsc{iv}\,\lambda1402  \else O\,\textsc{iv}\,$\lambda1402$\fi}
\newcommand{  \OIVIR    }{\ifmmode {\rm O}\,\textsc{iv}\,25.9\,\mu {\rm m} \else O\,\textsc{iv}\,$25.9\,\mu$m\fi}
\newcommand{  \ovi      }{\ifmmode {\rm O}\,\textsc{vi}   \else O\,\textsc{vi}\fi}
\newcommand{  \Ovi      }{\ifmmode {\rm O}\,\textsc{vi}\,\lambda1035 \else O\,\textsc{vi}\,$\lambda1035$\fi}
\newcommand{  \nei      }{\ifmmode {\rm Ne}\,\textsc{i}   \else Ne\,\textsc{i}\fi}
\newcommand{  \neii     }{\ifmmode {\rm Ne}\,\textsc{ii}  \else Ne\,\textsc{ii}\fi}
\newcommand{  \NeiiIR   }{\ifmmode {\rm Ne}\,\textsc{ii}\,12.8\,\mu {\rm m} \else Ne\,\textsc{ii}\,$12.8\,\mu$m\fi}
\newcommand{  \neiii    }{\ifmmode {\rm Ne}\,\textsc{iii} \else Ne\,\textsc{iii}\fi}
\newcommand{  \neiv     }{\ifmmode {\rm Ne}\,\textsc{iv}  \else Ne\,\textsc{iv}\fi}
\newcommand{  \nev      }{\ifmmode {\rm Ne}\,\textsc{v}   \else Ne\,\textsc{v}\fi}
\newcommand{  \NevIR    }{\ifmmode {\rm Ne}\,\textsc{v}\,24.3\,\mu {\rm m} \else Ne\,\textsc{v}\,$24.3\,\mu$m\fi}
\newcommand{  \nevi     }{\ifmmode {\rm Ne}\,\textsc{vi}  \else Ne\,\textsc{vi}\fi}
\newcommand{  \mgi      }{\ifmmode {\rm Mg}\,\textsc{i} \else Mg\,\textsc{i}\fi}
\newcommand{  \mgii     }{\ifmmode {\rm Mg}\,\textsc{ii} \else Mg\,\textsc{ii}\fi}
\newcommand{  \MgII     }{\ifmmode {\rm Mg}\,\textsc{ii}\,\lambda2798 \else Mg\,\textsc{ii}\,$\lambda2798$\fi}
\newcommand{  \sii      }{\ifmmode {\rm S}\,\textsc{ii} \else S\,\textsc{ii}\fi}
\newcommand{  \siii     }{\ifmmode {\rm S}\,\textsc{iii} \else S\,\textsc{iii}\fi}
\newcommand{  \siv      }{\ifmmode {\rm S}\,\textsc{iv} \else S\,\textsc{iv}\fi}
\newcommand{  \sili     }{\ifmmode {\rm Si}\,\textsc{i}   \else Si\,\textsc{i}\fi}
\newcommand{  \silii    }{\ifmmode {\rm Si}\,\textsc{ii}  \else Si\,\textsc{ii}\fi}
\newcommand{  \Siliv    }{\ifmmode {\rm Si}\,\textsc{iv}  \else Si\,\textsc{iv}\fi}
\newcommand{  \SilIVuv  }{\ifmmode {\rm Si}\,\textsc{iv}\,\lambda1400  \else Si\,\textsc{iv}\,$\lambda1400$\fi}
\newcommand{  \AlIII   }{\ifmmode {\rm Al}\,\textsc{iii}\,\lambda1857 \else Al\,\textsc{iii}\,$\lambda1857$\fi}
\newcommand{  \Aliii   }{\ifmmode {\rm Al}\,\textsc{iii} \else Al\,\textsc{iii}\fi}
\newcommand{  \caii     }{\ifmmode {\rm Ca}\,\textsc{ii} \else Ca\,\textsc{ii}\fi}
\newcommand{  \feii     }{\ifmmode {\rm Fe}\,\textsc{ii} \else Fe\,\textsc{ii}\fi}
\newcommand{  \feiii    }{\ifmmode {\rm Fe}\,\textsc{iii} \else Fe\,\textsc{iii}\fi}
\newcommand{  \Kalpha   }{\ifmmode {\rm K}\alpha \else K$\alpha$\fi}

\newcommand{ \Lhb   }{\ifmmode L_{\hb} \else $L_{\hb}$\fi}
\newcommand{ \Lha   }{\ifmmode L_{\ha} \else $L_{\ha}$\fi}
\newcommand{ \fwhb  }{\ifmmode {\rm FWHM}\left(\hb\right) \else FWHM(\hb)\fi}
\newcommand{\sighb  }{\ifmmode \sigma\left(\hb\right) \else $\sigma\left(\hb\right)$\fi}
\newcommand{ \ewhb  }{\ifmmode {\rm EW}\left(\hb\right) \else EW(\hb)\fi}
\newcommand{ \fwha  }{\ifmmode {\rm FWHM}\left(\ha\right) \else FWHM(\ha)\fi}
\newcommand{ \ewha  }{\ifmmode {\rm EW}\left(\ha\right) \else EW(\ha)\fi}
\newcommand{ \Lmg   }{\ifmmode L\left(\mgii\right) \else $L\left(\mgii\right)$\fi}
\newcommand{ \fwmg  }{\ifmmode {\rm FWHM}\left(\mgii\right) \else FWHM(\mgii)\fi}
\newcommand{ \Lciv  }{\ifmmode L\left(\civ\right) \else $L\left(\civ\right)$\fi}
\newcommand{ \fwciv }{\ifmmode {\rm FWHM}\left(\civ\right) \else FWHM(\civ)\fi}
\newcommand{ \fwhm  }{\ifmmode {\rm FWHM} \else FWHM\fi} 
\newcommand{ \voff  }{\ifmmode v_{\rm off} \else $v_{\rm off}$\fi} 
\newcommand{ \vmax  }{\ifmmode v_{\rm max} \else $v_{\rm max}$\fi} 

\newcommand{ \mumg  }{\ifmmode \mu\left(\mgii\right) \else $\mu\left(\mgii\right)$\fi}
\newcommand{ \fmg   }{\ifmmode f\left(\mgii\right) \else $f\left(\mgii\right)$\fi}
\newcommand{ \muciv }{\ifmmode \mu\left(\civ\right) \else $\mu\left(\civ\right)$\fi}
\newcommand{ \fciv  }{\ifmmode f\left(\civ\right) \else $f\left(\civ\right)$\fi}
\newcommand{  \auvo     }{\ifmmode \alpha_{\nu,{\rm UVO}} \else $\alpha_{\nu,{\rm UVO}}$\fi}
\newcommand{  \Ledd     }{\ifmmode L_{\rm Edd} \else $L_{\rm Edd}$\fi}
\newcommand{  \lamLlam  }{\ifmmode \lambda L_{\lambda} \else $\lambda L_{\lambda}$\fi}
\newcommand{  \lLl      }{\ifmmode \lambda L_{\lambda} \else $\lambda L_{\lambda}$\fi}
\newcommand{  \nuLnu    }{\ifmmode \nu L_{\nu} \else $\nu L_{\nu}$\fi}
\newcommand{  \nLn      }{\ifmmode \nu L_{\nu} \else $\nu L_{\nu}$\fi}
\newcommand{  \Luv      }{\ifmmode L_{1350} \else $L_{1350}$\fi}
\newcommand{  \Lop      }{\ifmmode L_{5100} \else $L_{5100}$\fi}
\newcommand{  \lLop     }{\ifmmode \log\left(\Lop/\ergs\right) \else $\log\left(\Lop/\ergs\right)$\fi}
\newcommand{  \Lthree   }{\ifmmode L_{3000} \else $L_{3000}$\fi}
\newcommand{  \lLthree  }{\ifmmode \log\left(\Lthree/\ergs\right) \else $\log\left(\Lthree/\ergs\right)$\fi}
\newcommand{  \Lsix      }{\ifmmode L_{6200} \else $L_{6200}$\fi}
\newcommand{  \lLisx     }{\ifmmode \log\left(\Lop/\ergs\right) \else $\log\left(\Lop/\ergs\right)$\fi}
\newcommand{  \Lxray    }{\ifmmode L_{\rm X} \else $L_{\rm X}$\fi}

\newcommand{  \Lhard    }{\ifmmode L_{\rm 2-10} \else $L_{\rm 2-10}$\fi}
\newcommand{  \Lsoft    }{\ifmmode L_{\rm 0.5-2} \else $L_{\rm 0.5-2}$\fi}

\newcommand{\Fthree}{\ifmmode F_{3000} \else $F_{3000}$\fi}
\newcommand{\fuv}{\ifmmode f_{\lambda}\left(1450{\rm \AA}\right) \else $f_{\lambda}\left(1450 {\rm \AA}\right)$\fi}
\newcommand{\fthree}{\ifmmode f_{\lambda}\left(3000{\rm \AA}\right) \else $f_{\lambda}\left(3000{\rm \AA}\right)$\fi}
\newcommand{\fH}{\ifmmode f_{\lambda}\left(1.65\micron\right) \else
$f_{\lambda}\left(1.65\micron\right)$\fi}

\newcommand{\fbol}{\ifmmode f_{\rm bol} \else $f_{\rm bol}$\fi}
\newcommand{\fbolwv}{\ifmmode f_{\rm bol}\left(\lambda\right) \else $f_{\rm bol}\left(\lambda\right)$\fi}
\newcommand{\fbolopt}{\ifmmode f_{\rm bol}\left(5100{\rm \AA}\right) \else $f_{\rm bol}\left(5100{\rm \AA}\right)$\fi}
\newcommand{\fbolthree}{\ifmmode f_{\rm bol}\left(3000{\rm \AA}\right) \else $f_{\rm bol}\left(3000{\rm \AA}\right)$\fi}
\newcommand{\fboluv}{\ifmmode f_{\rm bol}\left(1450{\rm \AA}\right) \else $f_{\rm bol}\left(1450{\rm \AA}\right)$\fi}

\newcommand{\fbolbat}{\ifmmode f_{\rm bol}\left(14-150\,\kev\right) \else $f_{\rm bol}\left(14-150\,\kev\right)$\fi}
\newcommand{\fbolhard}{\ifmmode f_{\rm bol}\left(2-10\,\kev\right) \else $f_{\rm bol}\left(2-10\,\kev\right)$\fi}

\newcommand{\fobs}{\ifmmode f_{\rm obs} \else $f_{\rm obs}$\fi}

\newcommand{  \mbh      }{\ifmmode M_{\rm BH} \else $M_{\rm BH}$\fi}
\newcommand{  \lmbh     }{\ifmmode \log\left(\mbh/\Msun\right) \else $\log\left(\mbh/\Msun\right)$\fi} 
\newcommand{  \lledd    }{\ifmmode L/L_{\rm Edd} \else $L/L_{\rm Edd}$\fi}
\newcommand{  \mmedd    }{\ifmmode \dot{m}/\dot{m}_{\rm \,Edd} \else $\dot{m}/\dot{m}_{\rm \,Edd}$\fi}
\newcommand{  \Lbol     }{\ifmmode L_{\rm bol} \else $L_{\rm bol}$\fi}
\newcommand{  \lbol     }{\ifmmode L_{\rm bol} \else $L_{\rm bol}$\fi}
\newcommand{  \lLbol    }{\ifmmode \log\left(\Lbol/\ergs\right) \else $\log\left(\Lbol/\ergs\right)$\fi} 
\newcommand{  \Lagn     }{\ifmmode L_{\rm AGN} \else $L_{\rm AGN}$\fi}
\newcommand{  \lagn     }{\ifmmode L_{\rm AGN} \else $L_{\rm AGN}$\fi}

\newcommand{\Lbha}{\ifmmode L\left({\rm b}\ha\right) \else $L\left({\rm b}\ha\right)$\fi}

\newcommand{  \tgrow     }{\ifmmode t_{\rm growth} \else $t_{\rm growth}$\fi}
\newcommand{  \tAD     }{\ifmmode t_{\rm acc} \else $t_{\rm acc}$\fi}
\newcommand{  \tacc    }{\ifmmode t_{\rm acc} \else $t_{\rm acc}$\fi}
\newcommand{  \tUni      }{\ifmmode t_{\rm Universe} \else $t_{\rm Universe}$\fi}

\newcommand{  \Mdotin	}{\ifmmode \dot{M}_{\rm infall} \else $\dot{M}_{\rm infall}$\fi}
\newcommand{  \Mdotbh	}{\ifmmode \dot{M}_{\rm BH} \else $\dot{M}_{\rm BH}$\fi}
\newcommand{  \Mdotad	}{\ifmmode \dot{M}_{\rm AD} \else $\dot{M}_{\rm AD}$\fi}
\newcommand{  \Mdotacc	}{\ifmmode \dot{M}_{\rm acc} \else $\dot{M}_{\rm acc}$\fi}
\newcommand{  \Mdotthin	}{\ifmmode \dot{M}_{\rm thin} \else $\dot{M}_{\rm thin}$\fi}
\newcommand{  \Mdotdisk	}{\ifmmode \dot{M}_{\rm disk} \else $\dot{M}_{\rm disk}$\fi}

\newcommand{  \Mindot	}{\ifmmode \dot{M}_{\rm infall} \else $\dot{M}_{\rm infall}$\fi}
\newcommand{  \Mbhdot	}{\ifmmode \dot{M}_{\rm BH} \else $\dot{M}_{\rm BH}$\fi}
\newcommand{  \Maddot	}{\ifmmode \dot{M}_{\rm AD} \else $\dot{M}_{\rm AD}$\fi}
\newcommand{  \Maccdot	}{\ifmmode \dot{M}_{\rm acc} \else $\dot{M}_{\rm acc}$\fi}
\newcommand{  \Mthdot	}{\ifmmode \dot{M}_{\rm thin} \else $\dot{M}_{\rm thin}$\fi}
\newcommand{  \Mdsdot	}{\ifmmode \dot{M}_{\rm disk} \else $\dot{M}_{\rm disk}$\fi}

\newcommand{  \as	}{\ifmmode a_{\rm *} \else $a_{\rm *}$\fi}
\newcommand{  \avec	}{\ifmmode \vec{a}_{\rm *} \else $\vec{a}_{\rm *}$\fi}
\newcommand{  \re	}{\ifmmode \eta      	 \else $\eta$\fi}
\newcommand{  \RISCO	}{\ifmmode R_{\rm ISCO}  \else $R_{\rm ISCO}$\fi}

\newcommand{  \mseed    }{\ifmmode M_{\rm seed} \else $M_{\rm seed}$\fi}
\newcommand{  \mbul     }{\ifmmode M_{\rm bulge} \else $M_{\rm bulge}$\fi} 
\newcommand{  \mstar    }{\ifmmode M_{*} \else $M_{*}$\fi} 
\newcommand{  \mgal     }{\ifmmode M_{*} \else $M_{*}$\fi} 
\newcommand{  \mhost    }{\ifmmode M_{\rm host} \else $M_{\rm host}$\fi}
\newcommand{  \mmsmall  }{\ifmmode M_{\rm BH}/M_{*} \else $M_{\rm BH}/M_{*}$\fi}
\newcommand{  \mmlarge  }{\ifmmode M_{*}/M_{\rm BH} \else $M_{*}/M_{\rm BH}$\fi}

\newcommand{  \mmdotlarge}{\ifmmode \dot{M}_*/\Mbhdot \else $\dot{M}_*/\Mbhdot$\fi}
\newcommand{  \mmdotsmall}{\ifmmode \Mbhdot/\dot{M}_* \else $\Mbhdot/\dot{M}_*$\fi}

\newcommand{  \mmwp     }{\ifmmode \left(M_{*}/M_{\rm BH}\right) \else $\left(M_{*}/M_{\rm BH}\right)$\fi}
\newcommand{  \ml       }{\ifmmode M_{*}/L_{*} \else $M_{*}/L_{*}$\fi}
\newcommand{  \mlwp     }{\ifmmode \left(M_{*}/L\right) \else $\left(M_{*}/L\right)$\fi}
\newcommand{  \mlk      }{\ifmmode \left(M_{*}/L_{K}\right) \else $\left(M_{*}/L_{K}\right)$\fi}
\newcommand{  \sigs     }{\ifmmode \sigma_{*} \else $\sigma_{*}$\fi}
\newcommand{  \Reff     }{\ifmmode R_{\rm e} \else $R_{\rm e}$\fi}
\newcommand{  \Rvir     }{\ifmmode R_{\rm vir} \else $R_{\rm vir}$\fi}
\newcommand{  \Rtwo     }{\ifmmode R_{200} \else $R_{200}$\fi}
\newcommand{  \Rfive    }{\ifmmode R_{500} \else $R_{500}$\fi}
\newcommand{  \Rgrp     }{\ifmmode R_{\rm grp} \else $R_{\rm grp}$\fi}
\newcommand{  \nser     }{\ifmmode n_{\rm s} \else $n_{\rm s}$\fi}
\newcommand{  \LSF      }{\ifmmode L_{\rm SF}  \else $L_{\rm SF}$\fi}
\newcommand{  \LFIR     }{\ifmmode L_{\rm FIR} \else $L_{\rm FIR}$\fi}
\newcommand{  \Lfir     }{\ifmmode L_{\rm FIR} \else $L_{\rm FIR}$\fi}
\newcommand{  \LTIR     }{\ifmmode L_{\rm TIR} \else $L_{\rm TIR}$\fi}
\newcommand{  \Ltir     }{\ifmmode L_{\rm TIR} \else $L_{\rm TIR}$\fi}

\newcommand{  \mdyn     }{\ifmmode M_{\rm dyn} \else $M_{\rm dyn}$\fi} 
\newcommand{  \mgas     }{\ifmmode M_{\rm gas} \else $M_{\rm gas}$\fi} 
\newcommand{  \mh       }{\ifmmode M_{\rm h} \else $M_{\rm h}$\fi}
\newcommand{  \mhalo    }{\ifmmode M_{\rm halo} \else $M_{\rm halo}$\fi}
\newcommand{  \sfr      }{\ifmmode {\rm SFR} \else SFR\fi}

\newcommand{ \Lcii     }{\ifmmode L_{\cii} \else $L_{\cii}$\fi}
\newcommand{ \fwcii  }{\ifmmode {\rm FWHM}\cii \else FWHM\cii\fi}

\newcommand  {\RBLR}        {\hbox{$ {R_{\rm BLR}} $}}

\newcommand{\bj}{\ifmmode b_{\rm J} \else $b_{\rm J}$\fi}

\newcommand{\iab}{\ifmmode i_{\rm AB} \else $i_{\rm AB}$\fi}

\newcommand{\jab}{\ifmmode J_{\rm AB} \else $J_{\rm AB}$\fi}
\newcommand{\hab}{\ifmmode H_{\rm AB} \else $H_{\rm AB}$\fi}
\newcommand{\kab}{\ifmmode K_{\rm AB} \else $K_{\rm AB}$\fi}

\newcommand{\jveg}{\ifmmode J_{\rm Vega} \else $J_{\rm Vega}$\fi}
\newcommand{\hveg}{\ifmmode H_{\rm Vega} \else $H_{\rm Vega}$\fi}
\newcommand{\kveg}{\ifmmode K_{\rm Vega} \else $K_{\rm Vega}$\fi}

\def\arcsec{\hbox{$^{\prime\prime}$}}

\newcommand{  \Chisq    }{\ifmmode \chi^{2} \else $\chi^{2}$}
\newcommand{  \nelec    }{\ifmmode n_{e} \else $n_{e}$\fi}     % electron density
\newcommand{  \nh       }{\ifmmode n_{\rm H} \else $n_{\rm H}$\fi}     % hydrogen density
\newcommand{  \Ncol     }{\ifmmode N_{\rm col} \else $N_{\rm col}$\fi} % column density
\newcommand{  \NH       }{\ifmmode N_{\rm H} \else $N_{\rm H}$\fi}     % column density

\def\sun{\hbox{$\odot$}}

\def\arcsec{\hbox{$^{\prime\prime}$}}
\def\ion#1#2{#1$\;${\small\rm\@Roman{#2}}\relax}

\newcommand{\zwar}{\texttt{ZWARNING}}
\newcommand{\zerr}{\texttt{ZERR}}
\newcommand{\pqf}{\texttt{PyQSOFit}}
\newcommand{\gd}{\textit{good}}
\newcommand{\bd}{\textit{bad}}
\newcommand{\skt}{Skewt\_QSO}
\newcommand{\guab}{\texttt{GUA-bright}}
\newcommand{\guad}{\texttt{GUA-dark}}
\newcommand{\hp}{\ifmmode hp \else \textit{hp}\fi}

\usepackage{amsmath}

\received{July 28, 2026}
\submitjournal{ApJ}

\shorttitle{New Large Quasar Samples in SDSS-V}
\shortauthors{Aviram et al.}

\graphicspath{{./}{Figs/}}

\begin{document}
\title{An Exploratory Analysis of New Large Gaia-informed Quasar Samples in SDSS-V}

\correspondingauthor{Shir Aviram}
\email{shiraviram@mail.tau.ac.il}

\author[0009-0008-0046-8064]{Shir Aviram}
\affiliation{School of Physics and Astronomy, Tel Aviv University, Tel Aviv 69978, Israel}

\author[0000-0002-3683-7297]{Benny Trakhtenbrot}
\affiliation{School of Physics and Astronomy, Tel Aviv University, Tel Aviv 69978, Israel}

\author[0000-0002-4459-9233]{Tom Dwelly}
\affiliation{Max-Planck-Institut f{\"u}r extraterrestrische Physik, Gie\ss{}enbachstra\ss{}e 1, 85748 Garching, Germany}

\author[0000-0002-6404-9562]{Scott F. Anderson}
\affiliation{Astronomy Department, University of Washington, Box 351580, Seattle, WA 98195, USA}

\author[0000-0002-6770-2627]{Sean Morrison}
\affiliation{Department of Astronomy, University of Illinois at Urbana-Champaign, Urbana, IL 61801, USA}

\author[0000-0002-3719-940X]{Michael Eracleous}
\affiliation{Department of Astronomy \& Astrophysics, The Pennsylvania State University, University Park, PA 16802, USA}
\affiliation{Institute for Gravitation and the Cosmos, The Pennsylvania State University, University Park, PA 16802, USA}

\author[0000-0003-1659-7035]{Yue Shen}
\affiliation{Department of Astronomy, University of Illinois at Urbana-Champaign, Urbana, IL 61801, USA} 
\affiliation{National Center for Supercomputing Applications, University of Illinois at Urbana-Champaign, Urbana, IL 61801, USA}

\author[0000-0001-7116-9303]{Mara Salvato}
\affiliation{Max-Planck-Institut f{\"u}r extraterrestrische Physik, Gie\ss{}enbachstra\ss{}e 1, 85748 Garching, Germany}

\author[0000-0001-7240-7449]{Donald P. Schneider}
\affiliation{Department of Astronomy \& Astrophysics, The Pennsylvania State University, University Park, PA 16802, USA}
\affiliation{Institute for Gravitation and the Cosmos, The Pennsylvania State University, University Park, PA 16802, USA}

\author[0000-0002-9508-3667]{Roberto J. Assef}
\affiliation{Instituto de Estudios Astrof\'isicos, Facultad de Ingenier\'ia y Ciencias, Universidad Diego Portales, Av. Ej\'ercito Libertador 441, Santiago, Chile}

\author[0000-0001-5609-2774]{Catarina Aydar}
\affiliation{Max-Planck-Institut f{\"u}r extraterrestrische Physik, Gie\ss{}enbachstra\ss{}e 1, 85748 Garching, Germany}
\affiliation{Excellence Cluster ORIGINS, Boltzmannsstra\ss{}e 2, 85748, Garching, Germany}

\author[0000-0002-8686-8737]{Franz E. Bauer}
\affiliation{Instituto de Alta Investigaci{\'{o}}n, Universidad de Tarapac{\'{a}}, Casilla 7D, Arica, 1010000, Chile}

\author[0000-0002-0167-2453]{W.N. Brandt}
\affiliation{Department of Astronomy \& Astrophysics, The Pennsylvania State University, University Park, PA 16802, USA}
\affiliation{Institute for Gravitation and the Cosmos, The Pennsylvania State University, University Park, PA 16802, USA}
\affiliation{Department of Physics, 104 Davey Lab, The Pennsylvania State University, University Park, PA 16802, USA}

\author[0000-0002-8725-1069]{Joel R. Brownstein}
\affiliation{Department of Physics and Astronomy, University of Utah, 270 S. 1400 E. E2108, Salt Lake City, UT 84112, USA}

\author[0000-0003-0426-6634]{Johannes Buchner}
\affiliation{Max-Planck-Institut f{\"u}r extraterrestrische Physik, Gie\ss{}enbachstra\ss{}e 1, 85748 Garching, Germany}

\author[0000-0003-2511-2060]{Jeremy Darling}
\affiliation{Center for Astrophysics and Space Astronomy, Department of Astrophysical and Planetary Sciences, University of Colorado, 389 UCB, Boulder, CO 80309- 0389, USA} 

\author[0000-0002-0900-9760]{Jos\'e G. Fern\'andez-Trincado}
\affiliation{Centro de investigaci\'on en Astronom\'ia, Facultad de Ingenier\'ia, Ciencia y Tecnolog\'ia, Universidad Bernardo O'Higgins, Av. Viel 1497, Santiago, 8370993, Chile}

\author[0000-0002-1763-5825]{Patrick B. Hall}
\affiliation{Department of Physics and Astronomy, York University, 4700 Keele St., Toronto, Ontario M3J 1P3, Canada}

\author[0000-0002-7386-944X]{Dong-Woo Kim}
\affiliation{Center for Astrophysics, Harvard \&\ Smithsonian, 60 Garden Street, Cambridge, MA 02138, USA}

\author[0000-0002-6610-2048]{Anton M. Koekemoer}
\affiliation{Space Telescope Science Institute, 3700 San Martin Drive,
Baltimore, MD 21218, USA}

\author[0000-0002-5907-3330]{Stephanie LaMassa}
\affiliation{Space Telescope Science Institute, 3700 San Martin Drive, Baltimore, MD 21218, USA}

\author[0000-0002-0761-0130]{Andrea Merloni}
\affiliation{Max-Planck-Institut f{\"u}r extraterrestrische Physik, Gie\ss{}enbachstra\ss{}e 1, 85748 Garching, Germany}

\author[0000-0001-5231-2645]{Claudio Ricci}
\affiliation{Department of Astronomy, University of Geneva, ch. d'Ecogia 16, 1290, Versoix, Switzerland}
\affiliation{Instituto de Estudios Astrof\'isicos, Facultad de Ingenier\'ia y Ciencias, Universidad Diego Portales, Av. Ej\'ercito Libertador 441, Santiago, Chile} 
\affiliation{Kavli Institute for Astronomy and Astrophysics, Peking University, Beijing 100871, China}

\author[0000-0002-6893-3742]{Qian Yang}
\affiliation{Center for Astrophysics, Harvard \& Smithsonian, 60 Garden Street, Cambridge, MA 02138, USA}

\author[0000-0002-7817-0099]{Grisha Zeltyn}
\affiliation{School of Physics and Astronomy, Tel Aviv University, Tel Aviv 69978, Israel}

\begin{abstract}
Quasars are extremely luminous objects that provide insights into the physics and evolution of supermassive black holes (SMBHs) and their accretion flows, galaxy evolution, and even cosmology. In this study, we present an exploratory study based on the ongoing fifth generation of the Sloan Digital Sky Survey (SDSS-V) and its unique dual-hemisphere, wide-field, and multi-object spectroscopic capabilities, with the aim of creating a comprehensive, all-sky quasar sample. The targets were selected through two novel methods, GUA and \skt, that rely primarily on data from WISE and Gaia, aiming to address gaps in previous large quasar samples. Our sample includes over 250,000 spectroscopically confirmed quasars reaching to $z\sim5$, with tens of thousands of newly identified quasars in the southern hemisphere. The selection methods are highly pure, with well over 80\% of the spectra collected being genuine quasars; the main contaminants are M-type stars. The detailed spectral decomposition procedure we employed shows that the quasars in the sample span a wide range of luminosities ($\Lbol \sim 10^{44}-10^{48}\,\ergs$), SMBH masses ($\mbh \sim 10^{6}-10^{10}\,\Msol$), and accretion rates ($\lledd\sim0.01-1$). The distributions of these properties are consistent with those of previous quasar catalogs, which are based on past generations of SDSS, once we account for potential selection biases related to the various survey depths. Our findings confirm that novel selection methods based on optical+IR colors and/or astrometry can yield a large, high-purity quasar sample over wide sky areas, including in cases where more nuanced multi-band photometry and/or multi-wavelength data in the X-ray or radio is not available. This SDSS-V sample, which is expected to grow in the near future, establishes a robust reference for future southern (time-domain) surveys, while enhancing and complementing our understanding of quasar demographics and SMBH evolution.
\end{abstract}


\section{Introduction}
\label{sec:intro}
Quasars, the high-luminosity manifestation of active galactic nuclei (AGNs), are powerful tools for exploring the Universe across cosmological distances. Quasars are powered by accreting supermassive black holes (SMBHs) that reside at the centers of massive galaxies, where a radiatively efficient accretion flow (i.e., a thin disk) releases large amounts of UV-optical radiation. When this primary radiation field is reprocessed by circumnuclear gas components, other forms of radiation may emerge. These primarily include 
broad emission lines that originate in fast-moving ($>1000\,\kms$) high-density gas called the broad-line region (BLR); 
narrow emission lines from lower-density gas moving at lower velocities ($\sim100-1000\,\kms$) known as narrow-line region (NLR);
infrared (IR) emission from a dusty axisymmetric structure (the so-called ``dusty torus'' and perhaps a dusty polar wind component); 
X-rays originating in a compact, hot and low-density ``corona''; as well as radio emission from relativistic jets in some objects \citep{Padovani17}. All these emission components can be used to identify quasars \citep[e.g.,][]{Alexander&Hickox2012} and other types of AGNs out to $z>7$ (\citealt{Wang+21z7.642}; see also \citealt{UHZ1}), and allow the study of the basic properties of the SMBHs that power them, the links between these SMBHs and their host-galaxies, and even probe cosmic reionization \citep{Fan+23review}.

Over the years, considerable effort has been dedicated to the search for quasars, i.e., luminous broad-line AGNs with limited obscuration along their lines-of-sight, as identified mostly in the rest-frame UV-optical regime.
The most recent version of the Million Quasars (Milliquas/v8;\, \citealt{milliquas}) compilation of AGN catalogs holds all the spectroscopically confirmed quasars, and other types of AGNs and candidates, and includes just over 900,000 spectroscopically confirmed quasars as of June 2023. The vast majority of this huge inventory (82.5\%) was made possible by the various generations of the Sloan Digital Sky Survey \citep[SDSS;][]{SDSS,SDSS_DR16}. 
More recently, the first data release of the Dark Energy Spectroscopic Instrument project \citep[DESI/DR1;][]{DESI_DR1} has added $\approx$1.2M new quasars \citep{DESI_DR1_QSO}.
Given the location of the relevant survey facilities, it is not surprising that most of the quasars surveyed to date are located in the northern celestial hemisphere. 

In contrast, the southern sky has been under-explored, with only a few notable surveys as exceptions. These are mostly the various wide-field spectroscopic surveys conducted at the Australian Astronomical Observatory, including the original Two-degree-Field Galaxy Redshift Survey \citep[2dF;][]{2dF}, the Six-degree-Field Galaxy Redshift Survey \citep[6dF;][]{6dF} as well as the quasar-focused surveys 2QZ and 2SLAQ (\citealt{2dF_2QZ} and \citealt{2dF_2SLAQ}, respectively).
While these programs provided valuable data for many thousands of southern-hemisphere quasars, most of the spectra obtained in these surveys lacked proper flux calibration, which naturally hindered some scientific inquiries. This imbalance between the celestial hemispheres is clearly evident in the Milliquas compilation, where only 13\% of the quasars are located at $\delta<0$. The relative scarcity of known quasars in the southern hemisphere is partly due to the historical lack of deep, wide-field optical imaging surveys. This is particularly striking given that this is the part of the sky that is accessible to many of the most advanced facilities available now (e.g., the VLT, ALMA, Rubin, 4MOST) and in the future (e.g., the ELT, SKA, Roman).

Another limitation of existing quasar samples lies in the way that candidates were selected, based on photometry and/or multi-wavelength data, for the spectroscopic observations that ultimately allow their confirmation and study. 
Specifically, most of the quasars in the legacy SDSS (i.e., up to DR7; \citealt{SDSS_DR7}, and DR7Q; \citealt{DR7Schneider}) were selected based on their optical colors \citep{Richards02}, which generally favored blue, point-like sources. A subset of quasars was targeted based on their radio emission, and a dedicated set of color selection criteria was used to select high-redshift ($z>3.5$) quasar candidates. The complex color selection introduced by \cite{Richards02} reaches 90\% completeness and a quasar selection purity (a.k.a., efficiency) higher than 65\%. That legacy SDSS color-based selection had limited capabilities in the $z{\approx}2.1-2.8$ regime, where quasar colors overlap with those of stars in optical color space, which motivated later generations of SDSS quasar searches to also incorporate a combination of more sophisticated selection methods, involving more nuanced (statistical) color- and variability-based selection criteria \cite[see][and references therein]{Bovy2011,Ross12,Myers15}. These resulted in a quasar selection purity exceeding $50\%$.
The focus of most quasar surveys on a high level of completeness ($\gtrsim$90\%), required to robustly determine key demographic measurements (e.g., the quasar luminosity function; e.g., \citealt{2dF_2QZ,2dF_2SLAQ}, often meant compromising the selection purity, even when surveying high Galactic latitudes. 
Samples of quasars observed at low Galactic latitudes remained modest in size, and thus in utility (see, e.g., \citealt{Fu2021,Werk2024,Huo2025} and references therein for recent progress).
The advent of Gaia astrometric measurements allowed to make significant progress in constructing large samples of quasar candidates that have both high completeness {\it and} improved purity, at all Galactic latitudes (see below and, e.g., \citealt{S19,Gaia23_exgal,YS23,quaia}).

Importantly, optical color-based selection methods may be biased against certain subpopulations of quasars, particularly those with redder spectral energy distributions (SEDs) \citep[though such quasars can still be found by those methods;][]{Krawczyk2013,Ross2015,Hamann17}, which may indeed be expected for a number of reasons. 
First, simple accretion disk theory predicts that the emerging SEDs should become redder with increasing BH mass (\mbh), and/or with decreasing BH spins \citep{ShakuraSunyaev}. Indeed, \citet{massiveBH_SED} argued that vigorously accreting SMBHs with low and/or retrograde spins might have colors consistent with red stellar sources. Second, some models of super-Eddington accretion flows predict that the innermost regions of the flow would become advection-dominated, thus ``trapping'' the UV radiation, again showing atypically red SEDs \cite[e.g.,][and references therein]{superEdd}. Finally, there is growing evidence that some of the most luminous AGNs, powered by high-mass SMBHs, are obscured by significant amounts of dust, likely on host-galaxy scales. Such heavily reddened sources, including red quasars \cite[e.g.,][]{Urrutia2009,Banerji2012,Glikman2012,Glikman2015, LaMassa2016,LaMassa2024}) and hot dust-obscured galaxies \cite[Hot DOGs][e.g.,]{Eisenhardt2012,W12drops}, can be identified thanks to their (mid-)IR emission. While they may appear to be rare, several studies reported they may trace a significant fraction of the highest-luminosity AGN population \citep{W12drops}, and/or a significant phase in SMBH growth \cite[e.g.,][and references therein]{Banerji2015_rQSOs,Vito2018,Glikman2024_rQSOs,Fawcett2023}.

The ongoing fifth generation of the SDSS project \citep[SDSS-V;][]{Kollmeier2026} holds the potential to alleviate some of the limitations of the currently available quasar samples. As a dual-hemisphere wide-field survey, it can target AGN and quasar candidates across the entire sky, and based on novel selection criteria. Specifically for the southern hemisphere, a cornerstone of SDSS-V is the optical spectroscopy of X-ray emitting AGN candidates drawn from the eROSITA surveys (i.e., the SPIDERS program; \citealt{SPIDERS_Dwelly}, \citealt{SPIDERS_Comparat}), with over 75\% confirmed as broad-line quasars \citep{Aydar25}. However, the SDSS-V survey strategy allows targeting of additional sets of AGN and quasar candidates, using other selection criteria. Whether these novel selection criteria are based on necessity or on physical motivation, their performance has to be empirically tested.

This paper focuses on two of the larger ancillary quasar searches within SDSS-V, enabled by novel candidate selection criteria that combine optical and mid-infrared (MIR) data, and which can be utilized for quasar searches over wide sky areas, including in the southern celestial hemisphere and/or low Galactic latitudes (e.g., \citealt{S19} and \citealt{YS23}). 
Together, these SDSS-V efforts have delivered new, high-quality optical spectroscopy for over 100,000 quasars, including over 70,000 that are newly identified---most of which are located in the southern celestial hemisphere.

The paper is structured as follows: Section~\ref{sec:obs} describes the two programs in terms of target selection and SDSS-V observations. Section~\ref{sec:analysis} details our analysis of the sample of reliable quasars collected so far, and the spectral measurements we conduct.  Section~\ref{sec:results} presents our main results in terms of the purity (efficiency) of quasar selection and the key properties of the quasars in our new sample(s). 
We summarize our conclusions in Section~\ref{sec:conc}. 
Throughout this work, we use the AB magnitude system (\citealt{ABsystem}; unless clearly stated otherwise) and assume a flat $\Lambda$CDM cosmology with $H_0 = 70\,\mathrm{km\,s^{-1}\,Mpc^{-1}}$, $\Omega_\mathrm{m} = 0.3$, and $\Omega_\Lambda = 0.7$.
We use the term ``quasars'' to broadly refer to broad-line AGNs whose optical continuum emission is dominated by the central accretion flow, regardless of their (bolometric) luminosity, including sources below the traditional limit of $\Lbol<10^{44}\,\ergs$.

%%%%%%%%%%%%%%%%%%%%%%%%%%%%%%%%%%%%%%%%%%%%%%%%%%%%%%%%%%%%%%%%
\defcitealias{S19}{S19}
\defcitealias{YS23}{YS23}

\newcommand{\papgua}{\citetalias{S19}}
\newcommand{\papdes}{\citetalias{YS23}}
%%%%%%%%%%%%%%%%%%%%%%%%

\section{Candidate Selection, SDSS-V Observations and Data} 
\label{sec:obs}

This Section describes the two main quasar selection techniques that were used to construct our sample, the way the corresponding target pools were refined for actual SDSS-V spectroscopic observations, and the observations and data themselves.

\begin{figure*}
    \includegraphics[width=\textwidth]{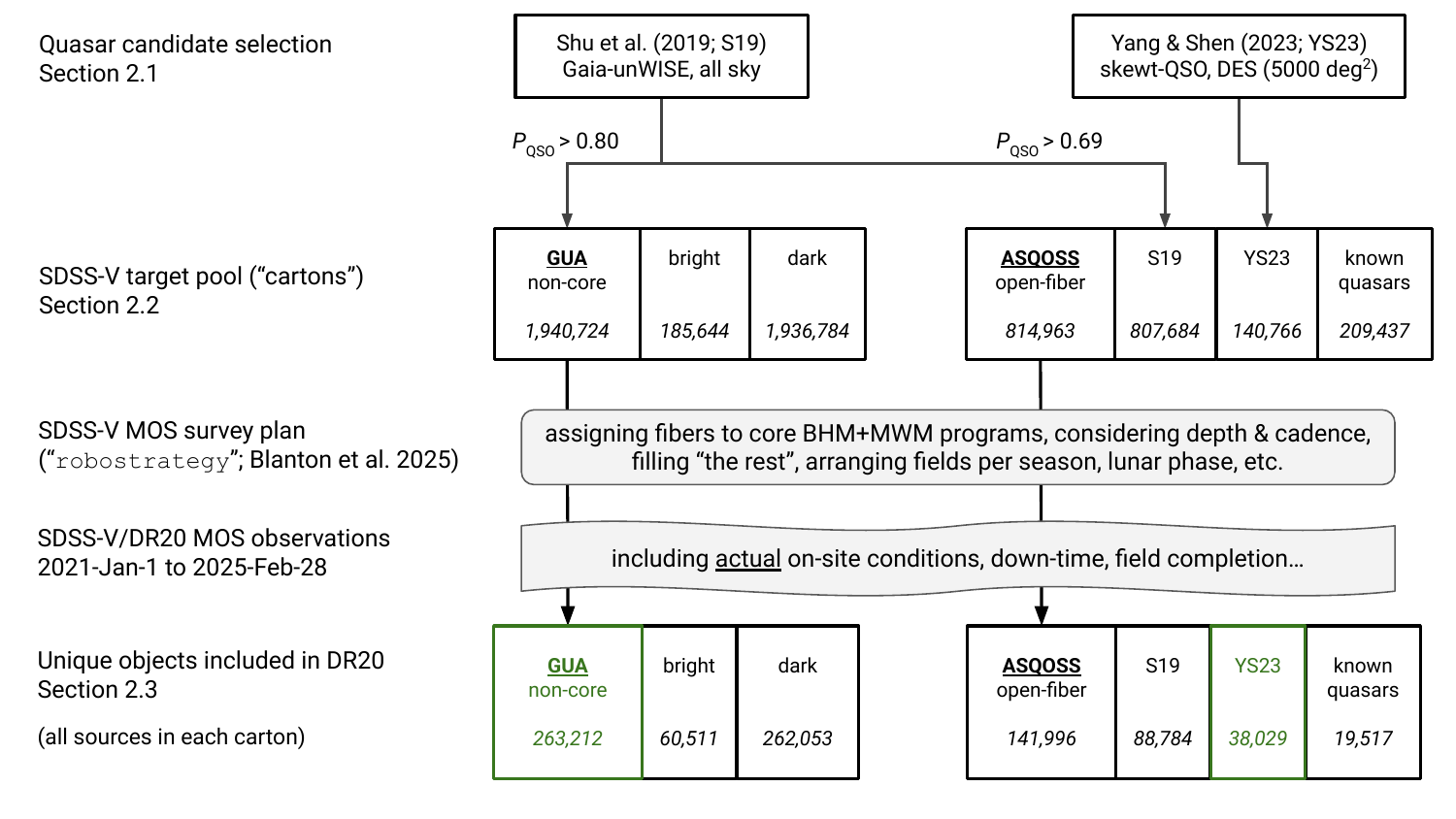}
    \caption{Parent sample construction. From top to bottom, the flowchart illustrates how quasar candidates are selected through the two main selection methods (based on \papgua\ and \papdes). The two techniques populate various, often non-unique SDSS-V target pools (``cartons''), which are used for actual MOS survey observations; and ultimately are made available for analysis as part of SDSS-V/DR20.
    We highlight in green the two final sets used for our analysis.}
    \label{fig:flowchart_trgt}
\end{figure*}

Given the complex nature of the SDSS-V Multi-Object Spectroscopy (MOS) survey design and observations, there is non-negligible overlap between various target pools (called ``cartons'' in SDSS jargon), and non-trivial considerations made during survey design and implementation. 
To summarize, any observed spectrum should, in principle, be considered as belonging to multiple cartons; it could be observed at various levels of priority; and under various observing conditions.
The various steps in the process of designing and implementing the SDSS-V MOS survey are explained in detail in several key SDSS-V publications, including 
\citet[][general survey overview]{Kollmeier2026},
\citet[][target pools and targeting philosophy]{DR18},
and \citet[][field and target assignment]{Blanton2026_rs}.
Here, we only briefly refer to the key steps in this process that are relevant for our analysis.
Figure~\ref{fig:flowchart_trgt} illustrates this process and provides guidance for the rest of this Section.

\subsection{Construction of Quasar Candidate Pools} 
\label{sec:target}

This work focuses on the construction and analysis of large samples of quasars, which were selected through two novel procedures. Both procedures aim to provide large, high-purity (or efficiency)\footnote{Throughout this work, we use the term purity (and, at times, efficiency) to denote the fraction of quasars, out of the total sample of sources targeted by the respective survey within which the search was conducted.} sets of quasar candidates across wide sky areas, by combining publicly available photometry (in the optical and MIR) and astrometry (based on Gaia; \citealt{Gaia_DR2}). 
The common challenge of both procedures is differentiating between quasar candidates and foreground interlopers (i.e., Milky Way stars), as both procedures are designed to be applicable also to low Galactic latitudes. 
In what follows, we summarize the key aspects of these two procedures; additional details can be found in the two respective studies (see below).

Both quasar selection methods rely on the comprehensive all-sky data provided by the Gaia (specifically DR2; \citealt{Gaia_DR2}) and WISE (specifically unWISE; \citealt{unWISE}) missions. 
Gaia provides not only high-quality three-band photometry over the entire sky (in the $G$, $BP$, and $RP$ bands), reaching depths of $G\lesssim21$ mag, but---importantly---astrometric measurements for the vast majority of celestial sources accessible to SDSS-V, which allows rejection of sources with proper-motion from any further consideration for extragalactic studies.
We note, however, that the precision of the Gaia (DR2) proper-motion measurements degrades significantly at faint magnitudes, from $\sim0.05\,\rm{mas}\,yr^{-1}$ at $G<15$ mag to $\sim0.2\,\rm{mas}\,yr^{-1}$ at $G=17$ mag and $\sim1.2\,\rm{mas}\,yr^{-1}$ at $G=20$ mag \citep{Lindegren2018}. Indeed, at $G\gtrsim20$ the astrometric uncertainty becomes comparable to the true proper motions of typical faint stellar contaminants if located at kpc-scale distances ($\lesssim$ few $\rm{mas}\,yr^{-1}$). Thus, the proper-motion based star--quasar discrimination loses much of its power for the faintest sources considered for the sample and analysis considered here, and one may expect that the quasar selection purity would decrease for such faint objects.

The four-band MIR photometry provided by WISE over the entire sky, the 
four bands,\footnote{Denoted W1-4, with effective wavelengths of 3.4, 4.6, 12, and 22\,\mic.} is widely used to robustly identify AGNs (and specifically, quasars), as their MIR continuum emission appears redder than that of inactive galaxies. 
\citet{SternW12} demonstrated that the simple criterion $W1-W2\geq0.8$ yields a highly robust AGNs selection, with completeness of 78\% and purity of 95\%. A higher $W1-W2$ color cut would enhance the purity at the expense of completeness. Later, \cite{Assef2013_MIR_selection} and \citet{AllWISE_Assef} introduced refined $W1-W2$ color cuts, optimized to improve selection reliability relative to the simpler \citet{SternW12} criterion.

\subsubsection{Gaia-unWISE based selection -- Shu et al.\ (2019)} 

The main pool of quasar candidates selected for ancillary SDSS-V wide-area targeting is based on the study of \citet[][S19 hereafter]{S19}, which combined Gaia and unWISE data to derive a large set of candidates across the entire sky, apart from areas encircling the Magellanic clouds and M31. 
While the Galactic plane is formally included, the highest-extinction sight-lines yield a drastically lower number, or indeed no, quasar candidates.
This selection method is based on ``AGN-like'' MIR (WISE) colors combined with optical Gaia data and no detectable proper-motion from Gaia DR2. 
The method leverages optical photometry from the G, BP, and RP bands of Gaia, as well as its astrometric data (Gaia DR2), and MIR photometry in the W1 and W2 bands (from the unWISE release). Sources are selected by cross-matching the Gaia DR2 catalog with the unWISE catalog (with a matching radius of 2\arcsec), focusing on objects with robust detections in both the W1 and W2 bands. 

\papgua\ describes a random forest (RF) algorithm developed for differentiating AGN and non-AGN sources, which is based on two large and correspondingly labeled training datasets. 
The (trained) algorithm can then predict the probability ($P_{\rm RF}$) that a given source is an AGN.
\papgua\ applied the trained algorithm to the entire Gaia–unWISE dataset and created two non-mutually-exclusive AGN candidate catalogs, dubbed C75 and R85, with 75\% completeness and 85\% efficiency,\footnote{\papgua\ refers to this as ``reliability'', hence the ``R'' in ``R85''.} respectively. 
Each of these catalogs assumes a different threshold of $P_{\rm RF}$ to balance completeness and efficiency, specifically $P_{\rm RF}\geq0.69$ for C75 and $P_{\rm RF}\geq0.94$ for R85. 
Each of the resulting catalogs contains over $2\times10^6$ AGN candidates across a calculated effective area that \papgua\ reports as ${\approx}$36,000 deg$^2$, with nearly half of them not appearing in prior AGN compilations. 
This effective area considers the low-candidate-density regions towards the highest-extinction regions in the Galactic plane.

\subsubsection{DES/\skt\ based selection}

The study of \citet[][YS23 hereafter]{YS23} presents a catalog of ${\approx}1.4\times10^6$ photometrically-selected quasar candidates in the southern extragalactic sky, covering about 5000 deg$^2$ of the footprint of the Dark Energy Survey (DES; \citealt{DES_DR1}). The selection process presented by \papdes\ combines optical photometry in the $griz$ bands from DES/DR2 \citep{DES_DR2}, NIR photometry from a collection of surveys (VIKING: \citealt{VIKING}, VHS: \citealt{VHS}, ULAS: \citealt{ulas}, UHS: \citealt{UHS} and 2MASS: \citealt{2MASS}), and MIR photometry from unWISE (\citealt{MIR-unWISE1}, \citealt{MIR-unWISE2}). Additionally, Gaia astrometric data were used to eliminate stars with detectable proper-motion, ensuring a cleaner selection of quasar candidates.

Similar to \papgua, \papdes\ refined their classification algorithm using two labeled training sets. The algorithm assigned each (non-training) source with a probability of being a quasar, galaxy, or star ($P_{\rm QSO}$, $P_{\rm Star}$, and $P_{\rm Galaxy}$, respectively),  based on parametric representations of the respective distributions in multi-color space(s).
A source was included in the quasar candidates sample only if $P_{\rm QSO}>P_{\rm Star}$ and $P_{\rm QSO}>P_{\rm Galaxy}$.\footnote{By definition, this requirement also implies that all such candidates have $P_{\rm QSO}>1/3$.}

The sample of quasar candidates selected by \papdes\ greatly overlaps with that of \papgua\ in the sky region that is common to the two works (i.e., the $\sim$5000 deg$^2$ of DES). Specifically, 97\%\ of the \papdes\ quasar candidates (136,222 of 140,766 objects) can also be found in the C75 catalog of \papgua. Nonetheless, in parts of our analysis, we opt to consider the two selection methods separately, in order to assess their effectiveness in selecting real quasars, and given the subtle difference in how various sub-programs were implemented in SDSS-V (see immediately below).

\subsubsection{Limitations of training sets}
\label{sec:limitations}

Both selection methods carry known limitations in their ability to recover the full range of quasar populations, inherited from their training samples. 
Specifically, \papgua\ was trained on the SDSS/DR14Q catalog \citep{SDSS_DR14}, whose foundation in the ugriz-color-selected DR7Q catalog \citep{Richards02,DR7Schneider} may have systematically under-represented heavily dust-reddened quasars. 
\papdes\ was trained on the SDSS/DR16Q catalog \citep{DR16Q} supplemented with confirmed quasars from the Milliquas compilation \citep{milliquasv6.4}, and further augmented with simulated $z>3.5$ quasars drawn from a distribution of {\it unobscured} SEDs. 
Both methods also rely on WISE-based aperture photometry that assumes point-source geometry, biasing against quasars with close companions. This may again complicate the selection of reddened quasars, as galaxy companions and/or mergers were shown to be linked with (enhanced) obscuration \cite[e.g.,][]{Ricci2017_mergers,Glikman2024_rQSOs}.
Thus, we may expect that heavily obscured/reddened quasars would be missed by the selection methods, even prior to considering the challenges related to obtaining high-quality optical spectra and identifying AGN signatures.
Other minority sub-populations of quasars whose colors may be under-represented in the training sets, and thus in any survey that relies on the corresponding selection methods, may include BAL quasars with particularly broad troughs.

\subsection{Surveying Quasars in SDSS-V}
\label{sec:sdssv}

The optical MOS component of SDSS-V utilizes two 2.5-meter wide-field telescopes \citep{Kollmeier2026}: the legacy Sloan Foundation Telescope located at Apache Point Observatory (APO) in New Mexico, USA \citep{APO_telescope}, and, since late 2022, the Ir\'en\'ee du Pont telescope at Las Campanas Observatory (LCO) in Chile \citep{LCO_telescope}, enabling full-sky coverage. SDSS-V operations at APO began in December 2020 with traditional fiber plug-plates. A year later, these were replaced by a robotic Focal Plane System (FPS; \citealt{FPS}), revolutionizing observing efficiency by automating fiber positioning.

Both the Black Hole Mapper (BHM) and Milky Way Mapper (MWM) programs within SDSS-V (Anderson \et, in prep.;  Johnson \et, in prep., respectively) provide lists of targets for fiber-fed spectroscopic observations. Given the dual-hemisphere nature of SDSS-V, the multitude of science goals \citep{Kollmeier2026}, and the varying observing conditions, the actual survey operations are designed to prioritize several core programs that drive field selection, fiber allocation, and exposure times, with a variety of progressively lower-priority programs being allocated additional fibers until each MOS field can be considered as delivering maximal science potential.
This means that the quasar candidates provided through either the \papgua\ and/or the \papdes\ selection methods can be observed as part of several cartons, with a variety of priorities, exposure times, and observing conditions.

Below we describe the various ways in which our quasars were observed, categorized chiefly by the priority of each program 
and by the additional information it may carry about the AGN nature of the target. We again stress that most AGN candidate cartons in SDSS-V have significant overlap, while for non-core programs, the effective survey depth depends on the requirements set for each visit of each field by the higher-priority core programs.
More details about the various SDSS-V programs and survey strategy can be found below, as well as in \cite{DR18}, and \cite{Kollmeier2026}.

\subsubsection{Gaia unWISE AGN (GUA)}
\label{sec:gua_carton}

The Gaia-unWISE AGN (GUA) program is the largest non-core, AGN-focused program within SDSS-V, designed to provide a large number of targets across the entire SDSS-V survey area and range of observation types.
Starting with the \papgua\ catalogs, the GUA selection focuses on sources within the C75 subset that have $P_{\rm QSO} \equiv P_{\rm RF} > 0.8$, $G>13$ mag (to avoid saturation). 
We stress that the availability of previous spectroscopy from other sources (e.g., LAMOST, 2dF/2QZ; about half of the sources in \papgua) was \textit{not} a consideration for GUA, thus potentially allowing a future use of new SDSS-V spectra obtained through GUA for long-term time-domain science.

When incorporating the GUA set of quasar candidates into the SDSS-V targeting efforts, it was further split into two subsets based on source brightness, making them suitable for either dark or bright (lunar) conditions. The \textit{GUA-dark} sub-sample consists of 2,156,582 targets with $16<G<21$ mag, while the \textit{GUA-bright} sub-sample includes 254,601 brighter targets, with $13<G<18$ mag, allowing them to reach an adequate signal-to-noise ratio ($S/N$) even during bright time. This separation enables optimal use of observing conditions for each subset, with the main distinction being that GUA-dark targets populate fields observed as part of the BHM program, which is pursued only during dark time, while the GUA-bright targets can also populate fields observed as part of the MWM program, which is pursued mostly during bright time. We stress that these two sub-samples are not mutually exclusive, as reflected by their overlapping $G$-band magnitude ranges. 

Overall, in its highly efficient regime, down to $G=20.5$, the GUA targeting reaches an all-sky averaged density of quasar candidates of ${\approx}37.7,{\rm deg}^{-2}$. 
This is slightly higher than the sky density of quasars observed through previous generations of the SDSS (e.g., ${\approx}24,{\rm deg}^{-2}$ in DR16, \citealt{DR16Q}). At the nominal GUA flux limit, $G=21$, these two sky densities are fully consistent.

\subsubsection{All Sky Quasar Optical Spectroscopic Survey (ASQOSS)}
\label{sec:asqoss_carton}

The All Sky Quasar Optical Spectroscopic Survey (ASQOSS hereafter) is a large ``open fibers'' program in SDSS-V\footnote{Internally designated with the carton name \texttt{openfibertargets\_nov2020\_27}.} aiming to complement the SDSS-V targeting pool so it can include the largest sample of quasar spectra across the entire sky, even if through relatively short exposures. As an open fibers program, ASQOSS targets can be selected to be observed in either dark or bright time, in either BHM- or MWM-led fields, generally with exposures as short as 15 minutes.

The overall ASQOSS target catalog contains 814,963 objects that are a superset of several distinct samples:
\begin{enumerate}
    \item Quasar candidates selected from the \papgua\ catalog, limited to $G<20$ mag. These 807,684 targets are distributed across the entire sky except for the regions surrounding the Magellanic Clouds and M31. 
    While this sample of targets is similar in spirit to GUA, here a threshold of $P_{\rm QSO}>0.69$ was assumed (c.f.\ $>0.8$ for GUA).
    
    \item Quasar candidates selected from the \papdes\ catalog, limited to $G<20$ mag. By construction, these 140,766 targets are located within the $\sim$5000 deg$^2$ DES footprint. 

    \item Previously-known quasars drawn from the SDSS/DR16Q catalog \citep{DR16Q}, the LAMOST/DR5 quasar catalog \citep{Yao19_LAMOST_DR45}, and the final 2QZ quasar catalog \citep{2dF_2QZ}. 
    These 209,437 targets are included in ASQOSS to maximize the time-domain-related science goals of BHM, namely searching for extremely variable and/or ``changing-look'' AGNs \citep{CLAGN}.

\end{enumerate}
All the targets in the ASQOSS carton are robustly detected by Gaia, have $G<20$ mag, and exhibit no detectable proper-motion (as of Gaia DR2). 
Regions with high extinction near the Galactic plane, as well as around the Magellanic Clouds and M31, are omitted.

In what follows, we only focus on those ASQOSS targets that are drawn from either the \papgua\ or \papdes\ selection procedures (i.e., the first two items above); whenever we use the term ``ASQOSS'' we refer only to these two types of targets. %within the larger ASQOSS pool.
This approach is driven by our focus on the efficiency of these novel quasar selection methods, and on the properties of the quasars they uncover when employed within SDSS-V. Moreover, targets for which there is clear prior knowledge about the nature of their nuclear activity (i.e., the previously-known quasars included in ASQOSS) would obviously skew any assessment of the efficiency of this open-fiber program. 

The ASQOSS/\skt\ targeting reaches an averaged density of quasar candidates of ${\approx}28.1,{\rm deg}^{-2}$ down to $G=20.5$. 
This is generally consistent with GUA and SDSS/DR16Q (see above).

\subsubsection{Other programs in SDSS-V and prioritization}
\label{sec:SDSS_other}

The target selection processes for the GUA and ASQOSS programs may overlap not only with each other but also with other observational initiatives. 
The SDSS-V DR18 and DR19 papers (\citealt{DR18} and \citealt{DR19}, respectively) provide a detailed discussion of the various SDSS-V (BHM) programs; here we only note those few BHM programs that are most relevant for the present study. 
For our purposes, the other BHM programs may be split based on whether they carry additional information about the AGN nature of the targets, and/or whether have higher or lower priority than GUA and/or ASQOSS during survey design and execution.

The high-priority core BHM programs include two key wide-field components, being the All-Quasar Multi-Epoch Spectroscopy program (AQMES; Eracleous et al., in prep.), which revisits previously known SDSS,   and the SPectroscopic IDentfication of ERosita Sources program (SPIDERS; Merloni \et, in prep.), which is based on X-ray observations conducted with the eROSITA mission \citep{eROSITA}. 
Moreover, the reverberation mapping program (BHM-RM; Trump \et, in prep.) monitors hundreds of known known quasars in a few distinct, discontiguous fields. The BHM-RM fields and targets are designed separately from the rest of the SDSS-V MOS survey, and their overlap with GUA and ASQOSS doesn't affect our analysis.
Another high-priority, though non-core program focuses on sources drawn from the Chandra Source Catalog Release 2 \cite[CSC 2.1][]{CSC_Evans24}, which naturally includes many AGN candidates.
When assigning fields and fibers/targets \citep{Blanton2026_rs}, GUA has a priority that is lower than any of the (BHM) core programs and than the CSC one. ASQOSS has a yet lower priority, being an open fiber program. 

Finally, there are many other open-fiber programs, some of which target various samples of AGNs and/or AGN candidates (see \citealt{DR18,DR19}). 
All these open-fiber programs have a priority that is identical to that of ASQOSS (and thus lower than that of GUA). 
One of these open fiber programs is designed to target quasar candidates from the Quaia catalog \citep{quaia}, selected based on Gaia and unWISE data; however, the vast majority of Quaia targets ($\approx$97\%) are already included in either GUA and/or ASQOSS. 

\subsubsection{Selection Overlap}
\label{sec:overlap}

The design of the SDSS-V MOS survey and the ultimate data acquisition considers all target cartons and their priorities, through a complex algorithm, which is described in detail in \cite{Blanton2026_rs}. 
The resulting dataset of quasar (candidate) spectroscopy incorporates a significant overlap between various target cartons, which should be considered when assessing the efficiency of any (quasar) selection method.
This complex target selection overlap is non-negligible for the samples under study.
Specifically, 80\% (652,278) of the objects in the ASQOSS carton are also part of the GUA carton (and 34\% targets in GUA are also in ASQOSS); moreover, 21\% (410,045) of the GUA targets and 27\% (220,584) of the ASQOSS targets are also selected as SPIDERS X-ray emitters. 
Appendix~\ref{app:overlap} and Figures therein, illustrate the full complexity of this target selection overlap.

In practice, when assessing the quasar selection efficiency of the \papgua\ and \papdes\ selection methods (in Section~\ref{sec:purity}), we consider several disjoint subsets of targets, based on their membership in the GUA, ASQOSS/\skt, and/or core+CSC cartons.
The subsets are conceptually illustrated in Appendix~\ref{app:overlap}. 
The detailed purity calculations are performed only over a restricted area within the high-Galactic-latitude part of the southern sky (see Appendix~\ref{app:overlap}).
Another point where the selection overlap is important is in our accounting of the number of new quasars detected through the GUA and ASQOSS/\skt\ programs (Section~\ref{sec:final_sample}). 
There, we focus on the subset of objects that are part of any of these two cartons, but are {\it not} part of any of the core programs or of the CSC carton. This ``GUA+\skt\ sole selection'' conceptually follows the subsets designated as $A$, $C$, and $E$ in Appendix~\ref{app:overlap}, employed over the entire footprint of the data in hand.

\subsubsection{SDSS-V spectroscopy} 
\label{sec:sdssv_spec}

For this study, we used the spectra that are part of the SDSS/DR20 Data Release (Griffith et al., in prep.)\footnote{\url{https://www.sdss.org/}}. All the technical details about spectral data acquisition, reduction, and classification can be found in the DR20 paper, and references therein. Here we only briefly mention a few key aspects.

The optical spectra analyzed in this paper were obtained with the Baryon Oscillation Spectroscopic Survey (BOSS) spectrographs \citep{BOSS}, which are fed by up to 500 fibers, each with apertures of either 2\arcsec\ or 1.3\arcsec\ (for APO and LCO, respectively), with a spectral resolution of $R \sim 2000$ over the wavelength range of $3600-10400$\,\AA. 
Each spectral observation consists of 15-minute-long exposures, repeated as required by the specifications of the (highest-priority) carton to which the target belongs. Typical total integration times vary between a single 15-minute exposure (e.g., for many \guab\ and ASQOSS targets) and 60 minutes (e.g., for \guad\ targets).
Many SDSS-V/BHM targets are visited multiple times, over a wide range of timescales, mainly through the AQMES and BHM-RM programs. 
The present work, however, does not go into such time-domain aspects.

The BOSS pipeline performs all the spectroscopic reduction steps, as described in more detail in the relevant SDSS-V publications (e.g., Sections 4.1.\ in \citealt{DR19} and 7.1.1.\ in \citealt{Kollmeier2026}) and in the dedicated publication by Morrison \et (in prep.).
Key calibrations are achieved using dedicated spectroscopy of standard stars and sky regions in each field, as well as calibration sources (i.e., arc lamps). Other parts of the pipeline provide spectral classification and, when applicable, redshift measurements for each source, using a variety of templates. Specifically, spectra are classified as those of a ``QSO'', ``GALAXY'', or ``STAR'' based on template-fitting.
All these classifications are based on (improvements to) the original BOSS pipeline \citep{Bolton2012}. While we imposed several quality criteria to further clean the resulting sample of optical spectra to be used for our analysis (see below), we stress that we made no effort to amend the pipeline-produced spectral classifications (and/or redshifts).
To assess the reliability of the pipeline-based quasar identifications, we make use of an internal visual inspection effort by the SDSS-V/BHM team, which is analogous to the analysis presented in \citet{DR12Q_2017} for the DR12Q catalog. Restricting to visually-inspected spectra which the pipeline classified as QSO with \zwar=0 and $S/N>2$ (23,986 spectra), this process confirms $\sim$98\% as genuine quasars, with only $\sim$1.4\% found to be misclassified stars or galaxies and a further $\sim$0.6\% deemed unclassifiable due to low $S/N$. The overall pipeline reliability inferred here is consistent with the $\sim0.3-1.3\%$ contamination reported for the SDSS-IV/eBOSS DR16Q catalog \citep{DR16Q}, which used the same BOSS pipeline.

Like all DR20-level BOSS data, the spectra we use were acquired up to MJD 60708 (February 2, 2025) and processed through version v6\_2\_1 of the BOSS pipeline. 

\subsection{Ancillary Photometric data}
\label{sec:photo_data}

We used several sources of data to augment the BOSS spectroscopy with relevant broad-band photometry.
The BHM database of reduced BOSS spectra (i.e., the \texttt{spAll} file) includes MIR photometry from the AllWISE Data Release \citep{AllWISE}, $W1-4$ bands in Vega magnitudes, which we used for searching Hot DOGs (see Section~\ref{sec:intro} and Figure~\ref{fig:W12_drops}).\footnote{The use of the AllWISE Data Release instead of the AllSky Data Release of WISE could have a minor effect when applying the Hot DOG selection function of \cite{Eisenhardt2012}.  See \cite{W12drops} for details.} In addition to the Gaia/DR2 \citep{Gaia_DR2} data used for targeting, the \texttt{spAll} file also includes cross-matched data from Gaia/DR3 \citep{Gaia_DR3} and even some data from Gaia/DR1 \citep{Gaia_DR1} (specifically, GAIA\_G\_MAG, BP\_MAG, RP\_MAG). Some of these data included extinction corrections, and others did not, so we could not directly use the magnitudes listed in the pipeline. To obtain the most up-to-date data from Gaia (DR3), we cross-matched our dataset through the Astro Data Lab Science Platform provided by NOIRLab\footnote{\url{https://datalab.noirlab.edu/}} \citep{Fitzpatrick2014_NOAO_Lab,Nikutta2020_DataLAb} with the Gaia DR3 catalog with a cross-matching radius of 3\arcsec. For extinction corrections, we followed \citet{Gaia_DR2} and used the color-free $k_{\rm G}$ term. 
We further retrieved the optical photometry of those sources in our dataset that are covered by the 10th data release of the DESI Legacy Imaging Surveys (LS/DR10; \citealt{DESI}).\footnote{Again using the Astro Data Lab Science Platform, and a cross-matching radius of 3\arcsec.} This dataset provided magnitudes in the \emph{griz} bands, as well as the corresponding extinction-corrected, PSF magnitudes in the AB system, for 230,678 sources.

We stress that these photometric data are used here \textit{not} to re-assess the target-selection techniques (which, at times, relied on slightly different photometric measurements), but instead to examine the multi-band properties of the GUA and \skt\ quasars under study, using some of the most up-to-date data that are available for (a large fraction of) our sources.

%%%%%%%%%%%%%%%%%%%%%%%%%%%%%%%%%%%%%%%%%%%%%%%%%%%%%%%%%%%%%%%%
\section{Analysis and Spectral Measurements}
\label{sec:analysis}

\subsection{Sample Construction and Refinement}
\label{sec:sample} 

As of the cut-off date for DR20 (February 2, 2025), the SDSS-V/BOSS pipeline provided over 8M calibrated spectra, including those of sources classified by the pipeline as stars, quasars, or galaxies. The meta-data for all the spectra are stored in a single \texttt{spAll} file. We used the epoch-coadded spectra, which aim at achieving some pre-defined $S/N$ criteria (for the core targets in each field), at times using integrations from adjacent nights, but without completely losing knowledge of possible longer-timescale source variability. 
As explained in Section~\ref{sec:sdssv}, the same object could be targeted for several different reasons, and moreover could have been observed multiple times (for either scientific or operational reasons). 
{Therefore, we first selected only objects that were targeted within either one of the desired cartons, that is \texttt{bhm\_gua\_dark}, \texttt{bhm\_gua\_bright}, or \texttt{openfibertargets\_nov2020\_27} (aka ASQOSS).
For those objects that appeared multiple times in the pipeline data products (with the same `SDSS\_ID'), we retained only the spectrum with the highest reported $S/N$ available, which resulted in a total of 282,076 unique objects.
For objects marked as `ASQOSS', we kept only the ones that were selected through the \skt\ or \papgua\ selection algorithm (275,287 objects).\footnote{The analysis presented in this paper focuses only on the GUA and \skt\ sources, and does not include the subset of ASQOSS targets selected through the broader \papgua\ selection.} 

As Fig.~\ref{fig:flowchart_trgt} shows, these selection steps provided spectra of 264,200 unique targets, of which 262,053 were part of the \texttt{bhm\_gua\_dark} carton, 60,511 of \texttt{bhm\_gua\_bright}, and 38,029 from ASQOSS/\skt. This breakdown further illustrates the overlap between the various cartons (see Appendix~\ref{app:overlap}).

Not all of these spectra are of sufficient quality for detailed spectral analysis, and in fact, not all of them have correct spectral classification and/or redshift determination. Given the scope of the data, we had to find ways to eliminate from the sample the problematic spectra via an automated process, based on experience gained through the visual inspection of a large set of BOSS spectra.

First, spectra with fatal issues were omitted from our analysis and will not be included in the forthcoming DR20. This identification was done by relying on the information contained within the \zwar\ flags (see Appendix~\ref{app:ZWARNING}), and includes cases where: 
(1) the fiber was a sky fiber; 
(2) the fiber was unplugged or damaged; 
(3) the targeting data had catastrophic issues (i.e., due to some mismatch between targeting catalogs); or (4) the reduced spectral files had no real data due to severe issues with the BOSS spectrographs. 

\begin{figure}[t]
    \centering
    \includegraphics[width=1\columnwidth]{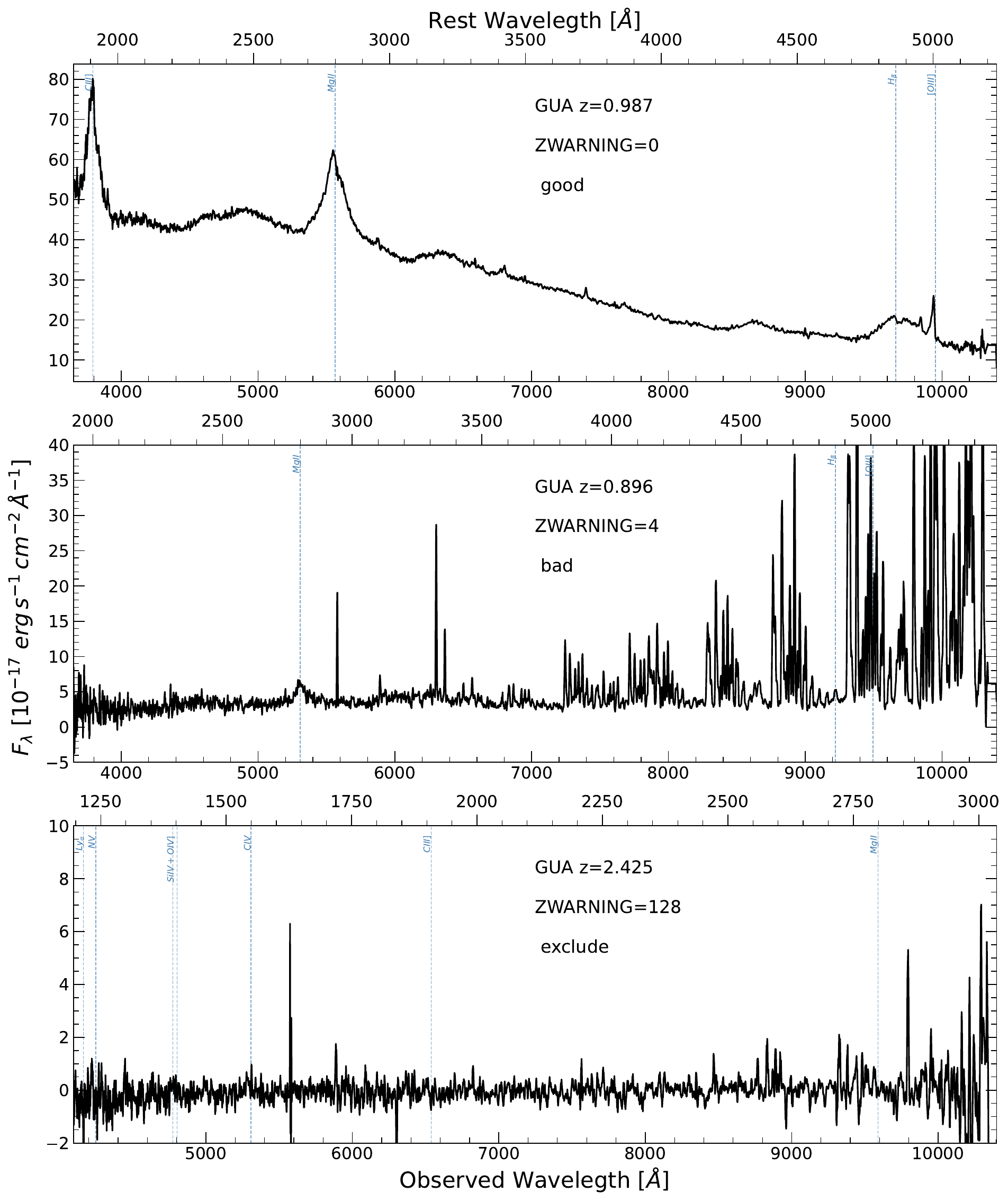}
    \caption{Examples of the three basic categories of SDSS-V spectra in our parent sample. From top to bottom:
    (1) a spectrum of a quasar that passed all the basic quality checks, including object classification and redshift assignment, and is therefore considered a \gd\ spectrum; 
    (2) a spectrum of a quasar where the BOSS pipeline has issued a warning, but is still worth considering for efficiency analysis (i.e., part of the \bd\ sample). 
    (3) a spectrum that was omitted from our analysis (and from DR20), based on its \zwar\ (which in this case suggests the fiber was unplugged).}
    \label{fig:quality}
\end{figure}

We further refined our sample, informed by the \zwar\ and \zerr\ data provided by the BOSS pipeline, and by our (partial) visual inspection.\footnote{$\zerr<0$ means that the 1D pipeline encountered a severe problem while measuring the redshift, and the reported redshift is likely unreliable.}
Specifically, we divided the GUA+\skt\ superset into two sub-samples: 
\begin{enumerate}
    \item The ``\gd'' sample: this sample contains all the spectra with $\zwar = 0$ (i.e., no warning), no catastrophic self-identified issues in redshift determination by the 1D pipeline ($\zerr>0$), a relative redshift error below 5\%, and a positive redshift ($z>0$). The \gd\ subsample comprises 213,528 objects (81\% of the total superset size).

    \item The ``\bd'' sample: contains spectra with non-negligible \zwar\ flags that are not catastrophic, as described in Appendix~\ref{app:ZWARNING}; spectra with $\zerr<0$, and sources with a negative redshift. The \bd\ subsample comprises 50,672 objects (19\% of the superset size).
\end{enumerate}  

We decided to retain these \bd\ spectra for parts of our analysis because, after visual inspection, most of them do show clear broad emission lines and are therefore spectroscopically confirmed quasars, albeit with problematic (pipeline-based) redshift determination.
Specifically, ${\approx}83\%$ of the visually-inspected \bd\ spectra which the BOSS pipeline classified as QSOs, indeed showed broad emission lines. Moreover, almost half of the visually-inspected \bd\ spectra classified as galaxies and ${\approx}16\%$ of those classified as stars---also showed such clear AGN spectral features.
While this \bd\ sub-sample is suitable for statistical studies of the SDSS-V quasar selection and content, only the high-quality (\gd) sub-sample is used for detailed spectral analysis (which assumes the pipeline redshifts are correct; as discussed in Section~\ref{sec:qsofit}). 

We compared our spectral quality cuts to those adopted by \citet{Aydar25}, which were based on a thorough analysis of the BOSS spectroscopy of the eFEDS field \citep{eFEDS}. 
Specifically, for sources with no visual inspection, \cite{Aydar25} adopted the combination of $S/N > 2$, $\zwar = 0$ and $0 < \zerr < 0.005$, to robustly select AGN based solely on the BOSS pipeline outcomes, which is comparable to our situation here.
Overall, the two sets of criteria result in highly consistent samples: ${\approx}$87\% of our \gd\ quasar sample would also pass the \cite{Aydar25} criteria, and almost all (${\approx}$99.9\%) of the \bd\ quasar sample would not pass them. 
However, based on our partial visual inspection, simply applying the \cite{Aydar25} criteria may leave many bona-fide broad-line AGNs out of the \gd\ sample used for spectral measurements (${\approx}$13\% of the \gd\ quasars).
We therefore preferred to only use our quality criteria for sample construction. In any case, choosing instead the \cite{Aydar25} criteria would not significantly change our main findings regarding the key properties of the quasars observed through the GUA and ASQOSS programs.

We present examples of spectra of various quality types in Figure ~\ref{fig:quality}. The top panel shows a \gd\ quasar spectrum and therefore was retained for detailed spectral measurements. The middle panel displays a spectrum with clear quasar-like broad emission lines, but also with significant residual sky features, which in turn greatly reduce the reliability of the measured redshift. Thus, the spectral fitting algorithm would have difficulty fitting a broad line in the sky-affected wavelength range. In this specific example, the \zwar\ flags suggest that the $\chi^2$ of the best fit is too close to that of the second-best one (so $\zwar\neq0$). Therefore, we keep this quasar for the parts of the analysis concerning quasar selection efficiency, as this is indeed a successful example of a quasar targeted by one of the programs under study (GUA in this particular case), but not for the parts of the analysis that rely on detailed spectral measurements. Finally, the bottom panel in Fig.~\ref{fig:quality} presents a completely different level of failure, where we cannot use the spectrum at all, as it is not possible to say whether this (GUA-selected) source is indeed a quasar, although the 1D pipeline classified the spectrum as such (and even assigned a redshift). In this example, the \zwar\ clearly indicates that the fiber was unplugged or damaged. 

Figure \ref{fig:bad_types} presents four examples from the \bd\ sample, each of which falls into this category for a different reason. While some spectra are indeed unsuitable for detailed spectral measurements, others still maintain relatively reliable quality. The top panel shows a quasar with no \zwar\ but with $\zerr = -1$, implying that the best fit is at the lowest or highest redshift tested for a template. Despite this result, it can be clearly seen that this is indeed a quasar with $z\simeq2.365$, as indicated by the 1D pipeline. Similarly, the second example also has a reliable redshift, although its \zerr\ suggests that the best-fit redshift is outside the fitting range. The third panel presents a quasar with a broad emission line, confirming its spectral classification, although the assigned redshift appears to be incorrect, and $z\simeq1.59$ would provide a better fit. The \zwar\ in this case suggests that the $\chi^2$ of the best fit is too close to that of the second-best. Finally, in the bottom panel, the \zwar\ stats that a quasar line exhibits negative emission, which is a warning issued only for objects classified as quasars, so as to assist in identifying potential cases of broad absorption line (BAL) quasars \citep[e.g.,][]{BAL_Weymann1991,BALQSO}. 
However, in this specific case, the redshift must be wrong, and we cannot perform any reliable spectral measurements (again, keeping in mind the enormous scope of the data in hand prohibited reassigning redshifts in such cases). Those four examples demonstrate that the \bd\ sample can contribute to the number of quasars identified through GUA and \skt, even if their currently available reduced spectra prohibit any reliable spectral measurements, including their most basic property -- their redshift. 

\begin{figure}[t]
    \centering
    \includegraphics[width=1\columnwidth]{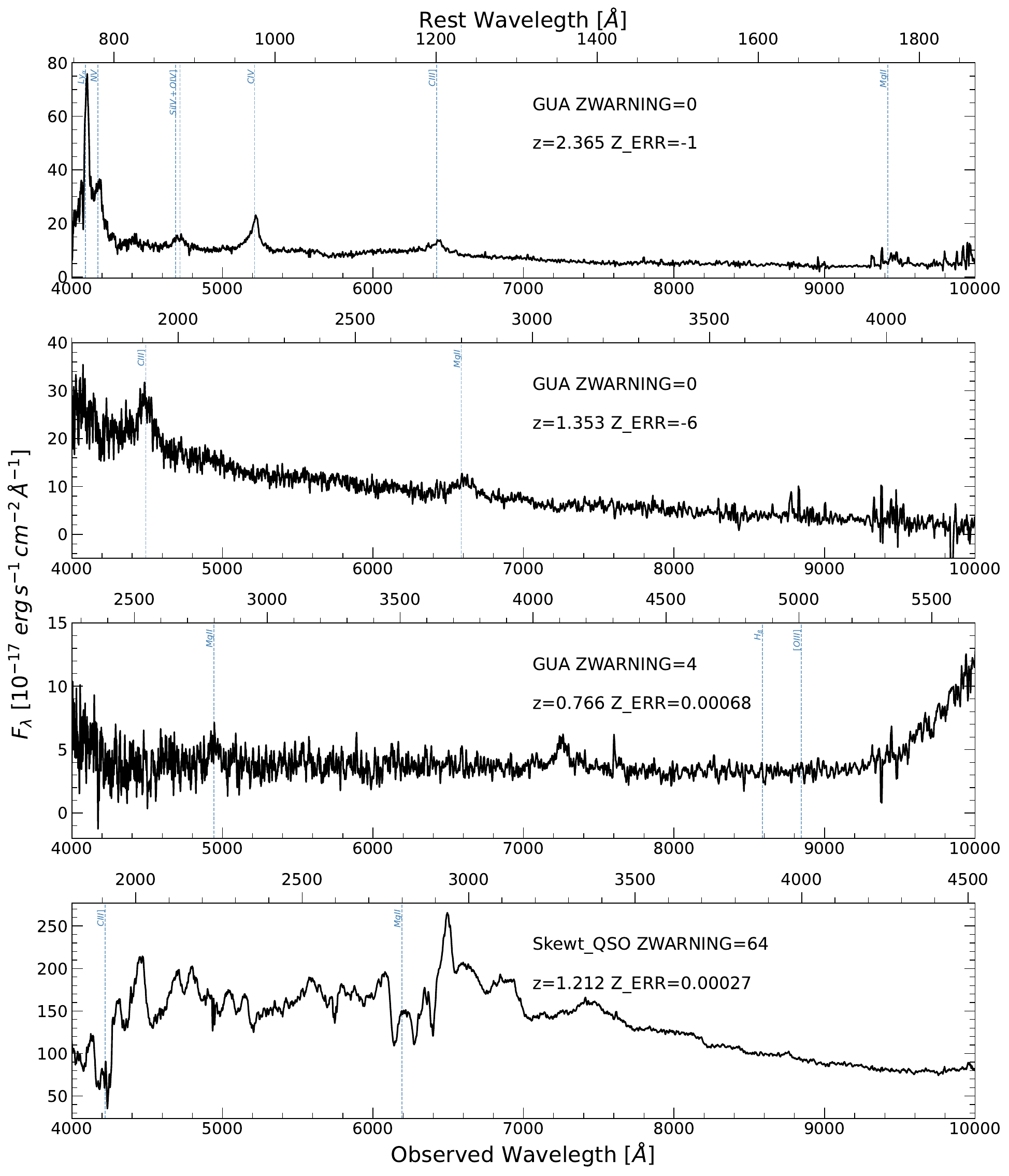}
    \caption{Examples of spectra classified into the \bd\ sample due to issues in the BOSS pipeline products. These classifications are based on pipeline diagnostics (e.g., \zwar and \zerr flags), and do not necessarily imply that the redshift is incorrect.
    From top to bottom: 
    (1) the best-fit is at the lowest or highest redshift tested for a template; 
    (2) the best-fit redshift lies outside the fitting range; 
    (3) the $\chi^2$ value of the best fit is too close to that of the second-best fit; and 
    (4) the spectrum exhibits negative emission (see Appendix~\ref{app:ZWARNING}).
    In all cases, the issues are not necessarily obvious from visual inspection alone, but are reflected in the pipeline quality flags.}
    \label{fig:bad_types}
\end{figure}

\begin{figure}
    \centering
    \includegraphics[width=1\columnwidth]{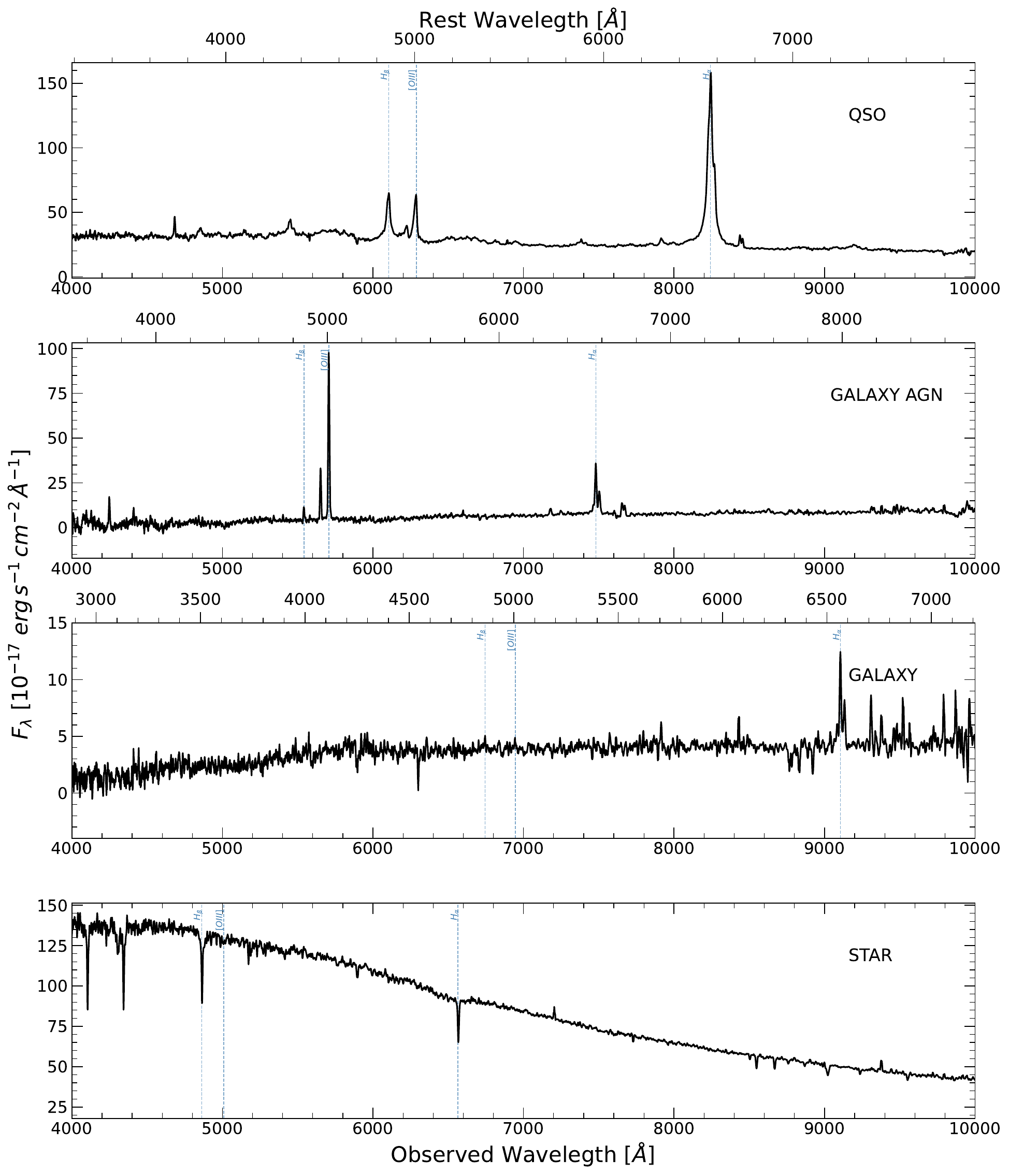}
    \caption{Examples of various types of celestial objects in the GUA- and \skt-based samples.  
    All four spectra are included in the \gd\ sample, as they are not affected by any severe issues (as indicate, e.g.,by the \zwar\ and \zerr). Stars and galaxies with reliable spectra must be part of the selection efficiency analysis, but are obviously not considered for the quasar sample itself (i.e., for spectral measurements).
    From top to bottom: a quasar at $z=0.255$, an AGN at $z=0.139$, a galaxy at $z=0.387$, and an F3/F5V star.}
    \label{fig:obj_types}
\end{figure}

Figure \ref{fig:obj_types} shows four different objects in the \gd\ sample. Although the above steps have significantly reduced the number of non-quasars in our sample, clearly some stars and galaxies remain. Some of the objects classified by the 1D pipeline as galaxies show prominent broad emission lines, and can thus be robustly identified as AGNs. Therefore, we added an AGN sample with galaxies that were sub-classified as `AGN', `BROADLINE', or `AGN BROADLINE'. Only 873 objects fit this definition, and only 797 of those %(455 in `Clean') 
are tagged as ``\gd''. 

%#--------------------------#
\subsection{Spectral Analysis} 
\label{sec:qsofit}

We now describe the detailed spectral decomposition we performed to all quasars in the \gd\ sample of GUA and \skt\ quasars.
For the spectral measurements, we used the publicly available package \pqf, which was introduced in \citet{pyqsofitOG} and \citet{pyqsofit_2}, and further modified as described in \citet{DR16WuShen}.\footnote{We used the \texttt{PyQSOFit\_wqy} version of the package: \url{https://github.com/QiaoyaWu/sdss4_dr16q_tutorial}}
To allow a comparison of our spectral measurements, and derived quantities, to the large catalog of SDSS/DR16Q quasar measurements by \cite{DR16WuShen}, we followed the \pqf\ setup introduced in that work as closely as possible. Below, we briefly summarize the key steps performed by \pqf, and the underlying model assumptions.

Prior to the fitting itself, \pqf\ first corrects each spectrum for Galactic dust attenuation using the maps from \citet[][based on \citealt{dust_map}]{dust_map_2}, and assuming the \citet{extinction_curve} extinction curve with $R_{\rm V} = 3.1$. The spectrum is then de-redshifted to the rest frame using the cataloged (i.e., pipeline-deduced) redshift. 

For the vast majority of sources, located at $z\geq0.3$ (200,749 objects; $\approx$95\% of the sample), we do not account for host-galaxy emission, as it is expected to be negligible at these redshifts and given the typical brightness (and thus luminosities) of our sources. Moreover, the host-galaxy decomposition technique implemented in \pqf\ is limited to $\lambda_{\rm rest}>3400$\,\AA, so at $z\geq0.3$ part of the spectrum would not be correctly treated.
For the minority of sources at $z<0.3$ (9,920 objects in total; $\approx$5\% of the sample), however, we did set up \pqf\ to account for host emission. This analysis is done using Principal Component Analysis (PCA), correlating with galaxy and quasar eigen-templates developed by \citet{Yip2004_gal_temp} and \citet{Yip2004_qso_temp}, respectively. We used the fraction of host emission at 5100\,\AA\ (\texttt{frac\_host\_5100}) from the \pqf\ output file to determine if the host decomposition succeeded or not. 
In some of the spectra, the host template that was fitted had a higher flux than the original data (i.e., host-galaxy + QSO, so a negative eigenvalue for the quasar template). Therefore, in the cases where $\texttt{frac\_host\_5100} \geq 1$ ($\approx$7.4\% of the low-$z$ subset), we reran \pqf\ but with no host decomposition. 

Next, \pqf\ fits the host-free continuum emission and the emission from blended iron features in several ``continuum windows''. The former component is fitted with a power-law and a third-order polynomial, while the latter is fitted with an empirically-derived template that can be broadened and shifted. 
\pqf\ then uses the continuum- and iron-subtracted spectrum to decompose several key emission line complexes, focusing on the spectral regions surrounding the \ha, \hb, \MgII, \CIII, \CIV, \SilIVuv, and \Lya\ lines. Broad and narrow emission lines are fitted with combinations of Gaussian profiles, with some key parameters being tied (e.g., shifts and widths of narrow lines in each spectral region; see \citealt{DR16WuShen}).
Here, we employed two choices that differ from those made in \citet{DR16WuShen}. 
First, we decided to remove the bluest default continuum window ($1150 - 1170$\,\AA) for all high-redshift quasars ($z>3.5$), in order to avoid problematic continuum placement in the spectral regime that is significantly affected by absorption from the intergalactic medium.\footnote{Note that \pqf\ is not (currently) designed to address IGM effects.} 
Second, we limited the widths of the broad Gaussian profiles to $\fwhm <15000\,\kms$. 
Absorption lines, including both intrinsic and intervening features, are common in quasar spectra, particularly at high redshift \citep[e.g.,][]{BALQSO}. To avoid the effect that such troughs may have on the continuum and emission-line fitting, we used the dedicated \pqf\ feature, which iteratively rejects spectral pixels falling $>3 \sigma$ below the original model fit. 
This approach is effective in reducing the influence of narrow absorption features, but is less effective for broad absorption troughs (e.g., BALs), and does not fully account for absorption from the intergalactic medium (IGM).
  
\begin{figure*}
    \centering
    \includegraphics[width=\textwidth]{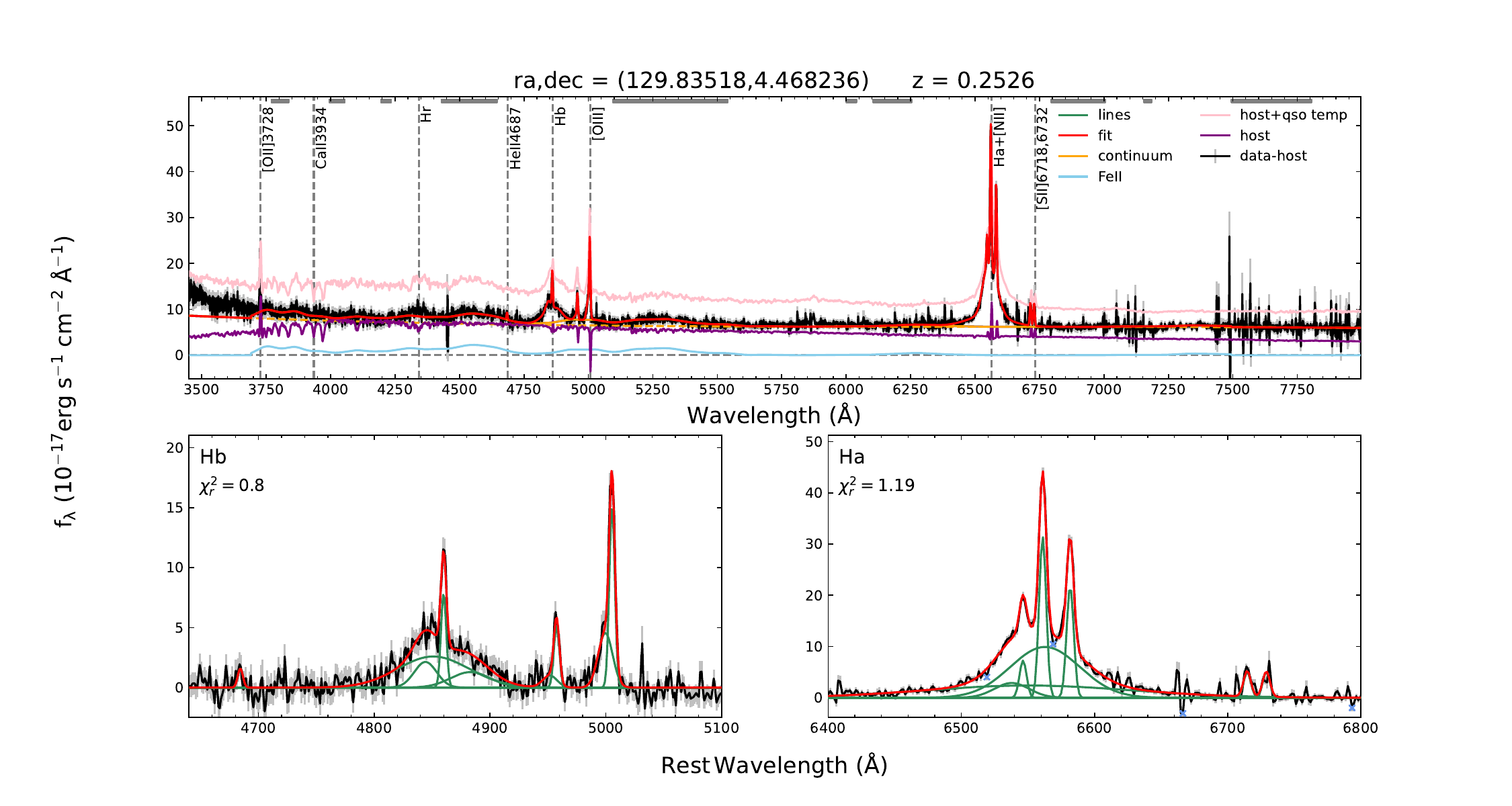}
    \caption{An example of a successful spectral fit using \pqf. The top panel shows the observed spectrum after host subtraction (black), the total best-fit model (red), and its various components: host-galaxy (purple), quasar continuum (orange), quasar and host templates (pink), and blended \feii\ emission (light blue). The gray stripes near the top of the panel mark the continuum fitting bands. The bottom panels display the emission line decomposition for the \ha\ and \hb\ spectral complexes, with the individual broad and narrow Gaussian profile components shown in green.}
    \label{fig:pyqsofit_examp}
\end{figure*}

The spectral measurements characterizing the continuum and line emission are derived from the best-fit model parameters. 
Uncertainties are estimated through a Monte Carlo (MC) procedure, which in turn is based on the error spectrum (i.e., errors on the flux densities). The algorithm generates a set of mock spectra, and each of them is fitted with the same model as the original spectrum. The uncertainty on each quantity is taken to be the half-width between the 16th and 84th percentiles of the corresponding distribution of measurements.
We chose to have 50 such iterations (i.e., 50 mock spectra) for each quasar. This choice is a compromise between the need for a statistical pool of mock fits and computational limitations for our large sample.
An example of how the spectral decomposition works for one of our quasars is shown in Figure \ref{fig:pyqsofit_examp}. 

We extracted several key quantities from the spectral fitting output of each of our quasars for estimating bolometric luminosity, BH mass, and Eddington ratios. 
Continuum and line luminosities are calculated from the model continuum flux densities in the corresponding continuum windows and from the integrated line fluxes.
The widths of the broad emission lines, in terms of \fwhm, are returned by \pqf\ for the fully reconstructed (broad) line profile, which consists of several (broad) Gaussians.
\pqf\ also provides uncertainties on these luminosities and FWHMs, deduced through the aforementioned MC procedure.

%#--------------------------#
\subsection{Bolometric Luminosities, Black Hole Masses, and Eddington Ratios}
\label{sec:LMLLEdd}

The spectral measurements described above are particularly important for estimating some of the key properties of the accreting SMBHs that power the quasars in our sample, including AGN bolometric luminosities (\Lbol), BH masses (\mbh), and Eddington ratios (\lledd), following widely-used and well-established procedures.

To this end, we rely on the measured monochromatic continuum luminosities at rest-frame wavelengths of 5100, 3000, and/or 1350 \AA, in terms of \lamLlam\ (\Lop, \Lthree, and \Luv\ respectively), as determined by \pqf. 
For the low-redshift AGNs in our sample, the measurement of the luminosities may be affected by host-galaxy emission, and/or by how \pqf\ was employed to address it (see above). We therefore also used the luminosity of the broad \ha\ line (\Lbha\ hereafter) for alternative estimates of \Lbol\ and \mbh\ (and thus, of \lledd).

We estimated \Lbol\ by applying simple, universal bolometric correction factors (\fbol) derived from the typical mean SEDs of quasars in \citet{Richards06_SEDs_BCs}. The specific correction factors used were $\fbol = 9.26, 5.15$, and 3.81, for \Lop, \Lthree, and \Luv\ respectively. These values were chosen to be fully consistent with \citet{DR16WuShen}. Some studies suggest a wide range of more elaborate corrections, either based on observed SEDs and/or physical models, often with some dependence on luminosity, \lledd, or other quasar properties \cite[see, e.g.,][and references therein]{Vasudevan2009_BCs,Runnoe2012_BCs,Netzer2019_BCs,Duras20}. Beyond the possible physical insights encoded in these variations, they highlight the inherent systematic uncertainty on any choice of \fbol, which is generally at least 0.2\,dex \cite[e.g.,][]{Gupta2024_BASS_SEDs}. 
For low-redshift quasars, we followed the work of \citet{GreeneHo} and estimated \Lbol\ through a combination of their $\Lbha-\Lop$ relation 
and the aforementioned $\fbolopt=9.26$.
We do not account for any beaming or lensing effects when estimating \Lbol\ for our quasars.

Our estimates of \mbh\ were also derived in the same manner as in \citet{DR16WuShen}, based on ``single-epoch'' prescriptions, as detailed in \citet{DR7Shen} \citep[see detailed discussion in][]{Shen13_mbh_rev}. 
In such prescriptions, the measured luminosity acts as a proxy for the BLR size, while the width of the broad lines is assumed to trace the velocity of the (virialized) BLR gas. 
The \mbh\ prescriptions we used are of the general form of: 
\begin{equation} \label{eq:MBH}
   \log \left(\frac{\mbh}{\Msun}\right) = 
    a + b\log \left(\frac{\lamLlam}{10^{44}\,\ergs}\right) + 2\log \left(\frac{\fwhm}{\kms}\right) \,\, ,
\end{equation}
where $a$ and $b$ are parameters derived from reverberation mapping campaigns. Following \citet{DR7Shen}, who in turn use the \cite{Vestergaard2006_mbh} calibrations, we assumed $(a,b)=(0.91,0.50)$, $(0.74,0.62)$ and $(0.66,0.53)$ when estimating \mbh\ from the \Hbeta, \mgii\ and \civ\ broad lines, respectively (i.e., when using \Lop, \Lthree, and \Luv\ as $L$, respectively). 
While there are many other calibrations available in the literature, our choice is, again, identical to what was used in \citet{DR16WuShen} and in previous large SDSS quasar catalogs. 

For the lower-redshift AGNs where a reliable broad \ha\ component (e.i., b\ha) is detected and measured, we employ a somewhat different prescription, following the one suggested by \citet{GreeneHo}, but re-scaling it slightly by $+0.125$\,dex following \citet{MejiaRestrepo2022_BASS}: 
\begin{equation} \label{eq:mbh_ha}
\begin{aligned}
   \log\left(\frac{\mbh}{\Msun}\right) = 6.43 
   +0.55 \log\left(\frac{\Lbha}{10^{42} \,\ergs}\right) \\
   +2.06 \log\left(\frac{\fwhm [{\rm b}\ha]}{10^3 \,\kms}\right) \,\, .
\end{aligned}
\end{equation}

This latter prescription naturally avoids the potential (residual) contribution of the host-galaxy to \Lop. Here, too, we prefer a \mbh\ prescription that is consistent with previous compilations of (low-redshift) AGNs over several alternatives \citep{Cho2023,DallaBonta2025}.

The significant uncertainties on such \mbh\ prescriptions, of order $\gtrsim$0.3\,dex, are dominated by systematics on all their sub-components (with an additional contribution from the intrinsic scatter in the underlying scaling relations): the $\RBLR-L$ relation(s), the assumption that the BLR itself is virialized, that the measured line width provides a robust probe of these dynamics, and that the overall scaling is correct. 
The reviews by \cite{Shen13_mbh_rev} and \cite{Peterson14_mbh_rev} elaborate on these issues.
Moreover, as extensively discussed in several studies \cite[e.g.,][and references therein]{DR7Shen,TrakhtenbrotNetzer12,ShenLiu12,Mejia-Restrepo16}, the \mbh\ estimates based on the \civ\ line are subject to greater uncertainty compared to those based on the lower-ionization lines (the Balmer and \mgii\ lines), likely due to non-virial gas dominating the \civ-emitting region. 
Alternatively, the \civ\ emission may arise from a combination of gas components with different kinematic properties (e.g., virialized and outflowing components), which can alter the reverberation behavior and introduce additional uncertainty in the inferred radius–luminosity relation and mass estimates \citep[e.g.,][]{Denney2012}. 
In practice, this uncertainty could be traced by blueshifted and/or highly asymmetric \civ\ profiles, or by the poor correlation between the width of \civ\ and the width of \hb\ and/or \mgii. Moreover, \civ\ is often observed to be narrower than the Balmer lines, in contradiction to what is expected from reverberation mapping and BLR models. 
However, in a large sample of quasars such as ours, it's essentially impossible to treat such issues on a case-by-case basis, particularly at $z\gtrsim1.5$, when none of the other broad emission lines we consider can serve as a benchmark.
%#--------------------------#

To obtain the most reliable estimates of \mbh\ and \Lbol, we utilized all the emission lines and continuum windows available for each of our spectra. We prioritized higher-quality spectral measurements, focusing on emission lines that are observed far from the edges of the spectral coverage. 
We first applied cuts on line measurement quality, utilizing the uncertainty assessments provided by \pqf. Spectral measurements were rejected if they had either \(\Delta \fwhm / \fwhm > 40\%\) or \(\Delta \lamLlam / \lamLlam > 20\%\). Additionally, we excluded all the invalid measurements (\pqf-based values were \texttt{nan}, \texttt{inf}, or zero).
Next, we implemented a set of redshift-dependent selection criteria for certain emission line complexes. For low-redshift AGNs ($z<0.4$) we prioritized \ha, while for higher redshifts we used \hb\ as the primary line for $0.4\leq z<0.7$, \mgii\ for $0.7\leq z<2$, and \civ\ for $z\geq2$. These choices are, again, similar to what was done in previous SDSS-related quasar catalogs \cite[e.g.,][]{TrakhtenbrotNetzer12,DR16WuShen}.
If the primary redshift-dependent emission line failed to meet the basic quality criteria, we adopted a predefined fallback sequence: \ha\ was replaced by \hb; \hb\ by \ha\ (preferred, for lower $z$) or \mgii\ (if \ha\ is unavailable, at $z\gtrsim0.4$); \mgii\ would be replaced by \hb\ (preferred) or \civ\ (at $z\gtrsim1.5$); and \civ\ by \mgii\ (at $2\lesssim z\lesssim2.3$). 
For each \mbh\ estimate retained, we also kept the corresponding \Lbol\ measurement\footnote{For example, if the \mgii\ line measurement was excluded, then \Lbol\ would {\it not} be based on \Lthree, but instead on the representative luminosity of the spectral region that was ultimately chosen (\Lop, \Luv, or \Lbha, depending on the redshift).} and kept only spectra with a valid \mbh\ estimate.

After estimating \mbh\ and \Lbol\ for our sources, we combined these quantities to estimate \lledd,  assuming 
\begin{equation} \label{eq:eddR}
  \lledd\equiv\frac{\Lbol}{\Ledd} = \left(\frac{\Lbol}{\ergs}\right) \left[1.3\times10^{38} \left(\frac{\mbh}{\Msol}\right)\right]^{-1} \,\,,
\end{equation}
which is appropriate for pure-hydrogen gas.
This choice is, again, identical to \cite{DR16WuShen}, but provides \lledd\ estimates that are higher (by 0.06\,dex) than those derived assuming a Solar metallicity gas, as is the case in many other studies.

%#--------------------------#

\subsection{Final Refinements} 
\label{sec:Final_Refinements}

\begin{figure*}[t]
    \centering
    \includegraphics[width=\textwidth]{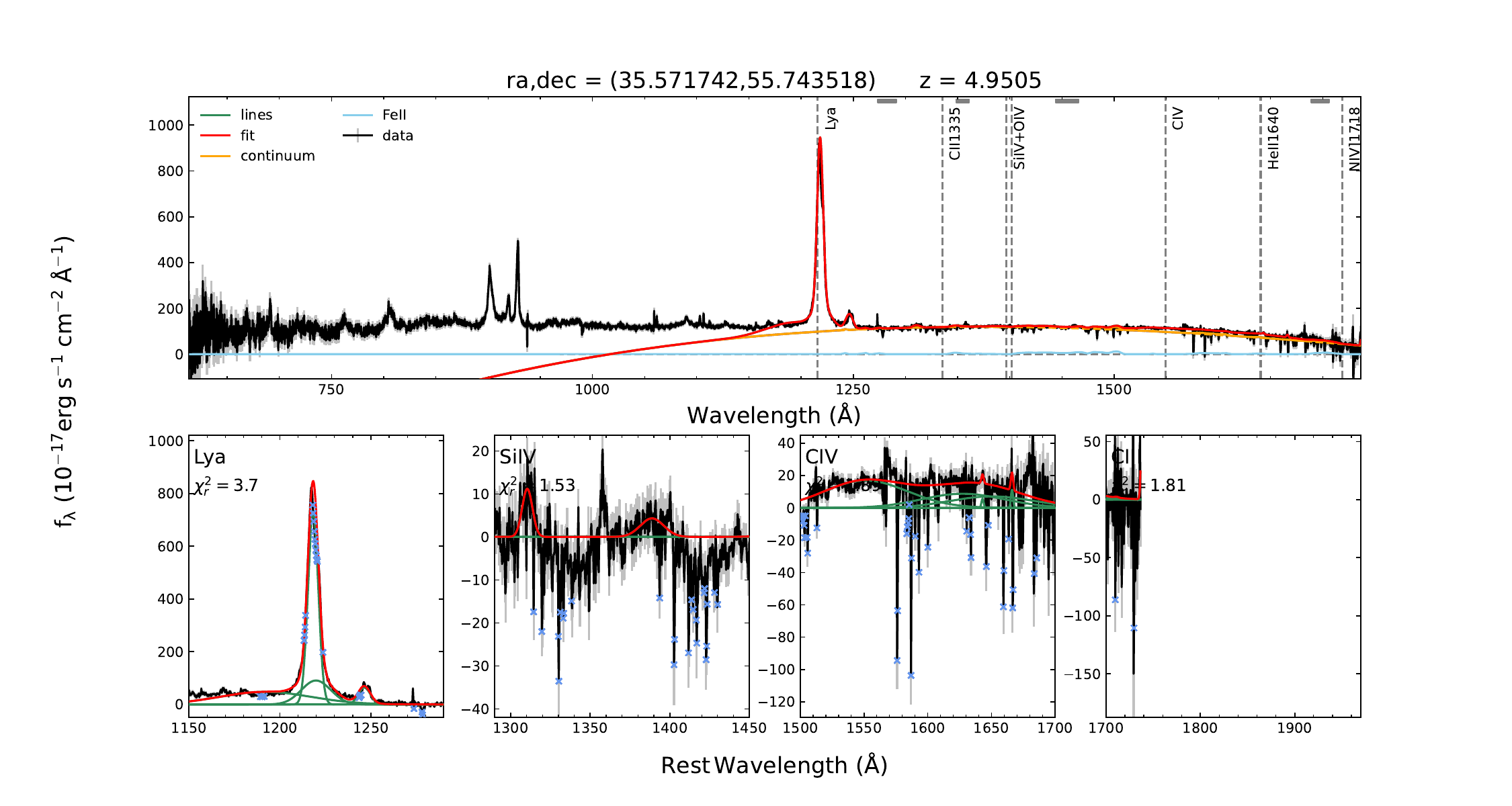}
    \caption{A source misclassified as a high-redshift quasar ($z=4.95$) by the BOSS pipeline, which we identified in our visual inspection of \pqf\ measurements yielding extreme quasar properties (in this case, extremely high \Lbol). The legend description is similar to Fig.~\ref{fig:pyqsofit_examp}.
    The erroneous redshift assignment is driven by the \ha\ line being mistaken for \Lya, however the broad \hb\ and narrow \oiii\ lines clearly indicate a redshift of $z \simeq 0.1$. }
    \label{fig:wrong_z}
\end{figure*}

While the procedures described in Sections~\ref{sec:qsofit} and \ref{sec:LMLLEdd} operated automatically over our entire sample of quasars, we recall that this sample spans a wide range of redshifts and data quality, and includes a variety of subclasses, including, e.g., BAL quasars and significantly reddened sources.
We therefore chose to further visually inspect several subsets of extreme cases (totaling $\approx$2,000 spectra), in order to identify recurring issues and to ultimately refine the sample and measurements used in our analysis. 
We specifically examined quasars with extremely high or low values of \Lbol\ or \mbh, or high redshifts. Upon visual inspection, we identified additional cases with incorrect redshift assignments, BAL quasars for which \pqf\ provided exceptionally large FWHMs (resulting in unrealistically high \mbh), and other issues. All those problematic cases (totaling $\approx$200 spectra) were excluded from our spectral analysis sample. 
However, spectra that did not fall into these extreme categories were not visually inspected, and therefore potential issues in such cases may remain undetected. For example, BAL quasars that yielded reasonable \mbh\ and \Lbol\ values were not flagged and likely remain in the sample.
An example of a misclassified redshift is shown in Figure \ref{fig:wrong_z}.
Notably, all quasars processed through \pqf\ had no \zwar\ or other quality flags indicating issues with redshift determination, meaning these problematic spectra could only be identified through spectral fitting and visual inspection. 

We thus consider our final set of spectral measurements to be as statistically reliable as other, similarly large compilations of quasar properties, which were analyzed with automatic procedures and could not be visually inspected in a comprehensive way.

%%%%%%%%%%%%%%%%%%%%%%%%%%%%%%%%%%%%%%%%%%%%%%%%%%%%%%%%%%%%%%%%
\section{Results and Discussion} \label{sec:results}

In this Section, we discuss the characteristics of the large quasar sample(s) selected through the novel selection methods utilized in SDSS-V/BHM, including the efficiency of the selection methods themselves, their photometric properties, and their distributions in key SMBH-related properties (\Lbol, \mbh, and \lledd).

\subsection{Final Sample} \label{sec:final_sample}

Our final sample of quasars can be considered in four different ways, as described in Table \ref{tab:samples_sum}. First, the \textbf{total sample} excludes the spectra with catastrophic errors, but includes all other spectra: both \gd\ and \bd\ spectra of quasars, as well as galaxies and stars. Second, the \textbf{quasars-only} subset focuses on all the quasars within the total sample, both \gd\ and \bd. 
These two samples are essential for getting a complete picture of all the objects selected through the GUA and \skt\ selection methods and are used for quantifying their efficiency. 
Next, the sample \textbf{\gd\ quasars} includes only those quasars for which the determination of redshifts and other basic (meta-)data is robust (see Section~\ref{sec:sample}). This sub-sample is used to investigate the photometric properties of the quasars, and---more importantly---as the input for the spectral measurement procedure (Section~\ref{sec:qsofit}). 
Finally, the sub-sample of \textbf{quasars with spectral measurements} includes those quasars for which we were indeed able to derive key AGN and SMBH properties (\Lbol, \mbh, \lledd), based on the detailed spectral decomposition (and subsequent visual inspection; see Section~\ref{sec:Final_Refinements}). 
Table~\ref{tab:samples_sum} further illustrates how each of these sub-samples can be further divided based on the selection method that is associated with each spectrum (i.e., mainly GUA or \skt).
This structured approach provides a comprehensive assessment of our sample while also focusing on the most reliable data for each aspect of the detailed analysis that we will present below.

To assess how many newly identified quasars are provided by the GUA and \skt\ samples within SDSS-V/DR20, we cross-matched the quasars-only sample(s) with the largest compilation of spectroscopically confirmed quasars (and AGNs) publicly available at the time of writing: a union of the Milliquas/v8 compilation \citep{milliquas} and the DESI/DR1 AGN/QSO Value-Added Catalog \cite[VAC;\footnote{See \url{https://data.desi.lbl.gov/doc/releases/dr1/vac/agngal/}}][]{DESI_DR1}.
Milliquas/v8 already contains the largest previous SDSS-based quasar catalog, based on SDSS/DR16Q \citep{DR16Q}. 
For our purposes, we down-selected only those Milliquas/v8 entries classified as type ‘Q’ or ‘A’ (i.e., having broad emission lines).\footnote{See \href{https://quasars.org/Milliquas-ReadMe.txt}{Milliquas-ReadMe}} 
The cross-match was done using a 3\arcsec\ threshold, and ignoring any potential mismatches in redshifts, sub-classifications, etc. (e.g., quasar vs. type-II AGN).
Our sample has a total of 151,972 newly spectroscopically confirmed quasars, with 76,840 of these belonging to the so-called ``GUA+\skt\ sole-selected'' subset, meaning they were not selected by neither the core programs or the CSC one (see Section~\ref{sec:overlap}).
For 116,487 (52,548 sole-selected) of our new quasars, we obtained satisfactory estimates of \Lbol\ and \mbh\ (and thus of \lledd). The majority of the newly identified quasars (112,512; or 50,628 sole-selected) are observed from the southern SDSS-V site (at LCO), and---importantly---more than half (120,879; or 55,933 among the sole-selected) are located in the southern celestial hemisphere ($\delta<0$), as can be seen in Figure~\ref{fig:maps}. In contrast, the corresponding fraction among the AGNs in the Milliquas/v8+DESI/DR1 super-set is merely ${\approx}$23\% (495,252 of 2,188,595 objects). 
This result clearly demonstrates the great impact that SDSS-V/BHM is having on the census of AGNs across the entire sky. 

The redshifts of the quasars in our sample span $0.0013\lesssim z \lesssim5.3$, and their full redshift distribution is shown in Figure~\ref{fig:z_dist}, compared to the well-established SDSS/DR7Q \citep{DR7Schneider} and SDSS/DR16Q \citep{DR16Q} quasar catalogs. Both the differential and cumulative distributions demonstrate that our sample spans this redshift range in a continuous manner, without any obvious drops or gaps. 
Moreover, among the newly discovered quasars, 575 (i.e., 0.22\%) have redshifts greater than 4, while the Milliquas/v8+DESI/DR1 super-set contains 8,301 quasars (0.38\%) with $z>4$. 
The number of $z>4$ quasars in our sample(s) is based on extensive visual inspection and filtering of all the sources initially flagged as being at $z>4$ (as explained in Section~\ref{sec:Final_Refinements}). 
Given the challenges involved in correctly identifying such high-redshift quasars (in general and in SDSS-V/DR20), we defer any further discussion of this population to a future publication.

\begin{figure*}[ht]
    \centering
    {All GUA and \skt}\hspace{4.9cm}{New and Previously Known Quasars}\\
    \includegraphics[width=1\columnwidth]{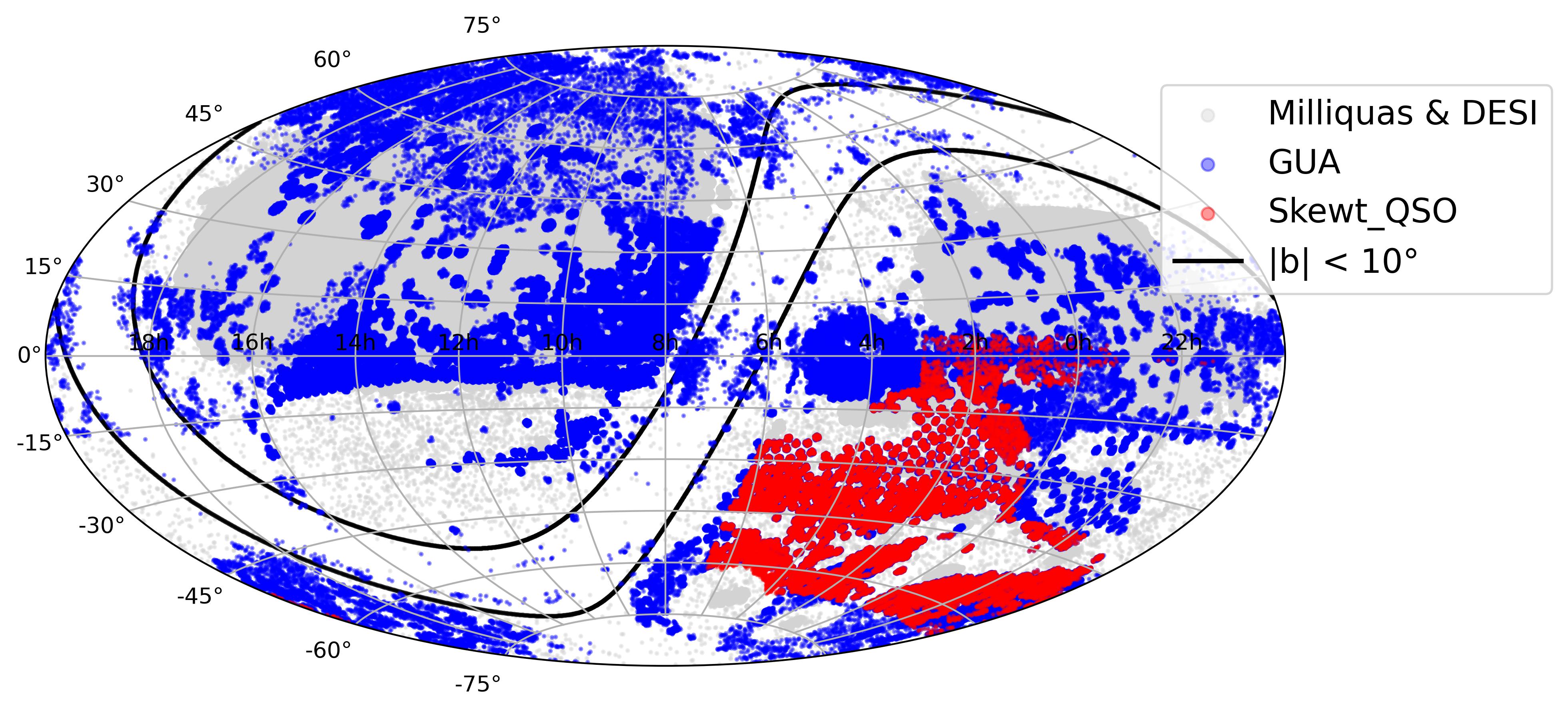}
     \includegraphics[width=1\columnwidth]{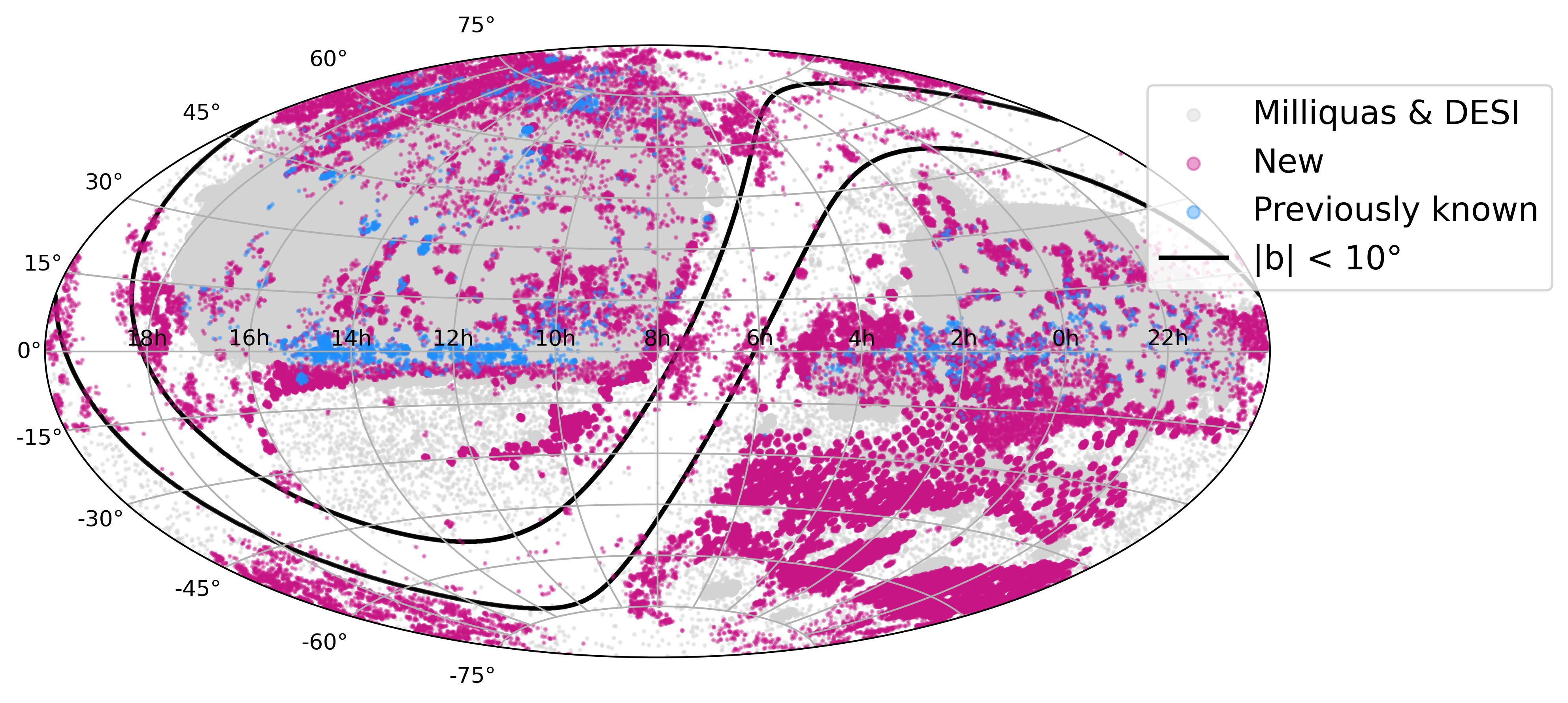}
    \caption{All-sky maps of the quasars observed within the GUA and \skt\ samples (including both \gd\ and \bd\ samples), compared with previously-known quasars  (gray; from Milliquas/v8 and DESI/DR1). {\it Left:} the GUA (blue) and \skt\ (red) samples. {\it Right:} the union of the two sole-selection subsets, separated to newly identified quasars (magenta) and previously-known ones (light blue).
    In both panels, the various symbols are (over-)plotted in the order indicated by the respective legend(s). In practice, there are many GUA quasars in the same part of the sky dominated by \skt\ quasars (i.e., blue points underneath the red points in the left panel), and many Milliquas \& DESI quasars in the regions that also have newly discovered quasars (i.e., gray points underneath magenta ones in the right panel).
    }
    \label{fig:maps}
\end{figure*}
 
\begin{deluxetable*}{l|c|ccc|c|c|l}
\tablecaption{Summary of the sample processing}
\label{tab:samples_sum}
\tablewidth{\columnwidth}
\tabletypesize{\scriptsize}
\tablehead{
\colhead{Sample} & \colhead{$N_{\rm tot}$\tablenotemark{\footnotesize a}} & 
\multicolumn{3}{c}{$N$\tablenotemark{\footnotesize b}} & 
\colhead{New} & \colhead{New} & \colhead{}\\
[-3ex]
\colhead{} & \colhead{} & 
\colhead{$N_{\rm GUA}$} & \colhead{$N_{\rm \skt}$} & \colhead{$N_{\rm S19}$} & 
\colhead{in SDSS\tablenotemark{\footnotesize c}} & \colhead{quasars\tablenotemark{\footnotesize d}} & \colhead{Sections}
}
\startdata
Total observed          & 275,287 & 263,212 & 38,029 & 88,784 &  \nodata & \nodata         & Sec.~\ref{sec:purity} \\
Quasars only            & 259,700 & 248,023 & 37,056 & 85,848 &  108,923 & {\bf 76,840}    & Sec.~\ref{sec:photometry} \\
Good quasars            & 220,976 & 209,876 & 33,818 & 77,043 &  83,495  & 58,237          &  \nodata \\
w/spec. measurements    & 207,892 & 197,158 & 32,595 & 73,890 &  75,903  & {\bf 52,548}    & Sec.~\ref{sec:M_L_dist} \\
\enddata
\tablenotetext{a}{The number of unique objects in the union of the GUA and ASQOSS programs (for ASQOSS: only \skt\ \& \papgua).}
\tablenotetext{b}{The number of objects selected by each of the selection methods (GUA, ASQOSS/\skt, and ASQOSS/\papgua).}
\tablenotetext{c}{The number of new quasars compared with SDSS/DR16Q \citep{DR16Q}.}
\tablenotetext{d}{The number of new quasars compared with the union of Milliquas/v8 and DESI/DR1.}
\tablecomments{Here, `S19' refers to the portion of ASQOSS selected from the original \papgua\ catalogs that is {\it not} included in either the GUA or \skt\ subsets.}
\end{deluxetable*}

\begin{figure}[t]
    \centering
    \includegraphics[width=1\columnwidth]{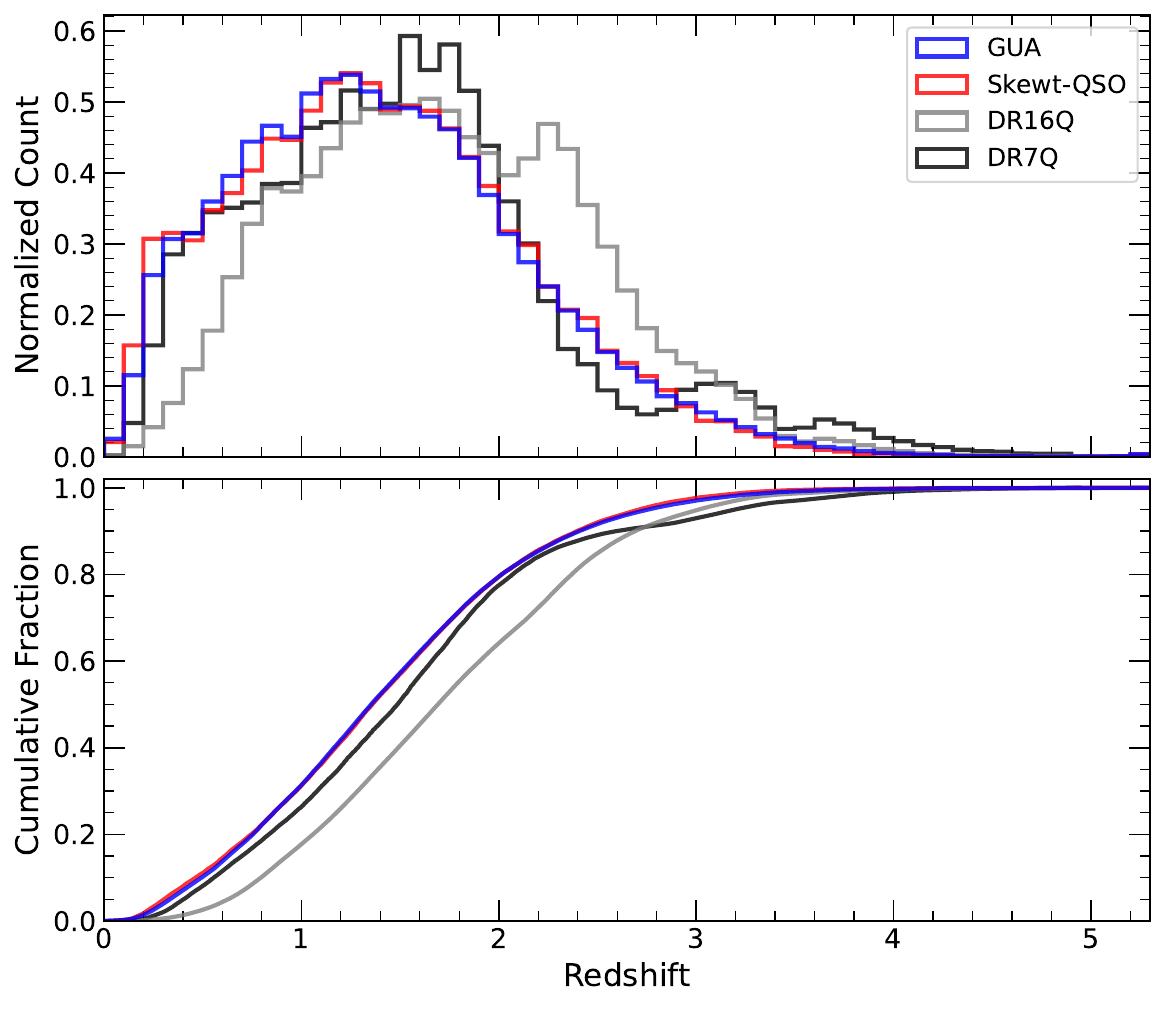}
    \caption{The differential (top) and cumulative (bottom) distributions of redshift for the quasars in our two samples (GUA and \skt), compared to the SDSS/DR7Q \citep{DR7Schneider} and SDSS/DR16Q \citep{DR16Q} reference quasar catalogs.}
    \label{fig:z_dist}
\end{figure}

%#--------------------------#
\subsection{Quasar Selection Purity} \label{sec:purity}

An essential aspect of evaluating the quality of the GUA and \skt\ samples is assessing their purity (or efficiency), i.e., the fraction of spectroscopically-confirmed quasars\footnote{Or, more precisely, broad-line systems with significant AGN contributions to their continuum emission.} identified among the larger set(s) of quasar candidates. 
We investigated the purity achieved by the selection methods and observing strategies of both the GUA and \skt samples. 
To account for the complex target selection and observing strategy, and in particular the significant overlap between the target pools of GUA, \skt, and/or higher-priority programs (e.g., SPIDERS), we compute and use a \textit{homogenized purity} (\hp\ hereafter) for each selection method.\footnote{Whenever we use ``purity'' in what follows, we refer to \hp, unless noted otherwise.} 
In this procedure, which is described in detail in Appendix~\ref{app:overlap}, we focus on a `test area'--located within the high-Galactic-latitude southern sky--where there are targets selected by GUA, \skt, and SPIDERS. 
We split all the targets of these three selection methods into seven mutually exclusive subsets based on co-targeting flags, which are (conceptually) presented in Figure~\ref{fig:venn}.
We then calculate the quasar selection purity within each such subset, and average over the (relevant) subsets weighing by their fractional contribution to the full target pool to derive \hp\ for GUA and \skt\ (see Eqs.~\ref{eq:hp_gua} and \ref{eq:hp_skt}, respectively).
In these calculations, we considered spectroscopically-confirmed quasars from both the \gd\ and \bd\ subsets as the success metric of the two selection methods (i.e., the numerator in the fraction that quantifies the purity).

\begin{figure}[t]
    \centering
    \includegraphics[width=1\columnwidth]{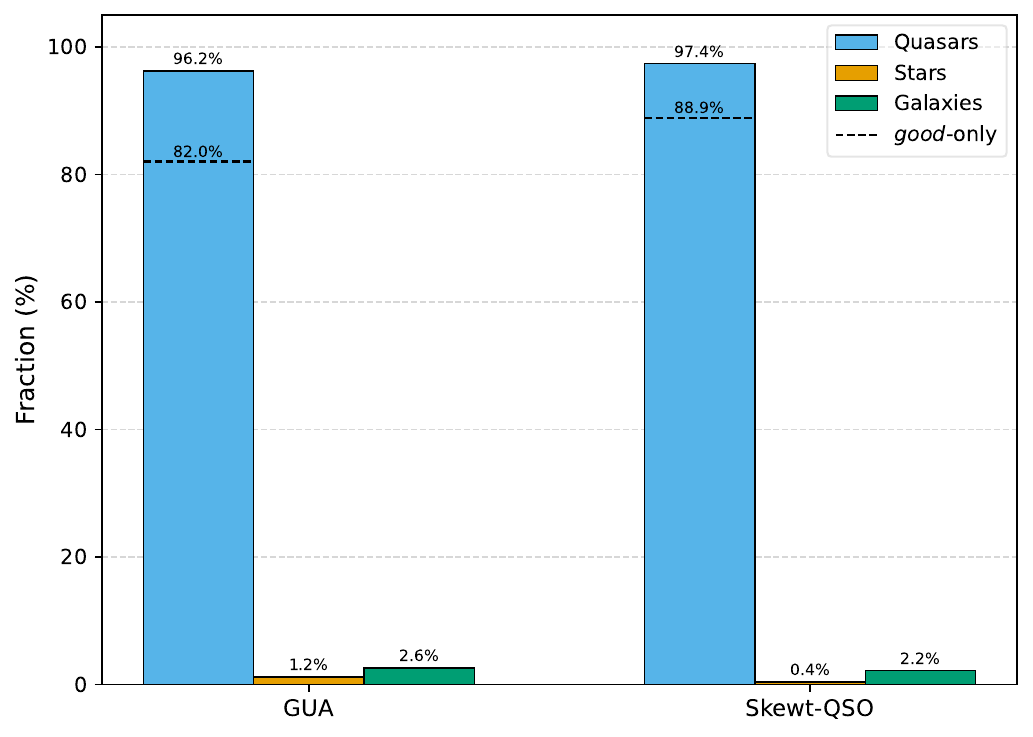}
    \caption{The purity (efficiency) of the quasar selection among the GUA and \skt\ samples. The two bar charts show the \hp\ of GUA and \skt\ samples. Notably, both of the methods achieved very high purity, over 96\%. The black dashed lines indicate the corresponding lower limits, obtained when only \gd\ quasars are counted as genuine quasars (see text).}
    \label{fig:bar}
\end{figure}

\begin{figure*}[ht]
    \centering
    \includegraphics[width=1\columnwidth]{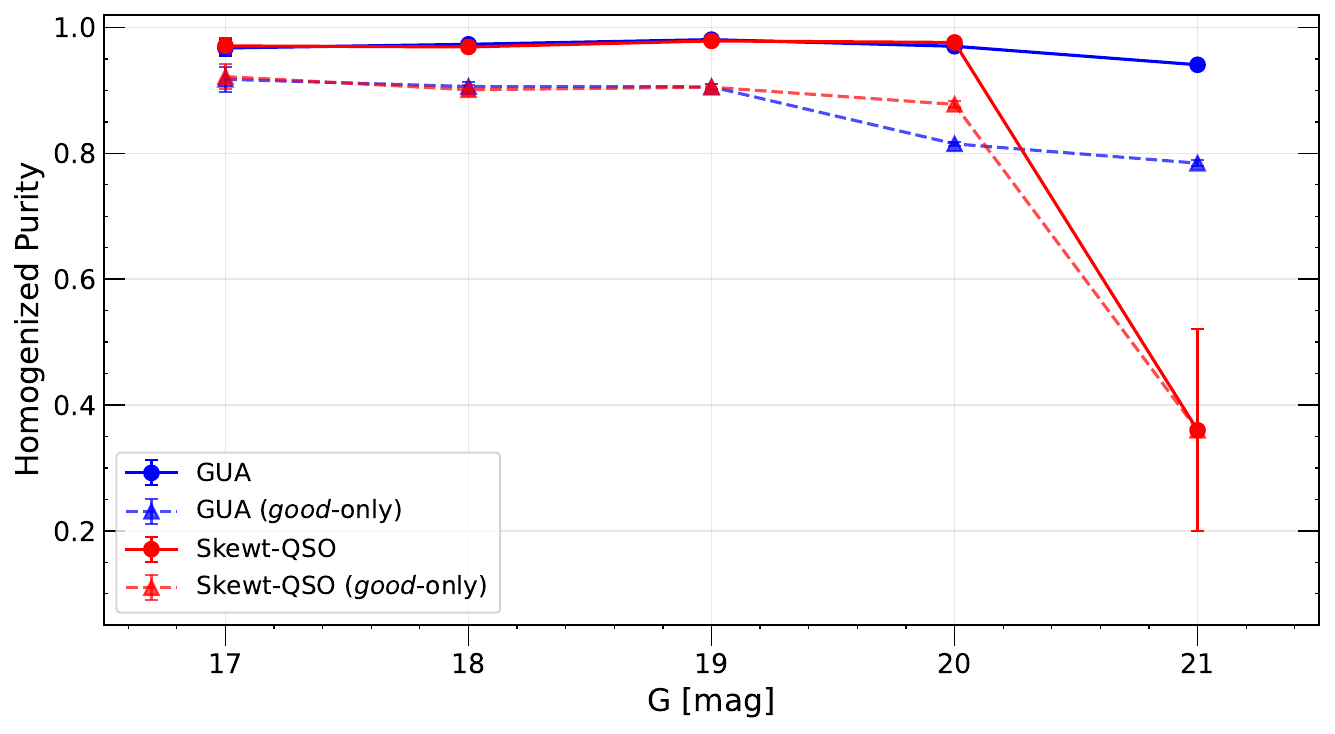}
     \includegraphics[width=1\columnwidth]{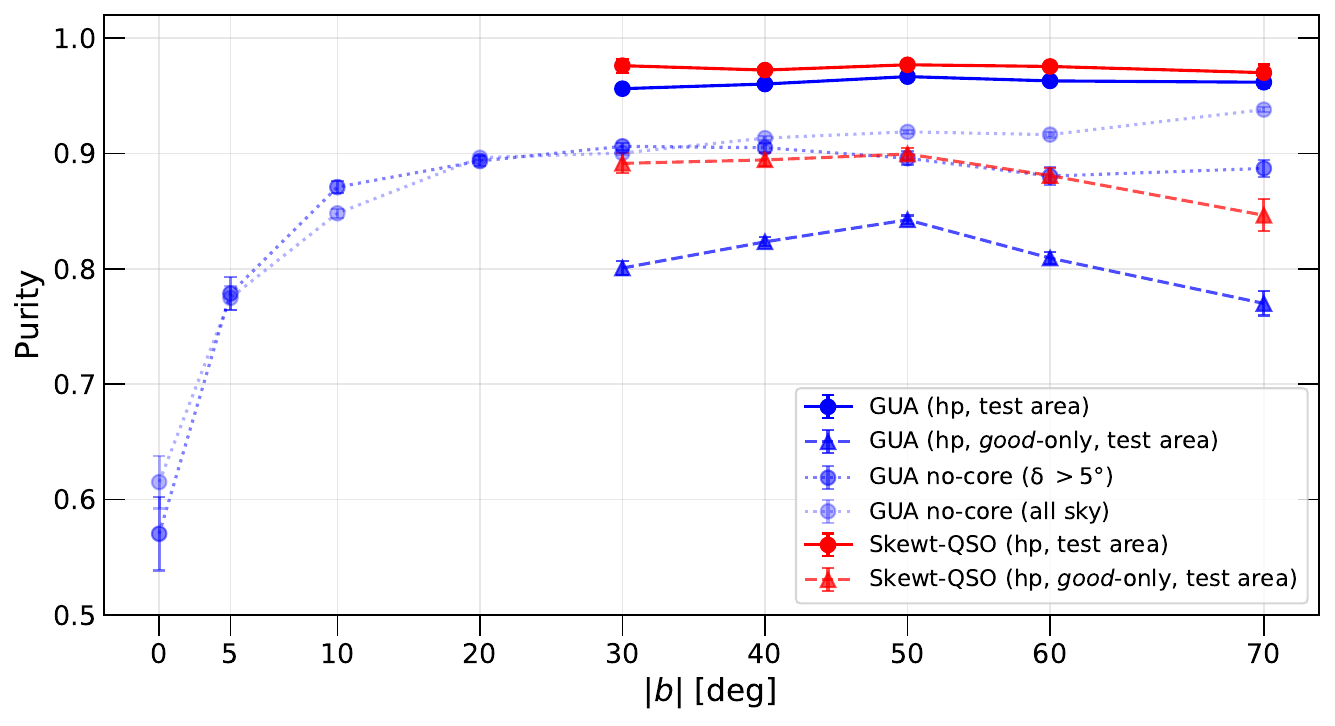}
    \caption{The quasar selection purity of the GUA and \skt\ samples as a function of source brightness (left) and Galactic latitude (right). 
    In both panels, the dashed lines with upward triangles show the corresponding lower limits, obtained by counting only \gd\ quasars as genuine quasars. Error bars represent standard binomial errors.
    {\it Left}: The quasar selection homogenized purity (\hp; explained in Appendix~\ref{app:overlap}) of the GUA and \skt\ samples as a function of Gaia $G$-band magnitudes. Both samples are highly pure for $G<20$ mag. 
    {\it Right}: The quasar selection \hp\ of the GUA and \skt\ samples as a function of Galactic latitude, $b$. Here, the blue and red solid lines are the \hp\ in the `test area' of the GUA and \skt, respectively. The blue doted and dashed lines are the overall (i.e., not homogenized) purities of the GUA sole-selection sample, for all-sky and for $\delta >+5^\circ$, respectively. The \papgua-based GUA program retains good quasar selection efficiency even when targeting low Galactic latitudes.}
    \label{fig:G_b_hp}
\end{figure*}

Figure~\ref{fig:bar} presents the \hp\ of the two main samples considered, GUA and \skt, including both the \gd\ and \bd\ subsets.
Both the GUA- and \skt-related selection methods (and survey strategies) achieved exceptionally high efficiencies of $\hp=96.2\%$ and 97.4\%, respectively. 
If only the \gd\ quasars are counted as genuinely identified quasars (i.e., excluding the \bd\ subset from the \hp\ nominators), the corresponding purities decrease to $\hp=82.0\%$ for GUA and 88.9\% for \skt, providing a conservative lower limit on the selection purity.

In Figure \ref{fig:G_b_hp} we present the \hp\ of both selection methods as a function of brightness ($G$-band magnitudes; left) and the \hp\ and standard purity as a function of Galactic latitude ($|b|$; right). 
The AGN selection purity (within the ``test area'') remains exceptionally high, $\hp>90\%$, across the range $17<G<20$ mag. For fainter sources, $20<G<21$, the purity of GUA remains high, while that of \skt\ declines to well below $60\%$. 
This is likely driven by the combination of two limiting factors.
First, at the target selection stage, the Gaia/DR2 astrometric uncertainty at $G\gtrsim 20$ mag is such that it becomes increasingly challenging to differentiate between Milky Way (MW) stars and quasar candidates based on proper motion (or lack thereof; see Section~\ref{sec:target} and \citealt{Lindegren2018}). 
Second, the SDSS-V/BHM observations, and particularly the shorter exposure times associated with many of our ASQOSS targets, naturally limit the quality of the acquired spectra, and thus of the ability to robustly classify such faint sources (see Section~\ref{sec:sdssv_spec}). 
Figure~\ref{fig:G_b_hp} also shows the corresponding lower limits, computed by counting only \gd\ quasars as genuine quasars. These lower limits closely track the fiducial \hp\ curves across most of the brightness and Galactic latitude ranges, indicating that the general trends seen across these quantities are robust to the choice of quality cuts.

The right-hand-side panel of Fig.~\ref{fig:G_b_hp} shows that the \hp\ of both GUA and \skt\ remains high across the entire high-Galactic-latitude range probed by the test area, $30^\circ \leq  |b| \leq 70^\circ$.
To further check how efficient the \papgua-inspired GUA selection is at lower Galactic latitudes, we considered the overall (i.e., not homogenized) quasar selection purities among two GUA-focused samples, going beyond the test area: (1) {\it all} the sources within our data that are part of the GUA target pool, but not part of any core program, and (2) the subset of such sources that are located at northern declinations ($\delta>+5^\circ$), thus eliminating any overlap with \skt. 
The quasar selection purities of these two large, GUA-focused samples are shown as dashed and dotted blue lines (respectively) in the right panel of Fig.~\ref{fig:G_b_hp}.
The purity of the GUA selection remains rather high even at low Galactic latitudes, dropping below ${\approx}80\%$ only at $|b| \lesssim 5^\circ$.
The high-Galactic-latitude extensions of these two GUA-focused samples again demonstrate the high efficiency of GUA to select quasars in parts of the sky that are not covered by other key quasar selection methods and/or (core) quasar/AGN-oriented programs within SDSS-V.

Comparing the overall efficiencies of the \papgua-inspired GUA method and the \papdes-based \skt\ one at face value, it appears that the latter is slightly more efficient in identifying quasars ($\hp = 96.1\%$ vs. 97.4\%, respectively). 
This may not be surprising given that the \skt\ selection incorporated more nuanced multi-band optical photometry, as well as NIR photometry, available the high-Galactic-latitude DES footprint (and thus our test area; Section~\ref{sec:target}). 
Indeed, Fig.~\ref{fig:G_b_hp} clearly shows that the purity of GUA approaches that of \skt\ in for $|b|> 30^\circ$.
  
The quasar census enabled by the two quasar selection methods we study and their implementation within SDSS-V is, in fact, extremely efficient compared to past efforts.
For reference, we first note that the legacy SDSS spectroscopic quasar selection procedure \citep{Richards02} was designed to achieve a purity of $\sim$65\%. Even under our conservative calculations, which focus only on \gd\ quasars, the purity of both GUA and \skt\ ($\geq82\%$) still substantially exceeds that of the legacy SDSS procedure of \cite{Richards02}.
That pioneering high-Galactic-latitude quasar selection methodology had to rely solely on 5-band optical photometry, with essentially no astrometric data, thus being challenged by a high incidence rate of foreground stars.

We next compare the performance of the two methods to the expectations set in the studies that designed them.
The original \papgua\ work estimated the efficiency of their C75 catalog at  ${\approx}79\%$ reliability, while \papdes\ estimated the efficiency of the \skt\ selection at ${\approx}80\%$. 
We note that even the conservative \gd-only lower limits meet (or indeed exceed) these original expectations.
It thus appears that the SDSS-V implementation of the \papgua\ approach, that is GUA, is outperforming the expectations in the original \papgua\ work, likely due to the more stringent cuts made in the former compared with the latter ($P_{\rm RF}\geq0.8$ vs.\ $\geq0.69$, respectively). 
As for the \skt\ method, the higher-than-expected purity of the \skt set (97\% vs. 80\%) clearly shows that this method is exceptionally efficient in identifying objects that are, indeed, quasars, even if our knowledge of their nature could perhaps be attributed to other programs and selection methods employed in SDSS-V (which also dictate longer exposures under preferable conditions).

To better understand the limitations of the two selection methods, we examined the types of sources that ``contaminated'' the quasar sample(s) the most. 
Based on the BOSS pipeline spectral classifications, MK-type stars are the type of  stellar contaminants in our data, comprising ${\approx}51\%$ of the stars in the GUA sample and ${\approx}29\%$ of the stars in the \skt\ sample. 
More particularly, M-type stars comprise ${\approx}30\%$ and ${\approx}12\%$ of the two samples (respectively).
The rest of the stellar contaminants are distributed across essentially all the spectral classes covered by the BOSS pipeline, including spectra classified as white dwarfs, cataclysmic variables, and more.
This dominance of MK-type stars is not surprising, given their intrinsically high abundance in the MW, as well as their intrinsically faint and red SEDs, which could be confused with those of (high-redshift) quasars. 
Indeed, their dominance is particularly pronounced within the GUA sample, which reaches fainter flux levels.
We note that, by definition, the MK-type stars in our sample have no obvious proper-motion or parallax as of Gaia/DR2, and moreover, many of them are found at high Galactic latitudes. Such stars could be of interest for studies of the structure and evolution of the MW.
In any case, more sensitive Gaia data (e.g., DR3 and 4), with more accurate proper-motion measurements, are expected to help with improving the purity of quasar selection efforts that rely on \papgua- or \papdes-like approaches. 
Such improvements are beyond the scope of the present analysis, and are {\it not} expected to be fully implemented in the final phases of SDSS-V operations.

A detailed analysis of how the selection purity may change with quasar redshift, which would require assessing a large input quasar sample with known redshifts combined with non-quasar contaminants, is beyond the scope of the present paper. 
We however note that neither of the two original quasar selection method studies (\papgua, \papdes), nor the two main samples of observed quasars considered here show significant drops or gaps across the redshift range we probe.
Moreover, the redshift distributions shown in Fig.~\ref{fig:z_dist} suggest that our quasar samples fare better than the legacy SDSS quasar selection method with the challenging $z{\approx}2.1-2.8$ regime \cite[see, e.g.,][]{Bovy2011,Ross12}.

We conclude that both the GUA and \skt\ selection methods, and their implementation within SDSS-V, provide an exceptionally pure way to construct large quasar samples, out-performing some legacy quasar surveys. Moreover, the \papgua-based GUA selection enables a high-purity identification of quasars over the entire sky, including at low Galactic latitudes. This is beneficial not only for quasar demographics and related science, but also for efforts directed at other types of astronomical objects (i.e., the \papgua\ and \papdes\ catalogs can be used to rule out the non-quasar nature of certain sources).
In what follows, we investigate the similarities and differences between the key properties of the quasars uncovered by these two selection methods and those provided by legacy (SDSS-based) quasar samples.
%#--------------------------# 

\subsection{Photometric Properties}\label{sec:photometry}

Here we present the basic photometric properties of the quasars in our sample(s). 
We utilize the multi-band photometry available for our sources either in the visual regime (i.e., SDSS-like $griz$ bands) or in the MIR (i.e., WISE-based), to compare our sample(s) to those from other well-established selection methods. The photometric data that we use here are described in Section~\ref{sec:photo_data}. In what follows, all photometric comparisons are performed on the full quasar sample, including both \gd\ and \bd\ subsets, unless otherwise noted.

Figure~\ref{fig:color-color} presents the distribution of our quasars in the  $r-i$ vs.\ $g-r$ (left) and the $i-z$ vs.\ $r-i$ (right) color-color spaces, compared with the SDSS/DR7 quasar sample \citep{DR7Schneider}. We note that the latter were selected mostly through a $ugriz$ multi-color selection procedure \citep{Richards02}, and therefore are meant to represent the legacy census of spectroscopically-confirmed quasars, rather than the ``intrinsic'' range of quasar SEDs.
As Fig.~\ref{fig:color-color} shows, most of our quasars occupy the same part of the color-color space as do SDSS/DR7Q, except that the GUA+\skt\ samples generally cover a wider range in both $r-i$ and $i-z$ colors (e.g., $r-i\gtrsim 0.65$ or $\lesssim-0.3$; $i-z\gtrsim0.7$, or $\lesssim -0.3$ when $r-i\gtrsim0.4$). Notably, these extended regions in color-color space, and particularly the redder regions ($r-i\gtrsim 0.65$, $i-z\gtrsim0.7$) are {\it not} dominated by high-redshift quasars (see colorbar). 
Thus, our quasar samples include some redder quasars, either due to their intrinsically redder SEDs or some form of line-of-sight extinction by dusty gas---located either on circumnuclear or host-galaxy scales. 
This demonstrates how the new, SDSS-V-enabled quasar samples can be used to extend quasar science beyond the regime dominated by intrinsically blue point-like sources.

\begin{figure}[t]
    \centering
    \includegraphics[width=1\columnwidth]{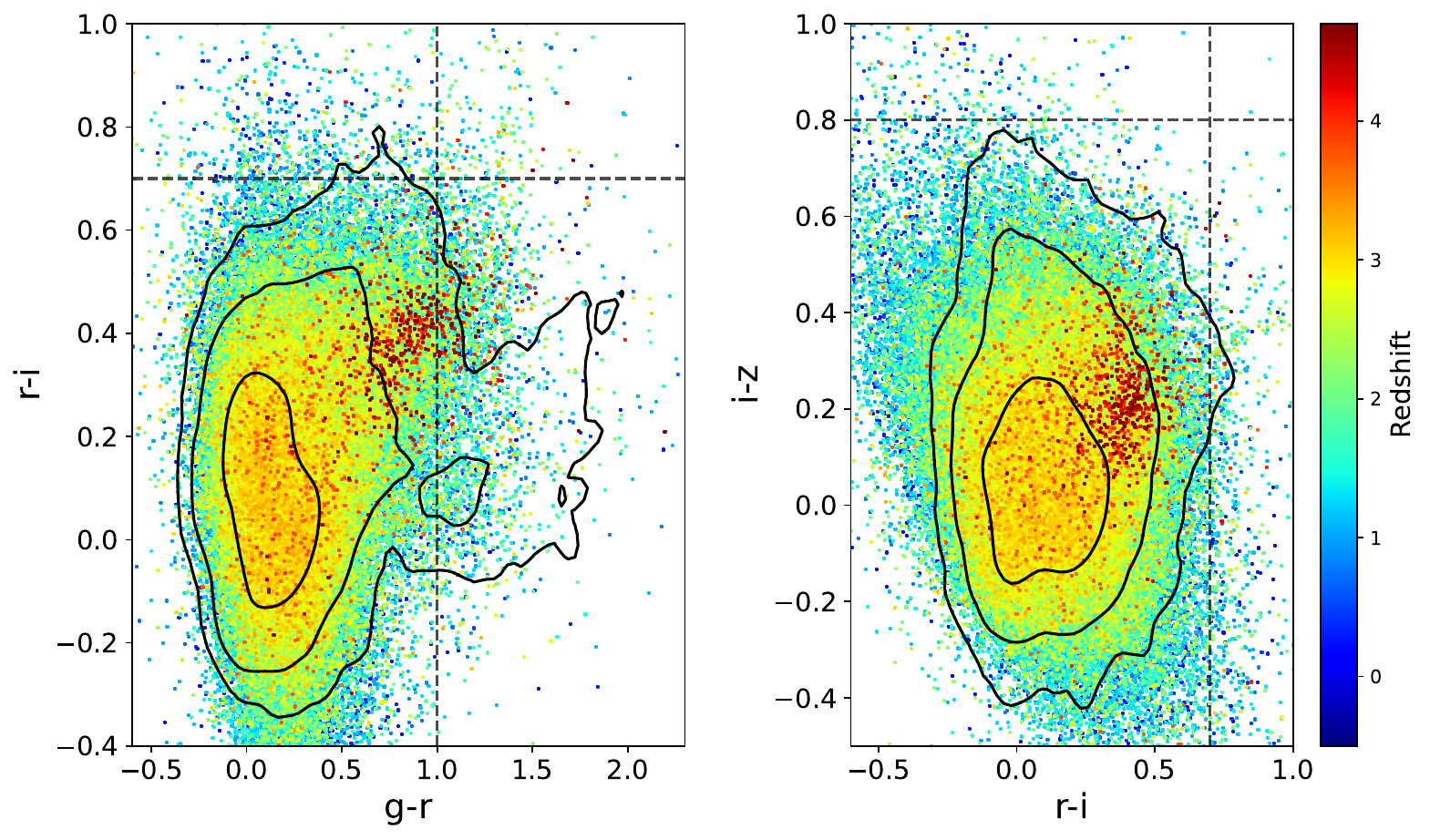}
    \caption{The distributions of our quasar samples in optical color-color space, specifically $r-i$ vs.\ $g-r$ (left) and $i-z$ vs.\ $r-i$ (right). In both panels, black contours represent quasars from the legacy SDSS/DR7Q \citep{DR7Schneider}, with enclosed regions corresponding to the 68\%, 95\%, and 99\% percentiles. The colored dots represent GUA and \skt\ data, color-coded by redshift (see colorbar). The GUA and \skt\ quasars are generally consistent with those of the legacy SDSS/DR7Q sample, however both GUA and \skt\ allow us to probe redder quasars, as expected from their selection based on NIR and MIR data. The black dashed lines indicate the color thresholds ($g-r>1$, $r-i>0.7$ and $i-z>0.8$) used to define a ``red quasar'' sub-samples analyzed in Section~\ref{sec:prelim_red_search}.}
    \label{fig:color-color}
\end{figure}

In Figure~\ref{fig:W1W2}, we show the distribution of the two main quasar samples in the $W1-W2$ vs.\ $W2$ space (as contours). 
\citet{SternW12} proposed that selecting sources with IR colors $W1-W2\geq0.8$ would yield an AGN sample with both a high completeness (78\%) and high-efficiency (95\%). 
Consistent with the findings of \citet{SternW12}, the vast majority of our quasars, $\simeq$84\% meet the $W1-W2\geq0.8$ threshold, while $\sim$16\% of the sources fall below this suggested cut, implying that adopting this criterion would miss $\sim$16\% of our quasars. 
Moreover, the sample completeness of 84\% is somewhat higher than that found by \citet[][of 78\%]{SternW12}. This is consistent with the DR7Q distribution, where 89\% of quasars satisfy the $W1-W2\geq0.8$ criterion. In addition, the efficiency we find for this MIR color cut, calculated based on the 176,592 quasars that have both $W1$ and $W2$ measurements,\footnote{Out of 184,558 objects in total.} is $\simeq$96\%. This is higher than the 95\% efficiency reported in \citet{SternW12}. 

To assess the efficiency of the \cite{SternW12} MIR cut, we also consider the stars identified through the GUA- and \skt-led SDSS-V observations, i.e. the contaminants of our sample (see points in Fig.~\ref{fig:W1W2}).
It is evident that the stars exhibit colors similar to those of the quasars in our sample, making it indeed challenging to distinguish between the two populations based solely on their MIR colors. Furthermore, most of the M-type stars---which are the most common contaminants of the spectroscopic sample (see Section~\ref{sec:purity})---are located below the $W1-W2=0.8$ cut-off line, suggesting that selecting objects above this threshold would result in fewer M-type stars contaminating our sample.
We stress that the target selection procedures for our quasar samples did not explicitly use neither the \cite{SternW12} cut, nor more elaborate MIR selection criteria \cite[e.g.,][and references therein]{Mateos2012_MIR,Assef2013_MIR_selection}. The resulting quasar sample and Fig.~\ref{fig:W1W2} show that adopting such a $W1-W2$ cut could have led to a smaller sample of quasars, with many fewer M-type stars contaminants, however, still with a non-negligible number of other stellar contaminants. 
Moreover, imposing the $W1-W2$ cut may limit the opportunity to access certain (mildly-)obscured, intermediate-redshift AGNs \cite[e.g.,][see below]{LaMassa2016}.

\begin{figure}
    \centering
    \includegraphics[width=1\columnwidth]{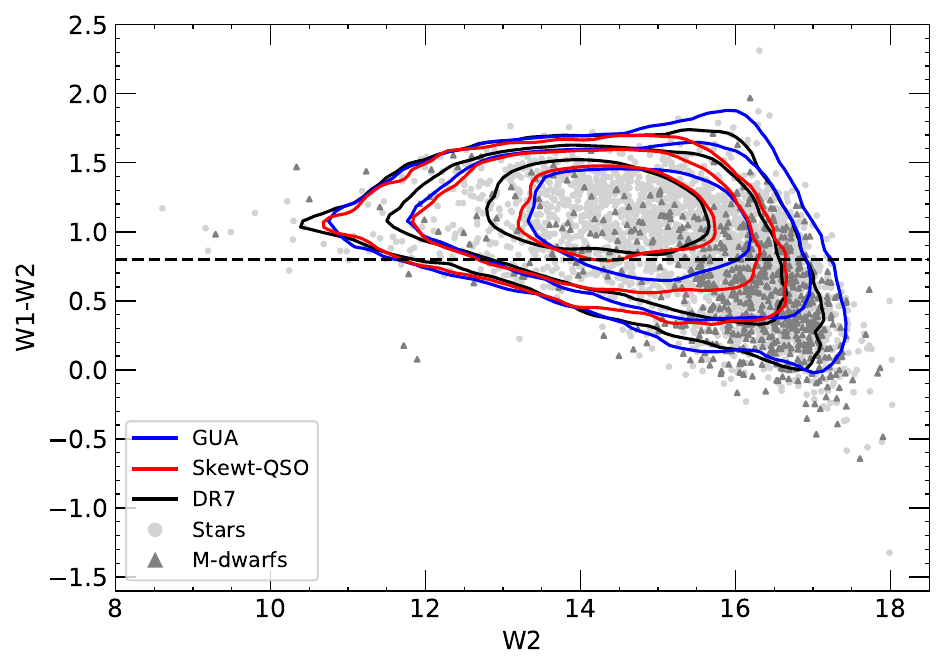}
    \caption{The mid-infrared $W1-W2$ vs.\ $W2$ diagnostic diagram for various sources from our quasar samples. 
    The blue, red and black contours represent the distributions for GUA, \skt\ and DR7Q, respectively (percentiles identical to those in Fig.~\ref{fig:color-color}). The black dashed line marks the $W1-W2>0.8$ threshold for AGN selection \cite[based on][]{SternW12}. Most of the quasars in our sample(s) (84\%) lie above the suggested cut-off, consistent with the DR7Q distribution (89\%). 
    Gray dots represent the stars in the sample(s), with M-type stars further highlighted as triangles.}
    \label{fig:W1W2}   
\end{figure} 

\subsubsection{Preliminary Search of Red Quasars}
\label{sec:prelim_red_search}

We can further use our sample of spectroscopically-confirmed quasars to assess the ability of the novel selection methods to identify significantly red sources, compared to previously-suggested (extreme) color cuts.
While there are several published selection criteria and samples of such objects \cite[e.g.,][and references therein]{Urrutia2009,Banerji2012,Glikman2012, Fawcett2020,LaMassa2024}, here we focus on the optical-to-MIR color diagnostics studied in detail by \cite{LaMassa2016} and \cite{Hamann17}, as well as the MIR-only diagnostic of \cite{W12drops} to search for hot dust obscured galaxies (Hot DOGs).

According to \citet{LaMassa2016}, the $R-W1$ color can be used to identify highly obscured AGNs, where the optical emission (measured by the $R$-band) is diminished, while the MIR emission ($W1$) remains strong as it is dominated by the torus, which reprocesses the absorbed (UV-optical) radiation from the central engine. \citet{LaMassa2016} show that broad-line, unobscured AGNs cover a wide range in $R-W1$, however highly obscured AGNs have $R-W1>4$. 
We followed \citet{LaMassa2016} and estimated the $R-W1$ color for our quasars, using the available $i$- and $r$-band measurements (see above) to derive $R$ (see Eqs. 1 \& 2 in \citealt{LaMassa2016}).

In Figure~\ref{fig:R_W1} we present the distributions of $R-W1$ for the quasars in our two main samples. 
Approximately 26\% of the quasars that have both $i$- and $r$-band measurements (44,024 of 166,440 \gd+\bd\ quasars) have $R-W1>4$. For comparison, Fig.~\ref{fig:R_W1} also shows the $R-W1$ distribution for X-ray selected, broad-line AGNs drawn from the more recent, highly complete Stripe-82X DR3 catalog, published by \citet{LaMassa2024}, as well as the DR7Q catalog \citep{DR7Schneider}. 
In \citet{LaMassa2024}, approximately 41\% of broad-line AGNs have $R-W1>4$, while only 17\% of DR7Q quasars exceed this threshold. Our GUA and \skt\ samples thus lie between these two cases. 
This suggests that the GUA and \skt\ selection methods recover a substantial fraction of red quasars, though they are slightly skewed towards bluer broad-line AGNs relative to the X-ray complete sample of \citet{LaMassa2024}.
The lower fraction of red quasars among DR7Q relative to our samples is consistent with the known color bias of the previous selection of SDSS quasars, which primarily targeted blue sources \citep[e.g.,][]{Richards02}. 
We also note that a significant fraction of the  $R-W1>4$ sources in our sample (${\approx}26\%$) have $W1-W2<0.8$, i.e. not satisfying the simple \cite{SternW12} AGN selection cut. This result, which is in qualitative agreement with, e.g., \cite{LaMassa2016}, further demonstrates how our sample probes a wider range in color and obscuration properties than previous samples of (optical) luminous AGNs.

The 26\% of quasars in the sample with $R-W1>4$ are not completely obscured, as they do---by definition---exhibit broad emission lines in their SDSS-V spectra. The exact reason for the enhanced MIR emission relative to the optical one, and/or the level of dust attenuation, remains unclear. Further analysis of this subset of GUA and \skt\ quasars is required to better understand the nature of these red quasars, but is beyond the scope of the present paper.
Nonetheless, we made a first step to assessing whether the redder quasars in our sample are driven by real, intrinsically redder colors (i.e., mildly-obscured systems) rather than by the redshifting of strong emission lines \cite[e.g.,][]{Richards03}. We selected a ``red quasar'' sub-sample to include the union of objects satisfying $g-1>1$, $r-i>0.7$ and/or $i-z>0.8$, within the redshift range $0.5<z<1.5$. This redshift range focuses on systems where emission line ``migration'' through the optical bands is significant, while avoiding host-galaxy dominated continua at very low redshifts. Figure~\ref{fig:R_W1} shows the $R-W1$ distribution of this sub-sample (dark-red line). Of the 1,302 quasars in this sub-sample, $\approx$75\% pass the $R-W1>4$ threshold, which is substantially higher than the $\approx$39\% baseline of the full quasar sample in the same redshift range. This confirms that the optically-red quasars in our sample are dominated by genuinely (mildly-to-moderately) obscured objects.

\begin{figure}[t]
    \centering
    \includegraphics[width=1\columnwidth]{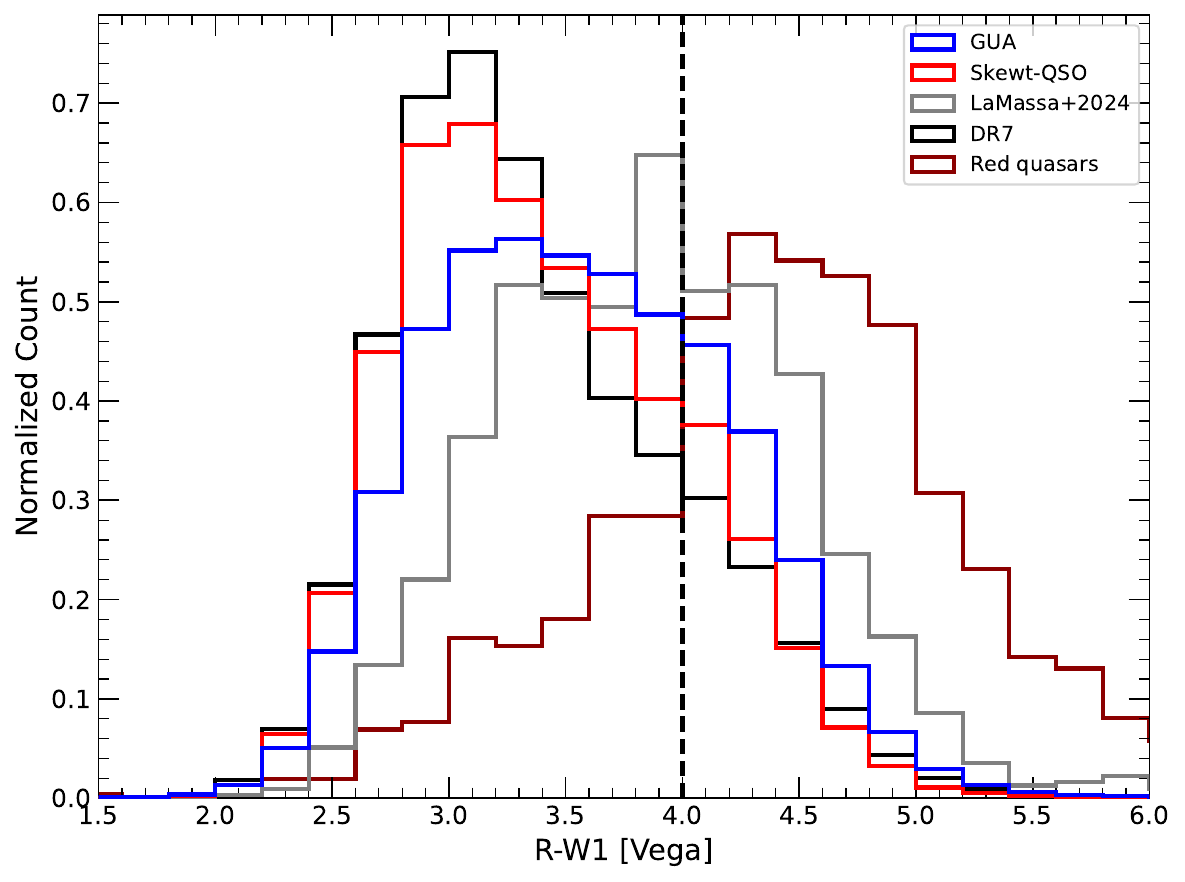}
    \caption{The distribution of the $R-W1$ MIR-optical color for our quasars compared to \citet{LaMassa2024}'s AGN sample.
    We show the GUA (blue) and \skt\ (red) samples, compared with the sample of broad-line AGNs from \citet{LaMassa2024} (gray) and DR7Q (black). The dark-red line shows the distribution of the ``red quasar'' sub-sample, defined as objects with $g-r>1$ and/or $r-i>0.7$ or $i-z>0.8$ within $0.5<z<1.5$. The dashed black line indicates the $R-W1>4$ threshold proposed by \citet{LaMassa2016} for selecting highly obscured (yet broad-line) AGNs. Approximately 26\% of the quasars in our combined dataset exceed this threshold.}
    \label{fig:R_W1} 
\end{figure}

\cite{Hamann17} presented a diagnostic for selecting extremely red quasars (dubbed ERQs), namely $i-W3\geq4.6$ (AB). Using this diagnostic alone they found an ERQ fraction of 0.095\%, based on a sample of 205 ERQs among a sample of 216,188 quasars at redshifts $1.53\leq z\leq 5.0$, drawn from the SDSS/DR12Q catalog \citep{DR12Q_2014,DR12Q_2017}. 
Based on their measured spectral features, these objects were interpreted as highly obscured quasars embedded in a dense, patchy medium, likely caught in a brief evolutionary phase of intense feedback and powerful outflows. 
We followed the same simple color cut, for all GUA \& \skt\ quasars at $1.53\leq z\leq 5.0$, that have valid $i$- and $W3$-band measurements (66,913 quasars). We find an ERQ fraction of 0.10\% (67 quasars)---fully consistent with, albeit slightly lower than, the findings of \cite{Hamann17}. 
In comparison, applying the same selection to the DR7Q quasar yields an ERQ fraction of 0.057\%.
This again demonstrates the enhanced completeness of the \papgua- and \papdes-based selection methods to quasars with SEDs that are (slightly) redder than those probed by the legacy SDSS color selection \citep{Richards02}.

As explained in Section~\ref{sec:intro}, Hot DOGs are among the brightest objects in the (mid-)IR sky. The GUA and \skt\ samples are large enough to warrant a preliminary search for such rare sources, which could in principle be selected for SDSS-V spectroscopy (unlike the blue-focused selection of legacy SDSS quasars). In Figure~\ref{fig:W12_drops}, we show the distribution of our quasars in two MIR color-magnitude spaces involving $W1-W2$ measurements. To identify potential Hot DOGs within our sample, we applied the ``W1W2 drop-outs'' selection criteria defined by \citet[][see Eqs. 1 \& 2 in \citealt{W12drops}]{Eisenhardt2012}.
However, we identified no robust Hot DOG candidates among our quasars. We note that a single candidate was rejected following a visual inspection of the WISE images, which revealed the MIR measurements were severely contaminated by a bright nearby star. 
While the selection method in the present study effectively identifies quasars across wide ranges in color space(s), it is less efficient at detecting Hot DOGs. These objects are so heavily obscured that their optical continuum is dominated by the stellar light from the host-galaxies, and not the accretion-driven AGN emission. Unlike the latter, the former is significantly fainter in the $G$-band (considering Hot DOGs are found at significant redshifts; see, e.g., \citealt{W12drops}), and thus challenging to be included in a relatively shallow optical spectroscopic survey like SDSS-V. 
It is still possible that (low-redshift) Hot DOGs \citep[e.g.,][]{Li2023,Li2025} may be revealed in ongoing SDSS-V observations, or alternatively, if deeper wide-field spectroscopic surveys would extend similar quasar selection methods to lower optical flux levels \cite[e.g., 4MOST;][]{deJong2019_4MOST}.

\begin{figure*}
    \centering
    \includegraphics[width=\textwidth]{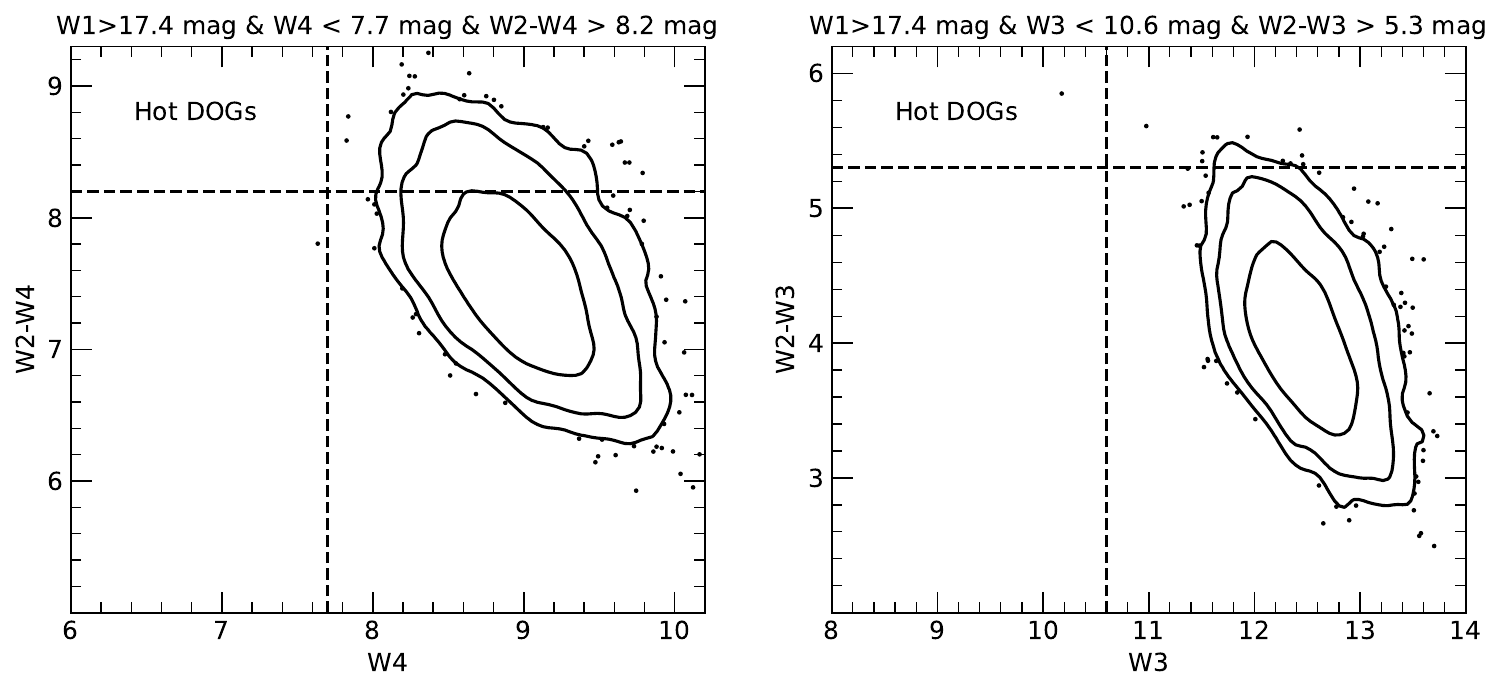}
    \caption{Preliminary search for Hot DOGs using mid-infrared diagnostics, following the ``W1W2 drop-outs'' method from \citet[][black dashed lines]{W12drops}. The magnitudes are given in the Vega system. The contours represent the distributions of the quasars in our sample that have $W1>17.4$ (percentiles are identical to those in Fig.~\ref{fig:color-color}), while the scattered points show the remaining 1\% of sources not included in the contours.  
    No Hot DOGs were found among the quasars in our sample; one candidate was discarded due to contamination from a bright nearby star.}
    \label{fig:W12_drops}
\end{figure*}

To summarize this part of the analysis, the comparisons we carried out suggest that our quasar samples are not biased toward particularly blue or red colors, and exhibit color properties broadly consistent with those of other well-known quasar catalogs, with only a slight enhanced completeness towards red (mildly obscured) systems.
As mentioned in Section~\ref{sec:limitations}, since the selection methods used to identify the samples we study may be somewhat biased against heavily obscured AGNs, the fact we have not identified a significant new population of such sources is not surprising.  

%#--------------------------#
\subsection{Distributions of \Lbol, \mbh, and \lledd} \label{sec:M_L_dist}

Here we present the distribution of AGN bolometric luminosity (\Lbol) and SMBH mass (\mbh) for the quasars in the sample(s). The bolometric luminosity range of the quasars is $41.6\lesssim \log(\Lbol/\ergs)\lesssim 48.2$, while the BH mass range is $5.7\lesssim \log(\mbh/\Msol) \lesssim 10.7$. The central 99\% of the sample spans $44.1\lesssim \log(\Lbol/\ergs)\lesssim 47.4$ and $6.8\lesssim \log(\mbh/\Msol)\lesssim 9.9$. As explained in Section~\ref{sec:Final_Refinements}, we verified that the extremes of these distributions are not driven by spurious measurements. In particular, we note that the lowest \mbh\ estimate in our sample, which in fact corresponds to the low-redshift galaxy UGC06728, is fully consistent with a robust, independent measurement (e.g., $\log(\mbh/\Msol)=5.85$ based on RM; \citealt{low_mass_BH}). 
Finally, essentially all the quasars in our sample have $0.01\lesssim\lledd\lesssim1$, as is found for many other large samples of broad-line AGNs drawn from highly complete surveys \cite[e.g.,][]{Schulze2010,DR7Shen,TrakhtenbrotNetzer12,Schulze2015,Ananna2022}.

Figure~\ref{fig:M_L_z} shows these distributions as a function of redshift, as well as the corresponding distributions of previous quasar sample (namely, SDSS/DR16Q; \citealt{DR16WuShen}), while Figure~\ref{fig:M_L} presents similar distributions in the mass-luminosity plane. Most of the quasars overlap with those from the previous large SDSS-based catalogs, except for a few low-mass, low-luminosity sources that we were able to identify among the GUA and \skt\ samples.
Figure \ref{fig:CDF_PDF} further compare the 1D distributions of the three key properties (\Lbol, \mbh, and \lledd), as well as the $G$-band magnitudes, for the various SDSS-V programs within our quasar sample, and the two main previous quasar catalogs, namely SDSS/DR16Q \citep{DR16WuShen} and the older DR7Q \citep{DR7Shen}. The bottom panel in each of the figures show the corresponding cumulative distribution functions (CDFs), while Table~\ref{tab:median} summarizes the corresponding key statistics, such as medians, averages, and standard deviations.

\begin{figure}
    \centering
    \includegraphics[width=1\columnwidth]{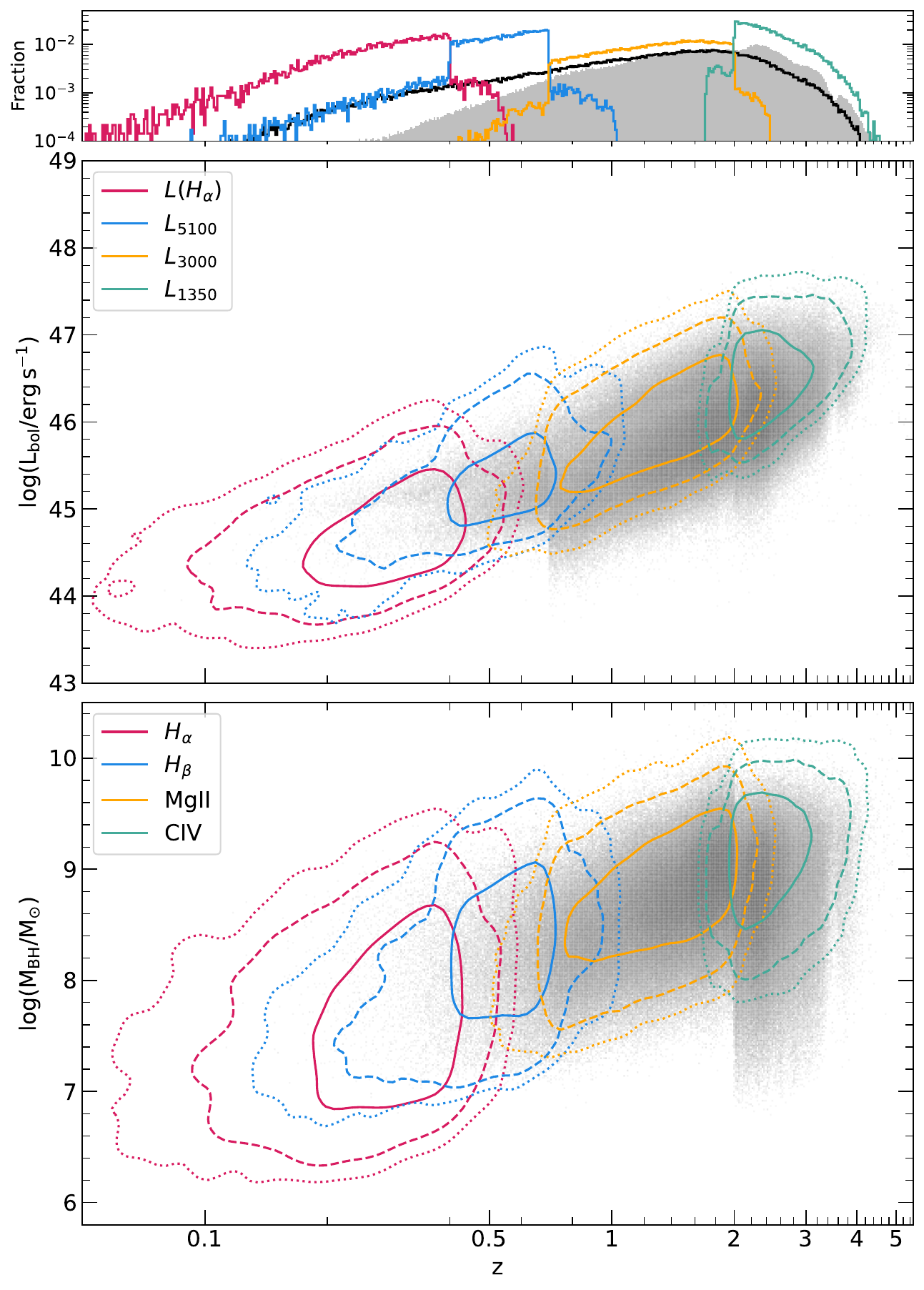}
    \caption{The distributions of our quasars in the $\Lbol-z$ (top) and $\log\mbh-z$ (bottom) planes. In both panels, the contours (defined as in Fig.~\ref{fig:color-color}) are color-coded according to the emission line and the continuum window used for their calculation. The small gray points represent measurements taken from \citet{DR16WuShen} for SDSS/DR16Q quasars. Our samples generally cover the same region in these parameter spaces as does the largest previous SDSS-based quasar catalog.}
    \label{fig:M_L_z}
\end{figure} 

\begin{figure}
    \centering
    \includegraphics[width=1\columnwidth]{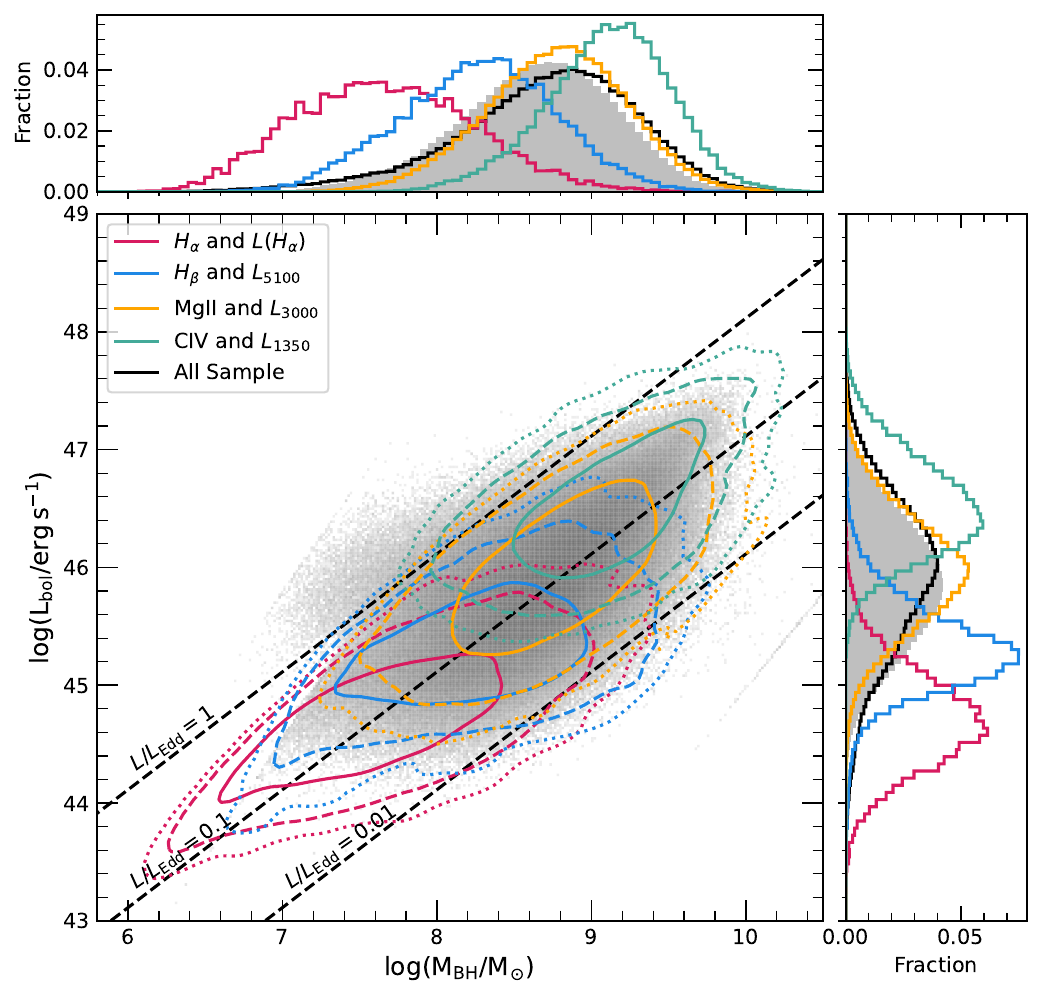}
    \caption{Similar to Fig.~\ref{fig:M_L_z}, but focusing on the $\log\Lbol-\log\mbh$ plane. The dashed diagonals indicate Eddington ratios of $\lledd\ = 1$, 0.1, and 0.01. 
    The additional panels on the top and to the right of the main one show the 1D distributions of the two properties, for the various line-dependent subsets (in color), and for their union (in black), compared with the SDSS/DR16Q catalog \cite[gray;][]{DR16WuShen}. The H$\alpha$-based measurements preferentially trace lower-\Lbol\ and/or lower-\mbh\ systems, likely reflecting their lower typical redshifts and the flux limit(s) employed during the survey planning.
    Our quasars cover $0.01\lesssim\lledd\lesssim1$, and show distributions of \Lbol, \mbh, and \lledd\ that are generally consistent with the largest previous SDSS-based quasar catalog.}
    \label{fig:M_L}
\end{figure} 

%#--------------------------#

There are several clear trends seen when comparing the various samples. First, the distribution of $G$-band magnitudes in \skt\ shows a concentration of objects within a certain brightness range due to increased selection at $19<G<20$ mag. Second, there are differences of $\lesssim1$ dex between \skt\ and DR16Q in both \Lbol\ and \mbh\, with \skt\ showing higher luminosities and masses. For \lledd\, we do not observe significant differences between the two samples. \skt, in fact, appears to be much more similar to the older DR7Q catalog. GUA, on the other hand, falls between \skt\ and DR16Q in terms of \Lbol\ and \mbh. In terms of brightness, the GUA sample exhibits two peaks, indicating the two different sub-selection samples of GUA bright and dark. For \lledd\ the distribution is similar to DR16Q and \skt.

To quantify the differences between the samples, we applied statistical tests. First, we applied the traditional 2-sample Kolmogorov–Smirnov (KS) test, which yielded extremely low $P$-values ($P_{\rm value}\ll 10^{-6}$) for all the CDFs under study, indicating that the distributions are different at very high significance levels. We note again that the differences themselves are not large, and the distributions of the various samples generally overlap. Since the KS test depends strongly on sample size and is mainly sensitive to differences near the median of the distributions, with relatively low sensitivity in the tails, we chose to also apply the Anderson-Darling (AD) test, which is known to better address the tails of the distributions that are compared. We obtained extremely low $P$-values in the AD test, too---again indicating that the distributions of \Lbol, \mbh, and \lledd\ differ between the samples we study. 

A closer inspection of Figure~\ref{fig:CDF_PDF} suggests that while the samples do differ from each other, these differences do not fully answer the question of whether the various selection methods uncover distinct populations of accreting SMBHs, as the derived properties depend on luminosities and thus on the flux distributions and limits of the surveys. 
Specifically, the left-most panels of Fig.~\ref{fig:CDF_PDF} clearly demonstrates how our sample(s) differ from the legacy SDSS ones in terms of visual-regime brightness ($G$-band magnitudes). 
This is not surprising, as the older SDSS-I/II-based DR7Q followed a rather rigid set of flux limit(s) \cite[e.g., $i<19.1$ or $<20.2$ mag; see][]{Richards02}, while the later SDSS-III/IV observations added in DR16Q used different selection criteria, aiming to enhance the number of (faint) quasars at $z\sim2-2.5$.
These differences in fluxes translate to differences in luminosities, in a way that also involves the distribution of redshifts (by itself potentially affected by selection criteria defined by colors and/or variability). The differences in luminosities then affect the distributions of \mbh\ and \lledd, either directly (i.e., through prescriptions like Eqn.~\ref{eq:MBH}, \ref{eq:mbh_ha}, and \ref{eq:eddR}), or less directly, i.e., when a survey misses AGNs that are not luminous enough to be selected (either low-\mbh\ limited to $\lledd\simeq1$ or high-\mbh\ systems with low \lledd).
To be able to make a more meaningful comparison between the various samples, a more nuanced approach is needed, which would remove these obvious biases.

\begin{figure*}
    \centering
    \includegraphics[width=\textwidth]{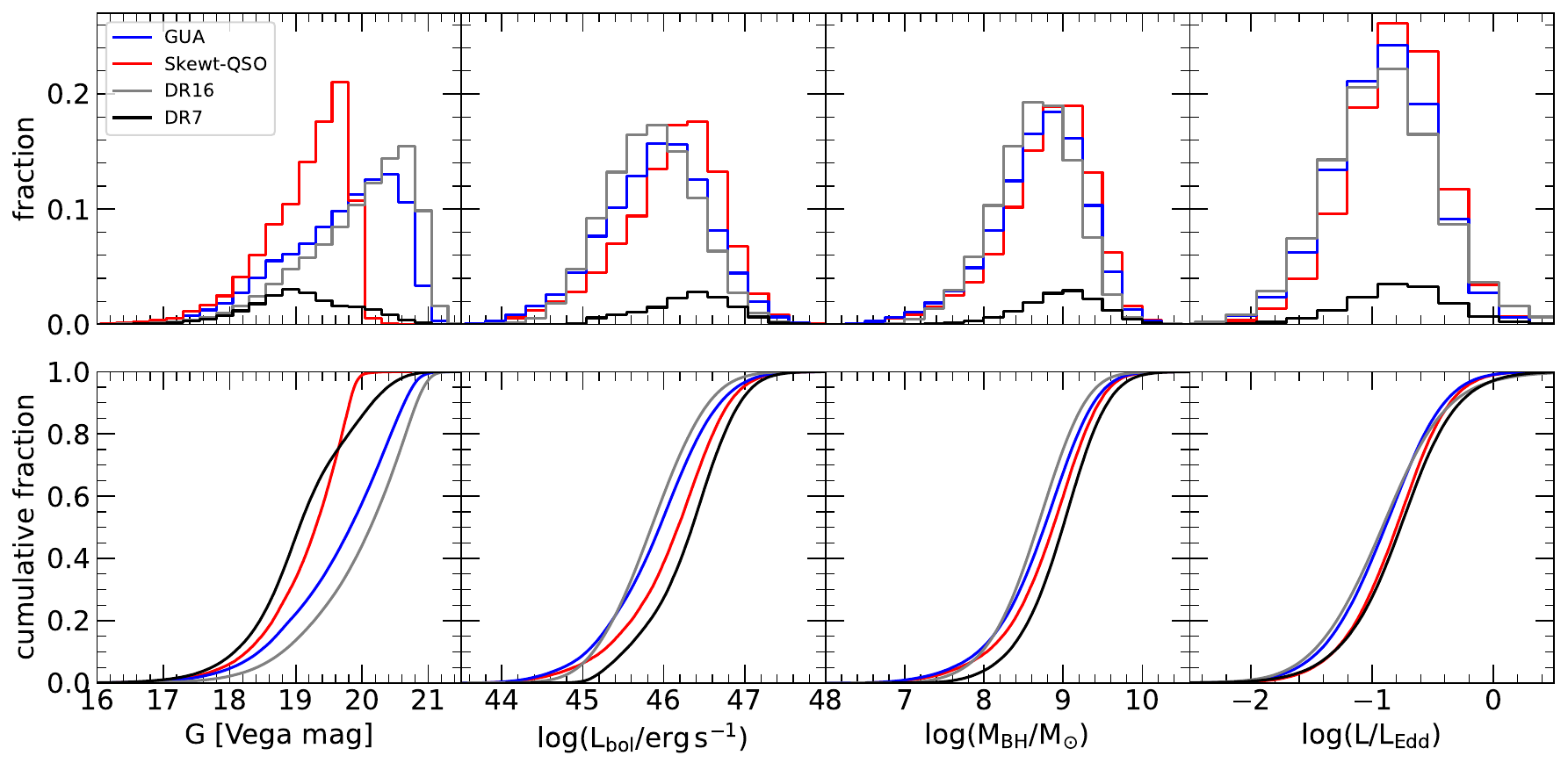}
    \caption{Distributions of key parameters for our two main samples, GUA and \skt, and two comparison samples (SDSS/DR7Q and DR16Q; see legend). 
    The upper panels show the probability density functions (PDFs) of, from left to right: $G$-band magnitudes, luminosities (\Lbol), black hole masses (\mbh), and Eddington ratios (\lledd).  
    The lower panels display the corresponding cumulative distribution functions (CDFs) for the same properties.  
    These plots highlight the differences in brightness between our samples and the comparison samples, driven by the various survey strategies, but which propagate to differences in the other properties (\Lbol, \mbh, and \lledd).} 
    \label{fig:CDF_PDF}
\end{figure*}

\begin{figure*}
    \centering   
    \includegraphics[width=\textwidth]{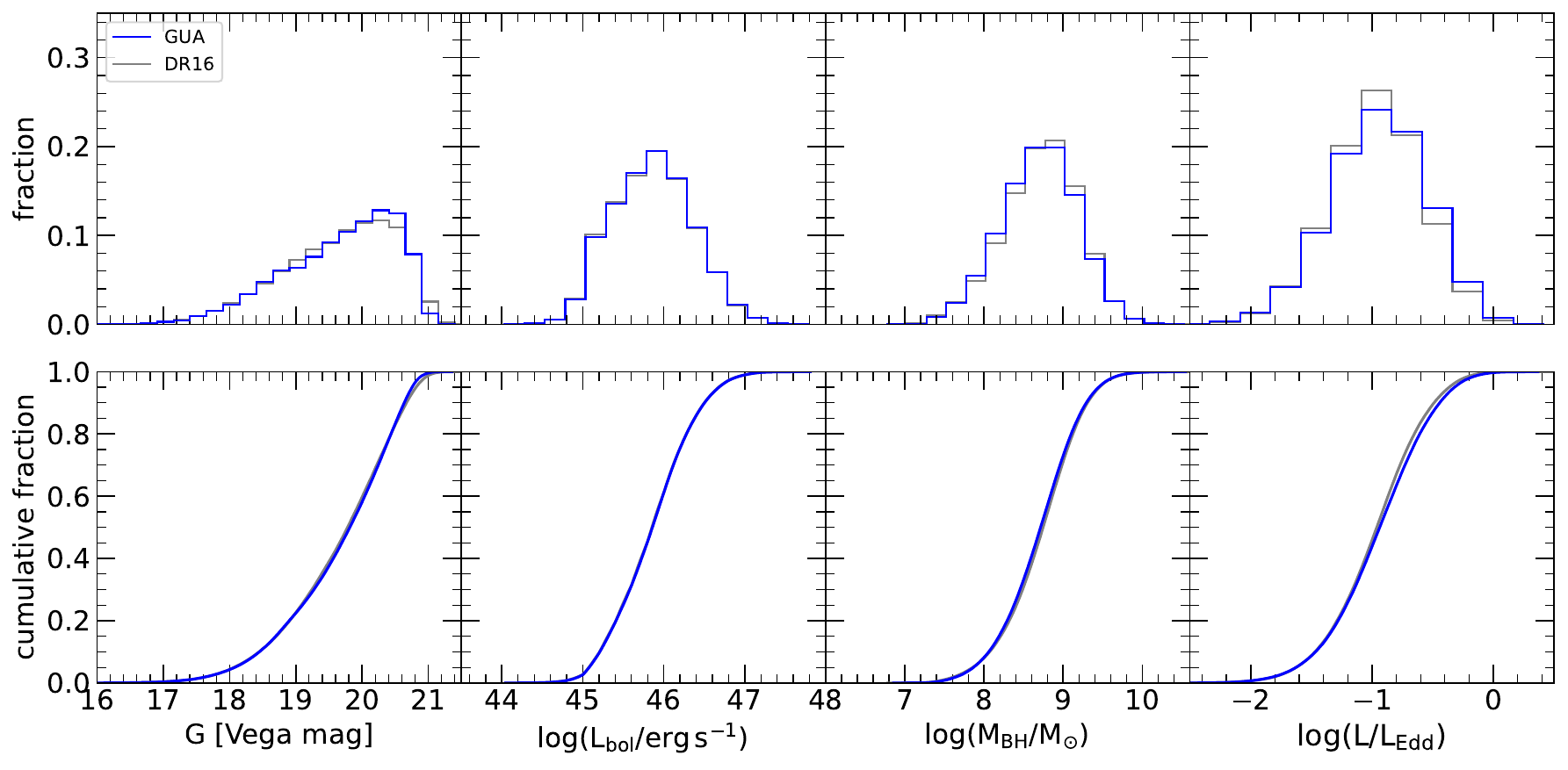}
    \caption{Results of the $z-\Lbol$ matching procedure, comparing GUA sample (as the test sample) and the SDSS/DR16Q sample (as the reference). The panels are identical to those of Fig.~\ref{fig:CDF_PDF}. 
    As expected, the two samples agree nearly perfectly in terms of both $G$ and \Lbol. 
    The distributions of \mbh\ and \lledd\ for the SDSS-V quasars are now in excellent agreement to that of largest previous SDSS-based catalog.}
    \label{fig:match_GUA}
\end{figure*}

\begin{figure*}
    \centering
    \includegraphics[width=\textwidth]{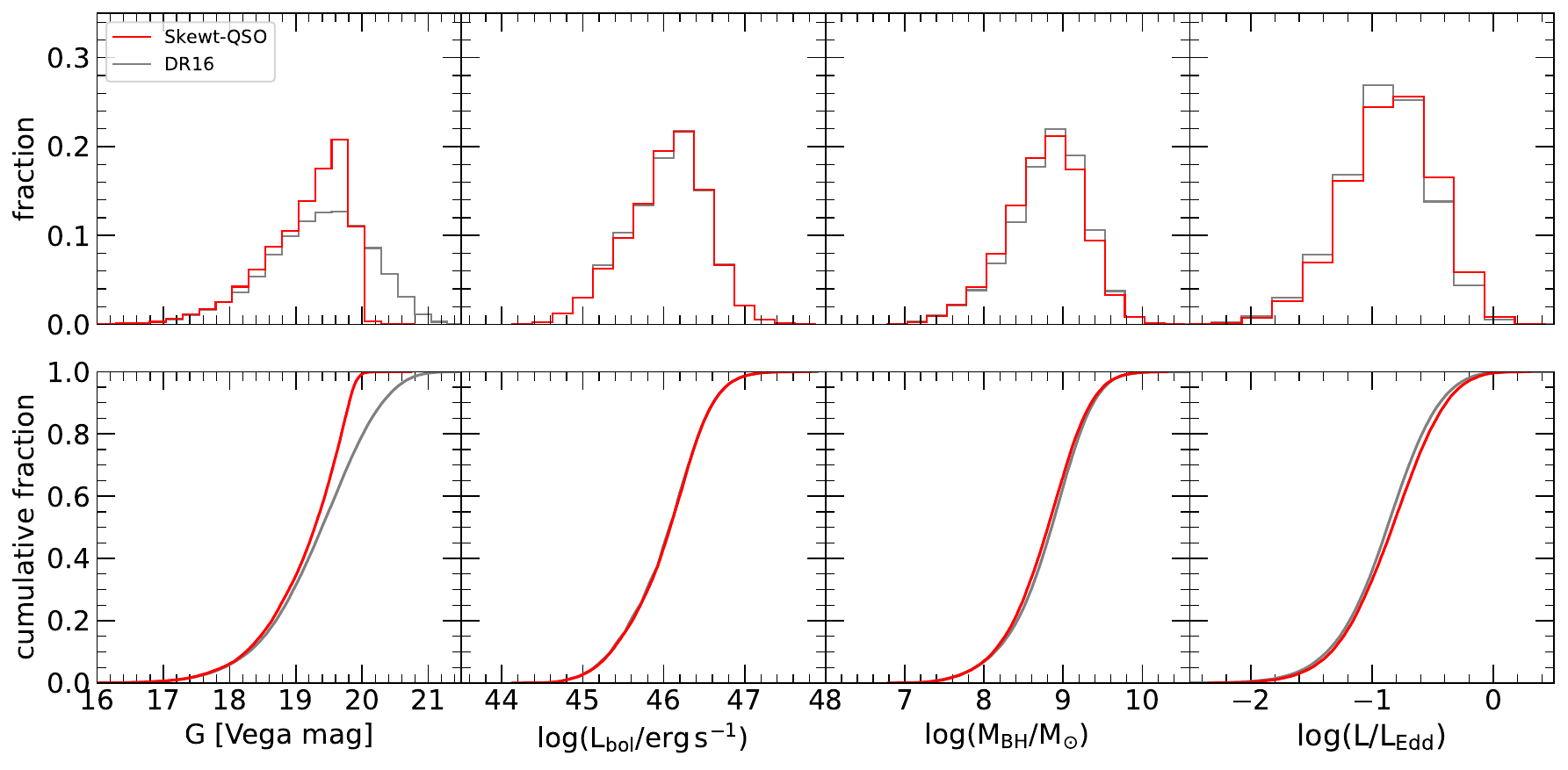}
    \caption{Similar to Fig.~\ref{fig:match_GUA}, but presenting the comparison between the matched \skt\ and SDSS/DR16Q samples.}
    \label{fig:match_skt}
\end{figure*}

\setlength{\tabcolsep}{2pt}
\begin{deluxetable*}{l|CCCCC|CCCCC|CCCCC}
\tablecaption{Distribution statistics for each sample}
\label{tab:median}
\tablewidth{0pt}
\tablehead{
\colhead{} &
\multicolumn{5}{c}{$\log(\Lbol/\ergs)$} &
\multicolumn{5}{c}{$\log(\mbh/\Msol)$} &
\multicolumn{5}{c}{$\log \lledd$} \\
[-2ex]
\colhead{Sample} &
\colhead{Median} & \colhead{Average} & \colhead{IQR} & \colhead{$\sigma_{STD}$} & \colhead{$\sigma_{MAD}$} &
\colhead{Median} & \colhead{Average} & \colhead{IQR} & \colhead{$\sigma_{STD}$} & \colhead{$\sigma_{MAD}$} &
\colhead{Median} & \colhead{Average} & \colhead{IQR} & \colhead{$\sigma_{STD}$} & \colhead{$\sigma_{MAD}$}
}
\startdata
GUA         & 45.95 & 45.90 & (45.48, 46.35) & 0.65 & 0.64 &
              8.76  & 8.70  & (8.36, 9.11)   & 0.59 & 0.55 &
             -0.89  & -0.91 & (-1.17, -0.63) & 0.41 & 0.40 \\
\skt        & 46.17 & 46.09 & (45.72, 46.53) & 0.63 & 0.58 & 
              8.87  & 8.80  & (8.48, 9.20)   & 0.57 & 0.52 & 
             -0.80  & -0.82 & (-1.06, -0.56) & 0.38 & 0.37 \\
SDSS/DR7Q   & 46.35 & 46.29 & (45.95, 46.67) & 0.54 & 0.52 & 
              9.00  & 8.96  & (8.66, 9.30)   & 0.50 & 0.47 & 
             -0.76  & -0.78 & (-1.03, -0.51) & 0.42 & 0.38 \\
SDSS/DR16Q  & 45.85 & 45.84 & (45.46, 46.23) & 0.55 & 0.57 &
              8.67  & 8.64  & (8.32, 9.00)   & 0.51 & 0.50 &
             -0.91  & -0.91 & (-1.22, -0.62) & 0.47 & 0.44 \\
\enddata
\tablecomments{The statistics listed here are for the distributions of the the GUA and \skt\ samples, and for the SDSS/DR7Q and DR16Q catalogs \cite[][,respectively]{DR7Shen, DR16WuShen}, without any matching---as shown in Figure~\ref{fig:CDF_PDF}. For each parameter we report the median, mean value, interquartile range (IQR), standard deviation ($\sigma_{STD}$), and the scaled median absolute deviation (MAD; $\sigma_{MAD}$).}
\end{deluxetable*}
%#--------------------------#

We therefore employed a more elaborate approach, of creating a ``matched comparison sample'' for each of our SDSS-V-enabled quasar samples.
The basic idea is to control for the differences in luminosity and redshift (which would also eliminate differences in fluxes), and then look more carefully into any remaining, intrinsic differences in the distributions of \mbh\ and \lledd. 
This scheme was adapted from the one described in the study of \citet{Zeltyn24}, which had to address similar challenges.

First, we restricted all the samples, both the SDSS-V and legacy SDSS ones, to the redshift range $0.3<z<2$ and to luminosities $\log(\Lbol/\ergs) > 44$, which is where the core of the quasars are (in all samples). This also means that most of the \civ-based estimates of \mbh\ (and \lledd)\footnote{Together with the corresponding \Luv-based estimates of \Lbol.} are omitted from what follows. We then divided the $z-\log\Lbol$ plane into several bins, with the goal of making the \emph{relative} distributions of each of the samples (``test samples'' hereafter) among these bins as close as possible to that of the reference samples \cite[i.e., DR7Q or DR16Q;][,respectively]{DR7Shen,DR16WuShen}. In practice, for each bin, we counted the number of quasars in both the test and reference samples and calculated the ratio between the two. Next, only the bins where the ratio between the two samples was greater than 1\% were retained (see Appendix~\ref{apx:matching example} for details). Using the bin with the highest ratio, we created a new reference sub-sample by randomly selecting objects without replacement in each bin so as to maintain this ratio across the samples, thus preserving a balanced comparison. In other words, following this procedure, the test sample and the reference sub-sample have a highly similar distribution in the $z- \log\Lbol$ plane. The only difference remaining is the fact that the reference sample can be significantly larger or smaller than the test sample, as a whole, but not in a way that is skewed toward any specific region of the $z-\log\Lbol$ plane. 
While this procedure can be applied to any choice of test and reference samples, in what follows we focus on comparing those samples that were also ``originally'' more similar to each other, in the broad sense of deeper or shallower flux limits. 

Figure~\ref{fig:match_GUA} presents matched comparisons between the GUA and DR16Q samples, while Figure~\ref{fig:match_skt} shows the corresponding distributions for the \skt\ sample. As expected, the pairs of \Lbol\ distributions are now essentially identical. Moreover, the distributions of $G$-band magnitudes are now much more similar to each other (pair-wise), compared with the original, unmatched comparisons (Fig.~\ref{fig:CDF_PDF}). This is expected, given that we have matched the samples in the $z-\log\Lbol$ plane and not simply in $\log\Lbol$. The same kind of matched comparisons were performed for GUA and \skt\ vs.\ DR7Q (see Figures \ref{fig:match_GUA_All_DR7} \& \ref{fig:match_skt_All_DR7} in Appendix~\ref{apx:matching example}).

Importantly, the corresponding distributions of \mbh\ and \lledd\ are now also much more similar to each other, and it is evident that the new GUA and \skt\ samples probe SMBHs with BH masses and accretion rates similar to those probed by previous large quasar catalogs (DR7Q and DR16Q), at least in the regions of the $z-\log\Lbol$ plane where they overlap. To further validate this result, we again applied the AD test to the redshift- and luminosity-matched distributions of \mbh\ and \lledd. 
Somewhat surprisingly, the AD test again yielded very low $P$-values ($P_{\rm AD} \ll 10^{-10}$), formally indicating that the samples are not drawn from identical parent distributions. However, given the very large sample sizes involved, even small differences between the distributions are expected to result in highly significant $P$-values. In this sense, the AD test is highly sensitive to minor discrepancies that may not be physically meaningful.  
Given the striking similarities between the distributions seen in Fig.~\ref{fig:match_GUA} and the large systematic uncertainties that affect \mbh\ and \lledd\ estimates, it is reasonable to conclude that the quasars in the samples are consistent with the previous SDSS catalogs.

The main conclusion drawn from this analysis is that GUA sample and the previous catalogs are quite consistent with each other. The similarities in the distributions of key quasar properties, such as luminosity, black-hole mass, and Eddington ratio, suggest that all the datasets could be considered as derived from the same parent population of quasars and SMBHs. This strong consistency across the different surveys affirms the reliability of our sample selection process and indicates that the quasar population we are studying aligns well with those identified in earlier SDSS catalogs, which focused on the northern celestial hemisphere and employed markedly different quasar selection techniques. GUA thus clearly demonstrates that we could extend our census of quasars and ability to study them across the entire sky, and without the need to employ selection methods that rely on nuanced multi-band selection and/or multi-wavelength data.

%%%%%%%%%%%%%%%%%%%%%%%%%%%%%%%%%%%%%%%%%%%%%%%%%%%%%%%%%%%%%%%%

\section{Conclusions} \label{sec:conc}
In this study, we presented an exploratory analysis of a new large sample of quasars selected for the dual-hemisphere SDSS-V project through two novel selection methods. We evaluated the efficiency of the selection methods, analyzed the key photometric and physical properties of the resulting quasar samples, and compared these findings with other previously established quasar catalogs. 
The key conclusions from this work are as follows.

\begin{itemize}
    \item SDSS-V enables optical spectroscopy of various samples of quasars and AGNs across the entire sky, providing comprehensive coverage for AGN studies and facilitating a more complete understanding of populations of accreting SMBHs (see Section~\ref{sec:sdssv}).
    
    \item The non-core GUA and ASQOSS (\skt) programs within SDSS-V, which are the focus of our study, have already provided more than 100,000 quasar spectra, of which $\approx$76,000 are newly identified sources. Most of these new quasars ($>$55,000) are located in the southern hemisphere ($\delta<0^\circ$; see Section~\ref{sec:final_sample} and Table~\ref{tab:samples_sum}).
    
    \item The quasar selection methods that produced our sample(s), which were based on previous works by \papgua\ and \papdes, have proven to be powerful ways to produce a large, highly efficient, all-sky quasar sample. The quasar selection efficiency of GUA, which relies solely on Gaia and unWISE data, exceeded 96\%, and remains high even at low Galactic latitudes (see Section~\ref{sec:purity} and Figure~\ref{fig:bar}). 

    \item While the photometric properties of the bulk of the GUA and \skt\ samples are broadly consistent with those of previous quasars catalogs (e.g., SDSS/DR7Q), they do cover a wider range in optical and/or IR color spaces, which allows to probe larger populations of red, mildly-obscured AGNs (see Section~\ref{sec:photometry}). A significant fraction (${\approx}26\%$) of our quasars satisfies the $R-W1>4$ obscuration criterion of \citet{LaMassa2016}, intermediate between samples based on optical color selection (i.e., DR7Q quasars) and on X-rays (i.e., \citealt{LaMassa2024}; see Fig.~\ref{fig:R_W1}).
    So far the GUA and \skt\ programs have not identified extremely obscured systems, such as Hot DOGs (mainly due to flux limits; see Fig.~\ref{fig:W12_drops}).
    
    \item The resulting sample includes quasars spanning a broad range of redshifts (reaching $z>4$), luminosities ($44 \lesssim\log(\Lbol/\ergs)\lesssim 48$), and black-hole mass ($6 \lesssim \log(\mbh/\Msol)\lesssim10$). This sample has similar properties to other well-established quasar catalogs (i.e., SDSS DR7Q and DR16Q), in terms of \Lbol, \mbh, and \lledd\ (see Section~\ref{sec:M_L_dist}).
    
    \item The sample uncovered several intriguing quasar types, including many BAL quasars, radio-emitting sources, and even sources showing extreme spectral variability, offering new opportunities for detailed follow-up studies. 
    
\end{itemize}

During this work, we encountered several challenges and limitations. First, a significant number of stars contaminated the sample, particularly M-type stars located far from the Galactic plane and with no proper-motion information in Gaia/DR2 (see Sec.~\ref{sec:purity}). This issue could potentially be addressed by incorporating more recent Gaia data (e.g., the available Gaia/DR3, \citep{Gaia_DR3}; or the upcoming Gaia/DR4), which should provide improved proper-motion measurements. This would lead to yet higher quasar selection efficiencies. We note that having refined high-efficiency quasar selection methods at low Galactic latitudes could also be beneficial for studies that aim to focus on (red) stars, avoiding sources that are most likely quasars.

With SDSS-V observations continuing into 2027, the GUA and ASQOSS programs are expected to yield yet larger numbers of yet-to-be-analyzed, and indeed yet-to-be-observed targets, eventually perhaps even doubling the sizes of the samples studied here.
These growing samples will be made public as part of upcoming SDSS DRs, and we expect that some aspects of our analysis will be occasionally revisited, particularly to better understand certain subsets of (rare) quasars. Specifically, further characterization of the red quasar subset ($R-W1>4$; see Fig.~\ref{fig:R_W1}) is essential in order to understand the role that such sources play in the cosmic evolution of SMBHs \cite[e.g.,][and references therein]{Banerji2015_rQSOs,Glikman2024_rQSOs}.
Moreover, our sample offers the opportunity to cross-match with previous spectroscopy of some (active) galaxies, in search of extremely variable and/or ``changing-look'' AGNs---including sources previously observed by southern-hemisphere surveys (e.g., 6dF, \citealt{6dF}; or GAMA, \citealt{GAMA}).

Thanks to the ongoing SDSS-V survey, we now have the capability to significantly expand the scope of quasar studies in the southern hemisphere, with both existing facilities (e.g., VLT, ALMA, and the Vera C. Rubin Observatory) and those still under construction (e.g., ELT, 4MOST, Roman). The large SDSS-V enabled samples we presented, constructed through robust selection methods, would join the large core eROSITA-based effort within SDSS-V (SPIDERS; Merloni \et, in prep.) to provide a valuable legacy dataset for future southern-hemisphere and time-domain surveys. 
The markedly high efficiency of the quasar selection methods we use can also be useful for future efforts to construct particular quasar samples, e.g., sources observed through the Galactic plane, in densely-monitored fields, etc.
We therefore hope that our exploratory study, and the GUA and \skt\ samples, would support many future studies of AGNs, and ultimately help expand our understanding of the SMBHs that power them, the galaxies that host them, and their cosmic evolution. 

%%%%%%%%%%%%%%%%%%%%%%%%%%%%%%%%%%%%%%%%%%%%%%%%%%%%%%%%%%%%%%%%
\section*{Data Availability}
The spectroscopic quasar sample analyzed in this work is a subset of the SDSS-V/DR20 spectroscopic dataset, which can be accessed through the SDSS-V/DR20 data release page at \url{https://www.sdss.org/}. The BOSS pipeline data products, including the \texttt{spAll} files used here, are distributed as part of DR20. The code used for spectral analysis in this work can be found at \url{https://github.com/QiaoyaWu/sdss4_dr16q_tutorial}.

\acknowledgments

A.S., B.T., and G.Z.\ acknowledge support from the European Research Council (ERC) under the European Union’s Horizon 2020 research and innovation program (grant agreement No. 950533) and the Israel Science Foundation (grant No. 1849/19).
R.J.A.\ was supported by FONDECYT grant number 1231718 and by the ANID BASAL project FB210003.
F.E.B.\ acknowledges support from ANID-Chile BASAL CATA FB210003 and FONDECYT Regular 1241005.
N.W.B.\ acknowledges support from NSF grant number AST-2407089.
J.G.F-T gratefully acknowledges the support provided by ANID Fondecyt Regular No. 1260371, ANID Fondecyt Postdoc No. 3230001 (Sponsoring researcher), the Joint Committee ESO-Government of Chile under the agreement 2023 ORP 062/2023 and the support of the Doctoral Program in Artificial Intelligence, DISC-UCN.
This research was supported by the Excellence Cluster ORIGINS and by the Munich Institute for Astro-, Particle and BioPhysics (MIAPbP), which are funded by the Deutsche Forschungsgemeinschaft (DFG, German Research Foundation) under Germany's Excellence Strategy - EXC 2094 - 390783311.
B.T.\ acknowledges the hospitality of the Instituto de Estudios Astrof\'isicos at Universidad Diego Portales, the Instituto de Astrof\'isica at Pontificia Universidad Cat\'olica de Chile, and the Institut d'Astrophysique de Paris, where parts of this study have been carried out.

Funding for the Sloan Digital Sky Survey V has been provided by the Alfred P. Sloan Foundation, the Heising-Simons Foundation, the National Science Foundation, and the Participating Institutions. SDSS acknowledges support and resources from the Center for High-Performance Computing at the University of Utah. SDSS telescopes are located at Apache Point Observatory, funded by the Astrophysical Research Consortium and operated by New Mexico State University, and at Las Campanas Observatory, operated by the Carnegie Institution for Science. The SDSS website is \url{www.sdss.org}.

SDSS is managed by the Astrophysical Research Consortium for the Participating Institutions of the SDSS Collaboration, including the Carnegie Institution for Science, Chilean National Time Allocation Committee (CNTAC) ratified researchers, Caltech, the Gotham Participation Group, Harvard University, Heidelberg University, The Flatiron Institute, The Johns Hopkins University, L'Ecole polytechnique f\'{e}d\'{e}rale de Lausanne (EPFL), Leibniz-Institut f\"{u}r Astrophysik Potsdam (AIP), Max-Planck-Institut f\"{u}r Astronomie (MPIA Heidelberg), Max-Planck-Institut f\"{u}r Extraterrestrische Physik (MPE), Nanjing University, National Astronomical Observatories of China (NAOC), New Mexico State University, The Ohio State University, Pennsylvania State University, Smithsonian Astrophysical Observatory, Space Telescope Science Institute (STScI), the Stellar Astrophysics Participation Group, Universidad Nacional Aut\'{o}noma de M\'{e}xico, University of Arizona, University of Colorado Boulder, University of Illinois at Urbana-Champaign, University of Toronto, University of Utah, University of Virginia, Yale University, and Yunnan University.

This research uses services or data provided by the Astro Data Lab, which is part of the Community Science and Data Center (CSDC) Program of NSF NOIRLab. NOIRLab is operated by the Association of Universities for Research in Astronomy (AURA), Inc. under a cooperative agreement with the U.S. National Science Foundation.

\software{astropy \citep{astropy}, PyQSOFit \citep{pyqsofitOG}}

\clearpage
\appendix

\section{Overlapping quasar/AGN selection methods and cartons}
\label{app:overlap}

As explained in Section~\ref{sec:SDSS_other}, we must consider the significant overlap between the various quasar/AGN selection methods under study, particularly when assessing the purity (or efficiency) of the GUA and \skt\ quasar selection methods.

In Figure~\ref{fig:upset} we show the \texttt{upset} diagrams \citep{upset}\footnote{\url{https://upset.app/}} that illustrate the overlap between key types of selection criteria relevant for our sample construction. Specifically, we split targets (top panel), observed spectra (middle) and quasars (bottom) by whether they're part of the cartons associated with the \papgua-based GUA selection; the \papdes-based ASQOSS/\skt\ selection; X-ray based  (SPIDERS+CSC) selection; and/or the two core BHM programs focusing on previously-known quasars (AQMES and BHM-RM). 
This grouping is meant to highlight different kinds of prior knowledge about the (potential) AGN nature of the sources.
The overlap between all these selection criteria, and specifically between GUA, \skt, and the other (higher-priority) BHM programs, is clearly evident.

In our analysis, however, we focus on a different grouping of sources, more closely linked to the the priority that each target had when considered for observations. Specifically, we consider whether sources are part of GUA, of \skt, and/or the superset of targets selected by any of the core BHM programs (SPIDERS, AQMES, and/or BHM-RM) combined with targets drawn from the CSC.
Sources targeted within this latter superset not only have higher observational priority within SDSS-V survey planning and operations, but carry additional information about the quasar/AGN nature of the target (see Section~\ref{sec:SDSS_other}). 
These three main sets of targets are illustrated in the Venn diagram shown as Figure~\ref{fig:venn}, which further denotes the seven disjoint subsets representing the various possibilities for overlap (or lack thereof) between the sets ($A, B,\dots, G$). 

To assess the purity of the GUA and \skt\ selection methods, we calculate for each of them a {\it homogenized purity} measure (\hp\ hereafter), calculated within a sky area where all three selection paths were applicable relevant. The quantity \hp\ is the weighted mean of the quasar selection purities of each of the various disjoint subsets relevant for each sample, with the weights representing the fraction among {\it targets} (not sources) selected through the particular selection path that belong to that subset.

\begin{figure}[ht]
    \centering
    \includegraphics[width=0.7\textwidth]{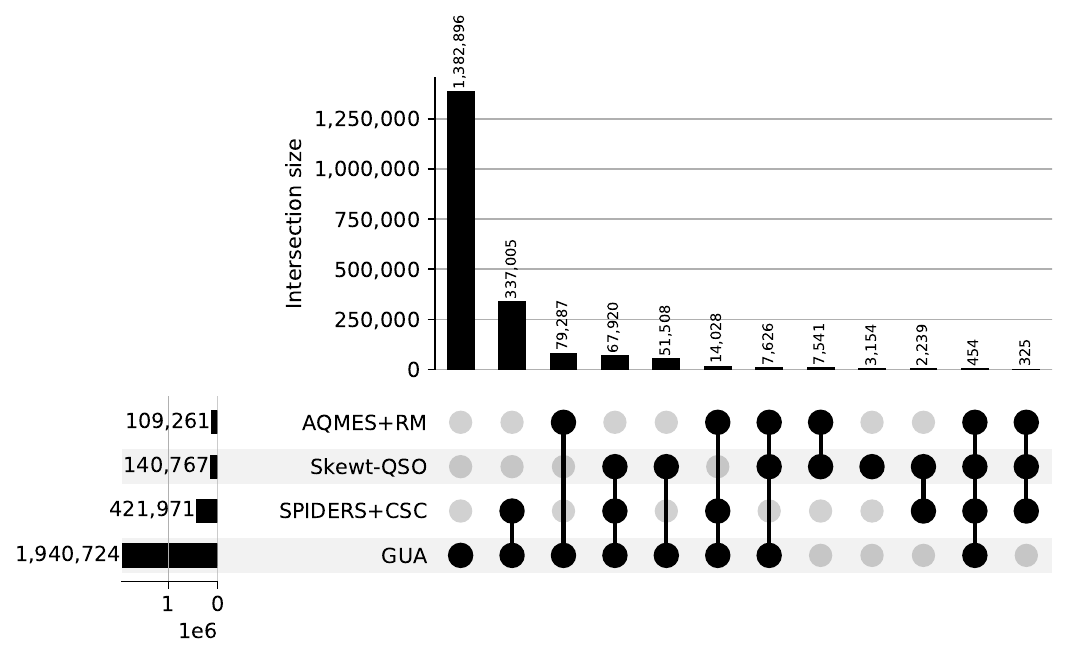}\\
    \includegraphics[width=0.7\textwidth]{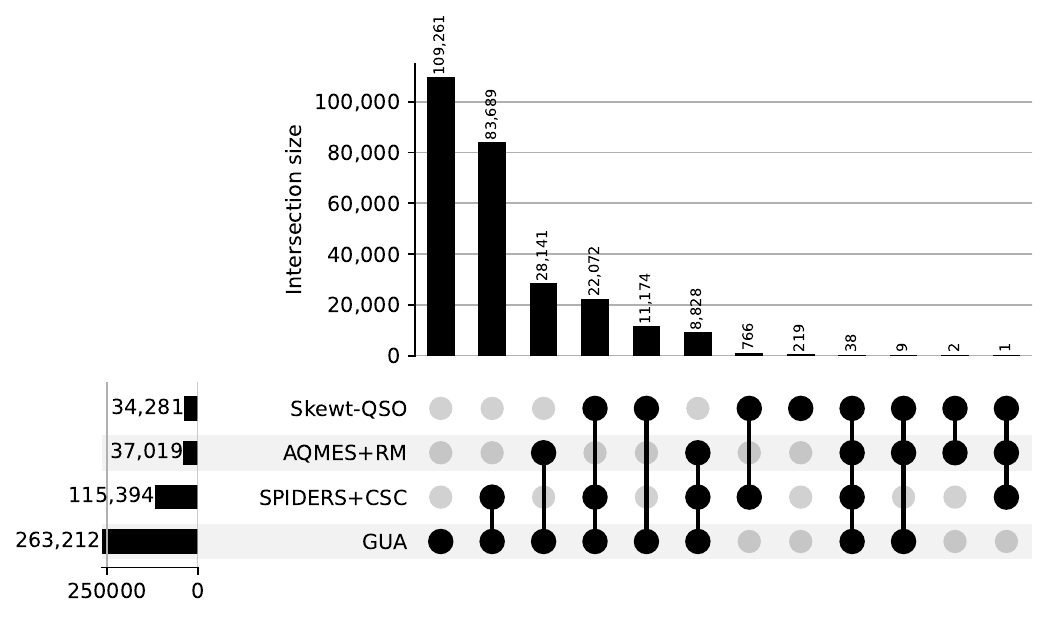}\\
    \includegraphics[width=0.7\textwidth]{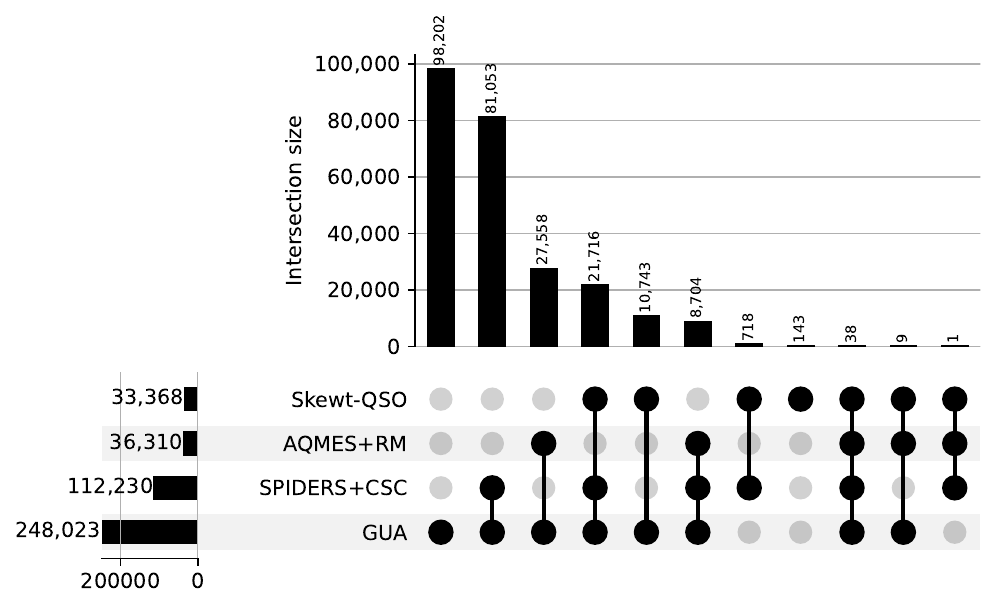}
    \caption{\texttt{Upset} diagrams illustrating the overlap between the GUA, \skt, and other main other higher priority BHM selection methods, grouped by the kind of prior knowledge they hold about the AGN nature of the sources. The top, middle and bottom panels show the overlap among targets (cartons), the observed sources (within DR20), and the spectroscopically confirmed quasars.}
    \label{fig:upset}
\end{figure}

\begin{figure*}[ht]
    \centering
    \includegraphics[width=0.48\columnwidth]{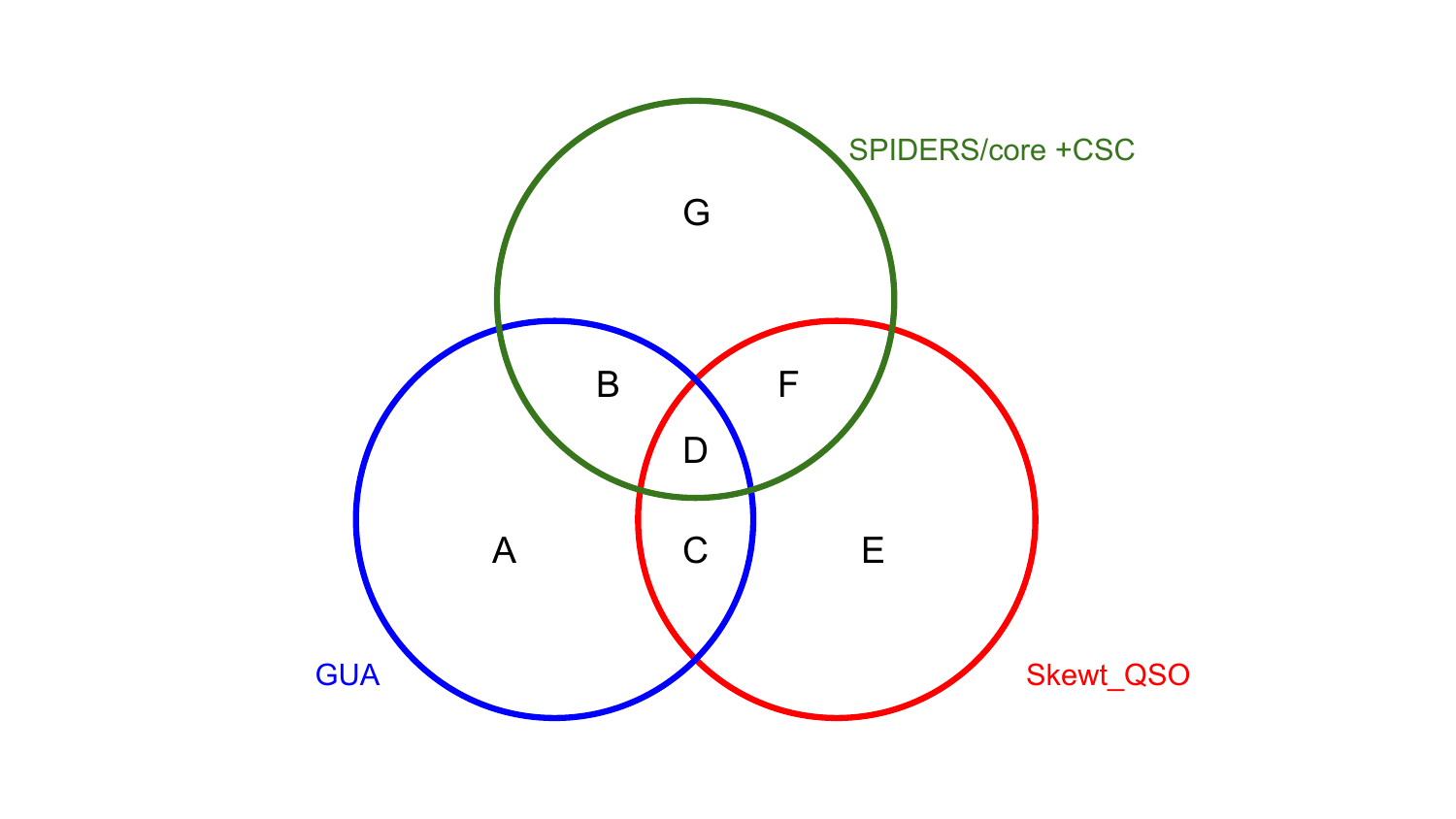}
     \includegraphics[width=0.48\columnwidth]{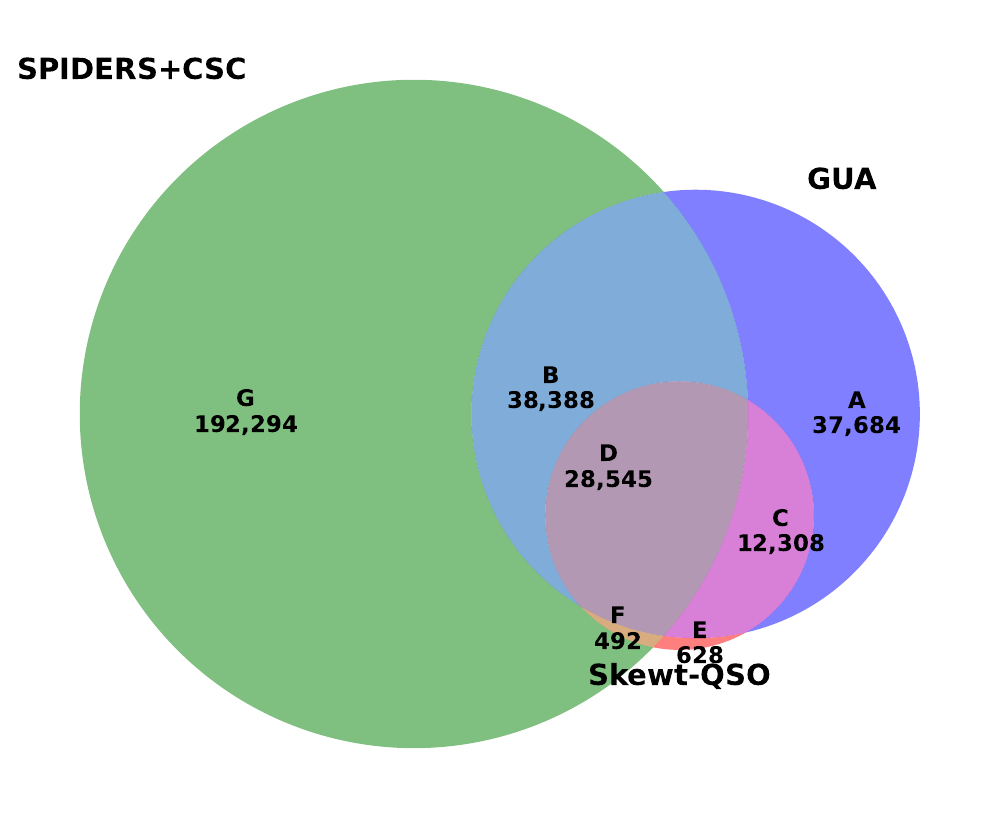}
    \caption{Venn diagrams presenting the selection and sample overlap considered in this work. {\it Left:} a conceptual illustration of the the seven disjoint subsets considered in our analysis of newly detected quasars (Section~\ref{sec:final_sample} and of selection efficiency (purity; this Appendix and Section~\ref{sec:purity}). {\it Right:} The same subsets, but now illustrating the actual degree of overlap between the various selection methods within the southern, high-Galactic-latitude ``test area''.}
    \label{fig:venn}
\end{figure*}

Specifically, we calculated \hp\ for GUA and \skt\ by first focusing on all the {\it targets} the corresponding cartons hold in the region enclosed with $2\,{\rm h} < \alpha < 5\,{\rm h}\, 20\,{\rm m}$ and $-60^\circ < \delta < -20^\circ$ (the ``test area'' hereafter).
Then, we calculate the fraction of targets selected by each method corresponding to each of the combinatorial subsets denoted in Fig.~\ref{fig:venn}. For example, $f_A$ is the fraction of GUA targets that were not selected by neither SPIDERS+CSC nor \skt, among all GUA targets in the test area. Likewise, $f_B$ is the fraction corresponding to targets selected by both GUA and SPIDERS+CSC, but not by \skt. 
In addition, $f_{C, \rm G}$ corresponds to sources selected by both GUA and \skt, but not by SPIDERS+CSC, while $f_{D, \rm G}$ corresponds to sources selected by all three approach (GUA, \skt, and SPIDERS+CSC).
The subscript `G' in both $f_{C, \rm G}$ and $f_{D, \rm G}$ stems from the fact that these fractions are calculated among all GUA targets in the test area. 
The corresponding four fractions for \skt\ would be $f_E$, $f_F$, and then $f_{C, \rm s}$ and $f_{D, \rm s}$, where the subscript `s' for \skt).
We note that the subset $G$, corresponding to sources selected only by SPIDERS+CSC, but not by GUA or \skt, is not used in our analysis analysis.
All the fractions, and the numbers used to calculate them, are listed in Table~\ref{tab:hp_calc}.

Next, we turn our attention to the {\it observed sources} corresponding to each of the logical selection subsets (${A, B,\dots}$), and again restricting ourselves to the test area. We calculate the quasar selection purity within each such subset as the fraction of quasars among all useful spectra (both \gd\ and \bd) of sources whose targeting information places them in the subset. Of the spectra obtained within this area, $\sim$85\% belong to the \gd\ subset and $\sim$15\% to the \bd\ subset.
For example, the quasar purity of subset $A$, $p_A$, is the fraction of quasars among all the spectra targeting sources that were flagged by the GUA method, but not by the \skt\ or SPIDERS+CSC targeting at the test area.
The quasar selection purities within each subset are also listed in Table~\ref{tab:hp_calc}.

Finally, the homogenized purity of GUA is calculated as:
\begin{equation}
    \hp_{\rm GUA} = (f_A\,p_A)+(f_B\,p_B)+(f_{C,\rm G}\, p_C) + (f_{D,\rm G}\, p_D) \,\, ,
    \label{eq:hp_gua}
\end{equation}
and that of \skt\ is calculated as:
\begin{equation}
    \hp_{\rm skewt} = (f_E\,p_E)+(f_F\,p_F)+(f_{C,\rm s}\, p_C) + (f_{D,\rm s}\, p_D).
    \label{eq:hp_skt}
\end{equation}

In Section~\ref{sec:purity} we also discuss trends of purity with source brightness (magnitude) and with Galactic latitude. In both cases, this entails repeating the entire \hp\ calculation presented above but restricted to targets (for fractions $f$) and spectra (for purities $p$) that belong to certain bin in either magnitude or Galactic latitude.

\begin{deluxetable}{|c|c|c|c|c|c|c|c|c|c|}
\tablecaption{Summary of the homogenized purity in the test area}
\label{tab:hp_calc}
\tablewidth{\textwidth}
\tabletypesize{\scriptsize}
\tablehead{
\colhead{Subset} & \colhead{GUA} & \colhead{\skt} & \colhead{SPIDERS+CSC} & 
\colhead{$N_{\rm targ,area}$} & \colhead{$f_{\rm GUA}$} & \colhead{$f_{\rm \skt}$} & 
\colhead{$N_{\rm obs,area}$} & \colhead{$N_{\rm obs-QSO,area}$} & \colhead{Purity} 
}
\startdata
A    & V & X & X & 37,684  &  0.322   & \nodata & 7,889   & 7,323   & 0.928   \\
B    & V & X & V & 38,388  &  0.328   & \nodata & 12,765  & 12,463  & 0.976   \\
C    & V & V & X & 12,308  &  0.105   & 0.293   & 3,035   & 2,943   & 0.970   \\
D    & V & V & V & 28,545  &  0.244   & 0.680   & 9,975   & 9,826   & 0.985   \\
E    & X & V & X & 628     & \nodata  & 0.015   & 22      & 13      & 0.591   \\
F    & X & V & V & 492     & \nodata  & 0.012   & 144     & 130     & 0.903   \\
G    & X & X & V & 192,294 & \nodata  & \nodata & \nodata & \nodata & \nodata \\
\enddata
\tablecomments{Columns 1--3 indicate whether a given subset is selected by GUA, \skt, and/or SPIDERS+CSC. $N_{\rm targ,area}$ is the number of targets in the subset within the test area, while $f_{\rm GUA}$ and $f_{\rm \skt}$ are the corresponding fractions among all GUA ($N_{\rm GUA}= 116,925$) or \skt\ ($N_{\rm \skt}= 41,973$) targets, respectively. $N_{\rm obs,area}$ is the number of observed sources with spectra in the subset, and $N_{\rm obs-QSO,area}$ is the number of quasars among them. The resulting homogenized purities are $hp_{\rm GUA}\simeq96\%$ and $hp_{\rm \skt}\simeq97\%$ (see Fig.~\ref{fig:bar}).
}
\end{deluxetable}

\section{ZWARNING Description} \label{app:ZWARNING}
Here, we present the \zwar\ flags, which indicate issues in the pipeline redshift fitting process. These flags are assigned by the SDSS pipeline to identify spectra where redshift determination may be unreliable due to various factors. The \zwar\ parameter is encoded as a bitmask integer, where different bits correspond to different warning conditions.\footnote{\url{https://www.sdss4.org/dr17/algorithms/bitmasks/\#ZWARNING}} Understanding these flags is essential for distinguishing between spectra with catastrophic errors that cannot be used for analysis and those that remain useful for quasar identification but have uncertain redshifts. The accompanying table provides a detailed list of \zwar\ flags and descriptions of the specific issues they signify.

\begin{deluxetable}{lll}
\label{tab:zwarning}
\tablecaption{Description of ZWARNING flags and their associated sample classification}
\tablewidth{\textwidth}
\tablehead{
\colhead{ZWARNING} & \colhead{Description} & \colhead{Sample}
}
\startdata
SKY                & Sky fiber & Excluded \\
LITTLE\_COVERAGE   & Too little wavelength coverage & Bad \\
SMALL\_DELTA\_CHI2 & $\chi^2$ of the best fit is too close to that of the second-best & Bad \\
NEGATIVE\_MODEL    & Synthetic spectrum is negative & Bad \\
MANY\_OUTLIERS     & Fraction of points more than $5\sigma$ away from the best model is too large & Bad \\
Z\_FITLIMIT        & $\chi^2$ minimum at the edge of the redshift fitting range (Z\_ERR set to -1) & Bad \\
NEGATIVE\_EMISSION & A QSO line exhibits negative emission. & \\
                   & Triggered only in QSO spectra if \civ, \ciii, \mgii, \hb, & \\
                   & or if \ha\ has LINEAREA + 3 $\times$ LINEAREA\_ERR $< 0$ & Bad \\
UNPLUGGED          & The fiber was unplugged or damaged; location of the spectrum is unknown & Excluded \\
BAD\_TARGET        & Catastrophically bad targeting data & Excluded \\
NODATA             & No data for this fiber (e.g., spectrograph was broken during the exposure) & Excluded \\
\enddata
\tablecomments{Not all of the quality flags listed here were actually identified in the internal SDSS-V spectral data, and not all are expected to be found in the public SDSS-V data products.}
\end{deluxetable}

\section{Line Configuration in PyQSOFit Analysis} \label{app:config}
In this appendix, we present the line configuration used in the PyQSOFit analysis, which follows a similar approach to that described in \citet{DR16WuShen}. This configuration defines how different emission line complexes are modeled within the spectral fitting process.

The selected line complexes include \ha, \hb, \mgii, C $\textsc{iii}$], \civ, Si $\textsc{iv}$, and \Lya. For each line complex, the algorithm fits both broad and narrow emission lines, ensuring that the components are properly constrained. Broad emission lines are modeled using a combination of 1 to 3 Gaussian components to capture their complex structure, while narrow emission lines are fit using a single Gaussian.

\begin{deluxetable}{ccc}
\tablecaption{\pqf\ configuration for emission line decomposition}
\label{tab:complex_table}
\tablewidth{\textwidth}
\tablehead{
\colhead{Complex Name} & \colhead{Lines} & \colhead{\# Gaussians}
}
\startdata
\ha     & broad \ha                                               & 3 \\
\ha     & narrow \ha                                              & 1 \\
\ha     & [N\,\textsc{ii}]\,$\lambda6548$                           & 1 \\
\ha     & [N\,\textsc{ii}]\,$\lambda6583$                           & 1 \\
\ha     & [S\,\textsc{ii}]\,$\lambda6716$                           & 1 \\
\ha     & [S\,\textsc{ii}]\,$\lambda6731$                           & 1 \\
\hb     & broad \hb                                               & 3 \\
\hb     & narrow \hb                                              & 1 \\
\hb     & [O\,\textsc{iii}]\,$\lambda4959$                          & 1 \\
\hb     & [O\,\textsc{iii}]\,$\lambda5007$                          & 1 \\
\hb     & broad He\,\textsc{ii}\,$\lambda4686$                    & 1 \\
\hb     & narrow He\,\textsc{ii}\,$\lambda4686$                   & 1 \\
\MgII   & broad \MgII                                             & 2 \\
\MgII   & narrow \MgII                                            & 1 \\
\CIII   & broad \CIII                                             & 2 \\
\CIII   & Si\,\textsc{iii}]\,$\lambda1892$                         & 1 \\
\CIII   & Al\,\textsc{iii}\,$\lambda1857$                         & 1 \\
\CIII   & Si\,\textsc{ii}\,$\lambda1816$                          & 1 \\
\CIII   & N\,\textsc{iii}]\,$\lambda1750$                          & 1 \\
\CIII   & N\,\textsc{iv}]\,$\lambda1716$                           & 1 \\
\CIV    & broad \CIV                                              & 3 \\
\CIV    & narrow He\,\textsc{ii}\,$\lambda1640$                   & 1 \\
\CIV    & broad He\,\textsc{ii}\,$\lambda1640$                    & 1 \\
\CIV    & narrow O\,\textsc{iii}]\,$\lambda1663$                   & 1 \\
\CIV    & broad O\,\textsc{iii}]\,$\lambda1663$                    & 1 \\
Si\,\textsc{iv} & Si\,\textsc{iv}\,$\lambda1403$                  & 1 \\
Si\,\textsc{iv} & Si\,\textsc{iv}\,$\lambda1394$                  & 1 \\
Si\,\textsc{iv} & C\,\textsc{ii}\,$\lambda1335$                   & 1 \\
Si\,\textsc{iv} & O\,\textsc{i}\,$\lambda1304$                    & 1 \\
\Lya    & broad \Lya                                              & 3 \\
\Lya    & N\,\textsc{v}\,$\lambda1240$                            & 1 \\
\enddata
\tablecomments{Line configuration used in PyQSOFit, including complex names, individual emission lines, and the number of Gaussian components fitted for each line.}
\end{deluxetable}

\clearpage
\section{Matching Method} \label{apx:matching example}
This appendix presents an example of the matching process between a test and reference sample, specifically demonstrating the comparison between GUA and DR16Q, as explained in Section~\ref{sec:M_L_dist}. Using the highest ratio (0.97 in this example) and all the green bins, a reference subsample was created to ensure a more statistically meaningful comparison between the two distributions. 

\begin{figure}[h!]
    \centering
    \includegraphics[width=\textwidth]{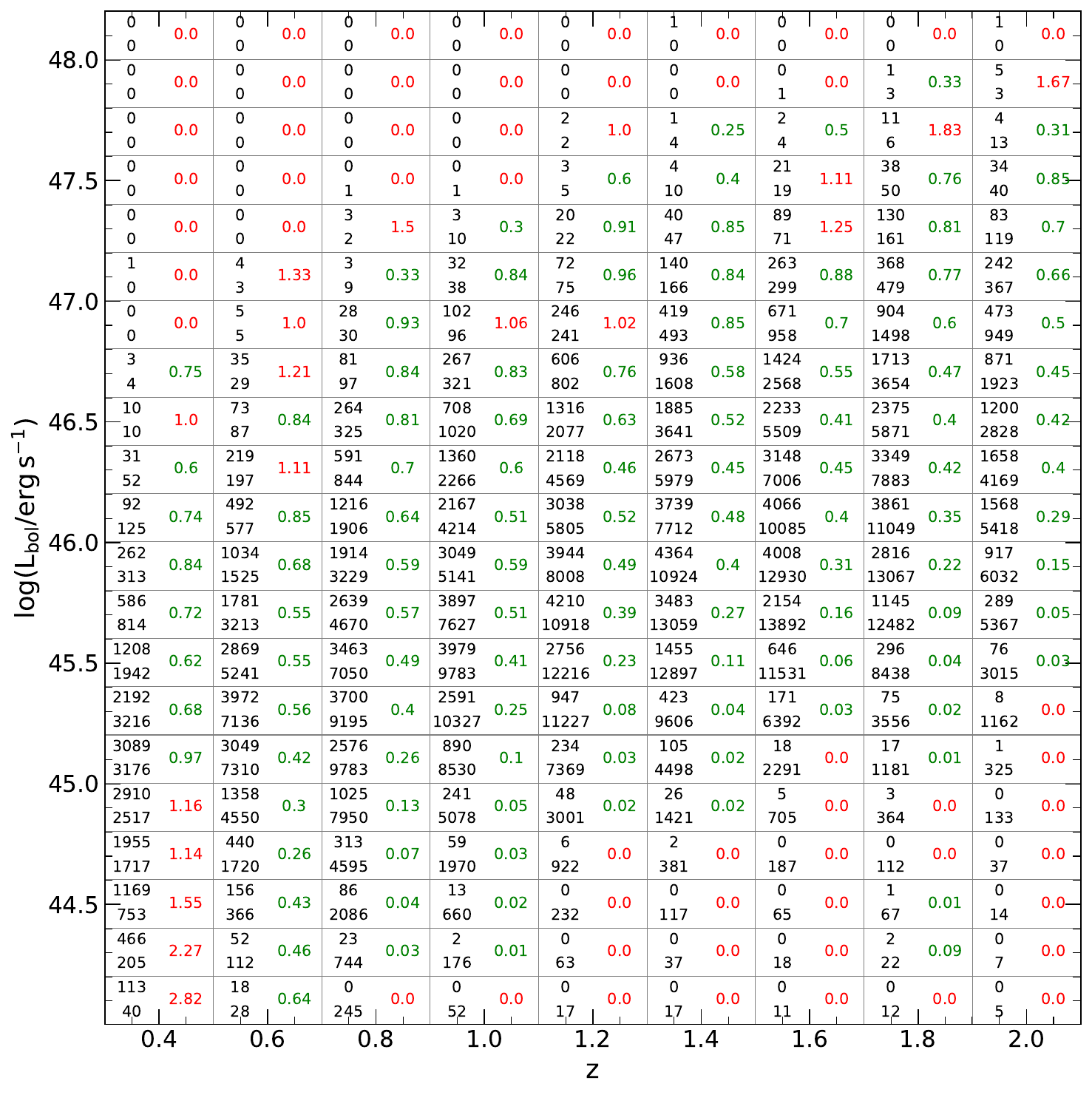}
    \caption{An example of the matching method, between GUA and DR16Q. Each bin in the \(\log \Lbol - z\) space represents the number of quasars in the test sample (top) and the reference sample (bottom). The ratio between the two samples is displayed within each bin, where bins with ratios above 1\% are shaded green, while those below are red.}
    \label{fig:match}
\end{figure}

\begin{figure*}
    \centering
    \includegraphics[width=\textwidth]{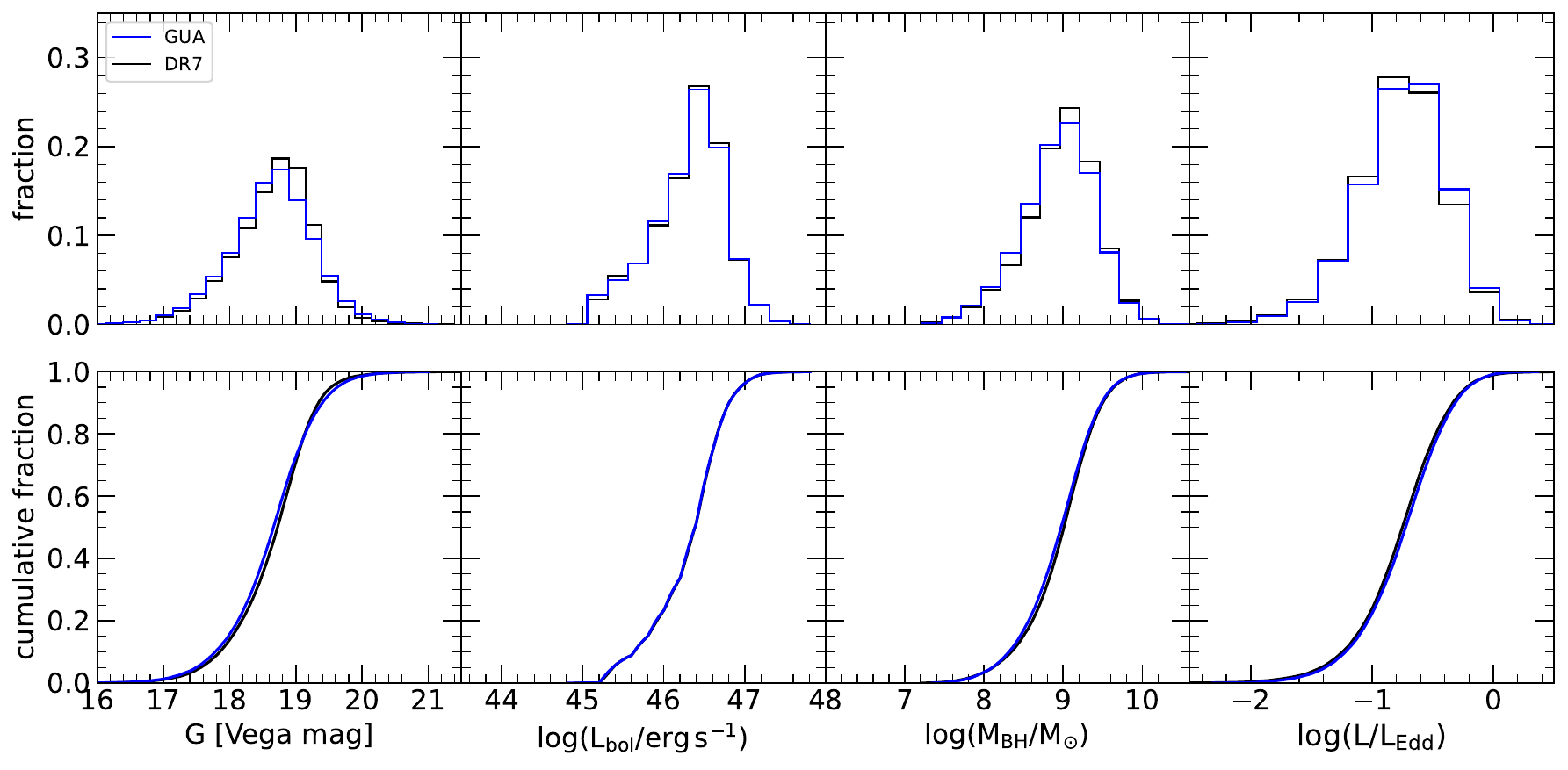}
    \caption{Similar to Fig.~\ref{fig:match_GUA}, but presenting the comparison between the matched GUA/`All' and SDSS/DR7Q samples.}
    \label{fig:match_GUA_All_DR7}
\end{figure*}

\begin{figure*}
    \centering
    \includegraphics[width=\textwidth]{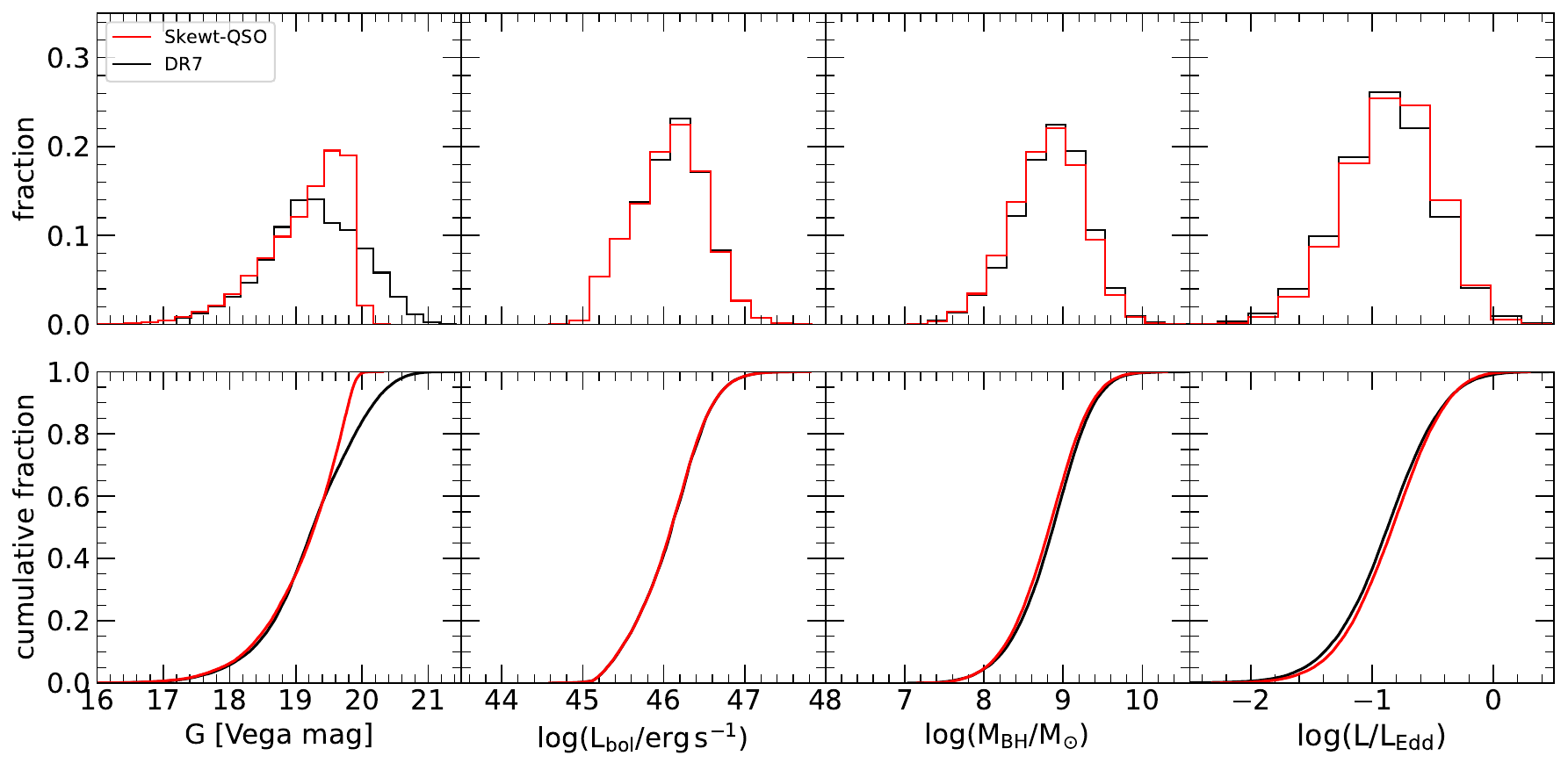}
    \caption{Similar to Fig.~\ref{fig:match_GUA}, but presenting the comparison between the matched \skt/`All' and SDSS/DR7Q samples.}
    \label{fig:match_skt_All_DR7}
\end{figure*}

\clearpage
\bibliographystyle{aasjournal}
\bibliography{gua_asqoss_bib}{}

\end{document}